\documentclass[twocolumn, astrosymb]{aastex631} % , linenumbers

\usepackage{graphicx}
\usepackage{amsmath}
\usepackage{amssymb}
\usepackage{dirtytalk}
\usepackage{threeparttable}
\usepackage{makecell}
\usepackage{caption}

\usepackage{savesym}
\savesymbol{tablenum}
\usepackage[binary-units]{siunitx}
\restoresymbol{SIX}{tablenum}

\shorttitle{Discovery of an Extreme Galaxy Overdensity at $z = 5.4$}
\shortauthors{Helton et al.}

\graphicspath{{./}{figures/}}

\begin{document}

\title{The JWST Advanced Deep Extragalactic Survey: \\ Discovery of an Extreme Galaxy Overdensity at $z = 5.4$ with JWST/NIRCam in GOODS-S}

%% Authors of the paper
\author[0000-0003-4337-6211]{Jakob M. Helton}
\affiliation{Steward Observatory, University of Arizona, 933 N. Cherry Ave., Tucson, AZ 85721, USA}

\author[0000-0002-4622-6617]{Fengwu Sun}
\affiliation{Steward Observatory, University of Arizona, 933 N. Cherry Ave., Tucson, AZ 85721, USA}

\author[0000-0001-5962-7260]{Charity Woodrum}
\affiliation{Steward Observatory, University of Arizona, 933 N. Cherry Ave., Tucson, AZ 85721, USA}

\author[0000-0003-4565-8239]{Kevin N. Hainline}
\affiliation{Steward Observatory, University of Arizona, 933 N. Cherry Ave., Tucson, AZ 85721, USA}

\author[0000-0001-9262-9997]{Christopher N. A. Willmer}
\affiliation{Steward Observatory, University of Arizona, 933 N. Cherry Ave., Tucson, AZ 85721, USA}

\author[0000-0002-7893-6170]{George H. Rieke}
\affiliation{Steward Observatory, University of Arizona, 933 N. Cherry Ave., Tucson, AZ 85721, USA}

\author[0000-0002-7893-6170]{Marcia J. Rieke}
\affiliation{Steward Observatory, University of Arizona, 933 N. Cherry Ave., Tucson, AZ 85721, USA}

\author[0000-0002-8224-4505]{Sandro Tacchella}
\affiliation{Kavli Institute for Cosmology, University of Cambridge, Madingley Road, Cambridge, CB3 OHA, UK}
\affiliation{Cavendish Laboratory, University of Cambridge, 19 JJ Thomson Avenue, Cambridge CB3 0HE, UK}

\author[0000-0002-4271-0364]{Brant Robertson}
\affiliation{Department of Astronomy and Astrophysics, University of California, Santa Cruz, 1156 High Street, Santa Cruz, CA 95064, USA}

\author[0000-0002-9280-7594]{Benjamin D. Johnson}
\affiliation{Center for Astrophysics $|$ Harvard \& Smithsonian, 60 Garden St., Cambridge MA 02138, USA}

\author[0000-0002-8909-8782]{Stacey Alberts}
\affiliation{Steward Observatory, University of Arizona, 933 N. Cherry Ave., Tucson, AZ 85721, USA}

\author[0000-0002-2929-3121]{Daniel J. Eisenstein}
\affiliation{Center for Astrophysics $|$ Harvard \& Smithsonian, 60 Garden St., Cambridge MA 02138, USA}

\author[0000-0002-8543-761X]{Ryan Hausen}
\affiliation{Department of Physics and Astronomy, The Johns Hopkins University, 3400 N. Charles St.,
Baltimore, MD 21218, USA}

\author[0000-0001-8470-7094]{Nina R. Bonaventura}
\affiliation{Steward Observatory, University of Arizona, 933 N. Cherry Avenue, Tucson, AZ 85721, USA}

\author[0000-0002-8651-9879]{Andrew Bunker}
\affiliation{Department of Physics, University of Oxford, Denys Wilkinson Building, Keble Road, Oxford OX1 3RH, UK}

\author[0000-0003-3458-2275]{Stephane Charlot}
\affiliation{Sorbonne Universit\'e, CNRS, UMR 7095, Institut d'Astrophysique de Paris, 98 bis bd Arago, 75014 Paris, France}

\author[0000-0002-2678-2560]{Mirko Curti}
\affiliation{Kavli Institute for Cosmology, University of Cambridge, Madingley Road, Cambridge, CB3 OHA, UK}
\affiliation{Cavendish Laboratory, University of Cambridge, 19 JJ Thomson Avenue, Cambridge CB3 0HE, UK}
\affiliation{European Southern Observatory, Karl-Schwarzschild-Strasse 2, D-85748 Garching bei Muenchen, Germany}

\author[0000-0002-9551-0534]{Emma Curtis-Lake}
\affiliation{Centre for Astrophysics Research, Department of Physics, Astronomy and Mathematics, University of Hertfordshire, Hatfield AL10 9AB, UK}

\author[0000-0002-3642-2446]{Tobias J. Looser}
\affiliation{Kavli Institute for Cosmology, University of Cambridge, Madingley Road, Cambridge, CB3 OHA, UK}
\affiliation{Cavendish Laboratory, University of Cambridge, 19 JJ Thomson Avenue, Cambridge CB3 0HE, UK}

\author[0000-0002-4985-3819]{Roberto Maiolino}
\affiliation{Kavli Institute for Cosmology, University of Cambridge, Madingley Road, Cambridge, CB3 OHA, UK}
\affiliation{Cavendish Laboratory, University of Cambridge, 19 JJ Thomson Avenue, Cambridge CB3 0HE, UK}
\affiliation{Department of Physics and Astronomy, University College London, Gower Street, London WC1E 6BT, UK}

\author[0000-0002-4201-7367]{Chris Willott}
\affiliation{NRC Herzberg, 5071 West Saanich Rd, Victoria, BC V9E 2E7, Canada}

\author[0000-0002-7595-121X]{Joris Witstok}
\affiliation{Kavli Institute for Cosmology, University of Cambridge, Madingley Road, Cambridge, CB3 OHA, UK}
\affiliation{Cavendish Laboratory, University of Cambridge, 19 JJ Thomson Avenue, Cambridge CB3 0HE, UK}

\author[0000-0003-4109-304X]{Kristan Boyett}
\affiliation{School of Physics, University of Melbourne, Parkville 3010, VIC, Australia}
\affiliation{ARC Centre of Excellence for All Sky Astrophysics in 3 Dimensions (ASTRO 3D), Australia}

\author[0000-0002-2178-5471]{Zuyi Chen}
\affiliation{Steward Observatory, University of Arizona, 933 N. Cherry Ave., Tucson, AZ 85721, USA}

\author[0000-0003-1344-9475]{Eiichi Egami}
\affiliation{Steward Observatory, University of Arizona, 933 N. Cherry Ave., Tucson, AZ 85721, USA}

\author[0000-0003-4564-2771]{Ryan Endsley}
\affiliation{Department of Astronomy, University of Texas at Austin, 2515 Speedway Blvd Stop C1400, Austin, TX 78712, USA}

\author[0000-0002-4684-9005]{Raphael E. Hviding}
\affiliation{Steward Observatory, University of Arizona, 933 N. Cherry Ave., Tucson, AZ 85721, USA}

\author[0000-0003-3577-3540]{Daniel T. Jaffe}
\affiliation{Department of Astronomy, University of Texas at Austin, 2515 Speedway Blvd Stop C1400, Austin, TX 78712, USA}

\author[0000-0001-7673-2257]{Zhiyuan Ji}
\affiliation{Steward Observatory, University of Arizona, 933 N. Cherry Ave., Tucson, AZ 85721, USA}

\author[0000-0002-6221-1829]{Jianwei Lyu}
\affiliation{Steward Observatory, University of Arizona, 933 N. Cherry Ave., Tucson, AZ 85721, USA}

\author[0000-0001-9276-7062]{Lester Sandles}
\affiliation{Kavli Institute for Cosmology, University of Cambridge, Madingley Road, Cambridge, CB3 OHA, UK}
\affiliation{Cavendish Laboratory, University of Cambridge, 19 JJ Thomson Avenue, Cambridge CB3 0HE, UK}

\correspondingauthor{Jakob M. Helton}
\email{jakobhelton@arizona.edu}

%% Abstract of the paper.

\begin{abstract}
We report the discovery of an extreme galaxy overdensity at $z = 5.4$ in the GOODS-S field using JWST/NIRCam imaging from JADES and JEMS alongside JWST/NIRCam wide field slitless spectroscopy from FRESCO. We identified potential members of the overdensity using HST+JWST photometry spanning $\lambda = 0.4-5.0\ \mu\mathrm{m}$. These data provide accurate and well-constrained photometric redshifts down to $m \approx 29-30\,\mathrm{mag}$. We subsequently confirmed $N = 81$ galaxies at $5.2 < z < 5.5$ using JWST slitless spectroscopy over $\lambda = 3.9-5.0\ \mu\mathrm{m}$ through a targeted line search for $\mathrm{H} \alpha$ around the best-fit photometric redshift. We verified that $N = 42$ of these galaxies reside in the field while $N = 39$ galaxies reside in a density around $\sim 10$ times that of a random volume. Stellar populations for these galaxies were inferred from the photometry and used to construct the star-forming main sequence, where protocluster members appeared more massive and exhibited earlier star formation (and thus older stellar populations) when compared to their field galaxy counterparts. We estimate the total halo mass of this large-scale structure to be $12.6 \lesssim \mathrm{log}_{10} \left( M_{\mathrm{halo}}/M_{\odot} \right) \lesssim 12.8$ using an empirical stellar mass to halo mass relation, which is likely an underestimate as a result of incompleteness. Our discovery demonstrates the power of JWST at constraining dark matter halo assembly and galaxy formation at very early cosmic times.

\end{abstract}

%% Defines the keywords.

%% Start of section one.
\section{Introduction}
\label{SectionOne}

In the local Universe, galaxy clusters represent the largest and most massive gravitationally bound structures, consisting of up to thousands of individual galaxies contained within a virialized or virializing dark matter halo, and representing the most extreme matter overdensities allowed by the standard cosmological paradigm of hierarchical structure formation \citep{White:1978}. In the early Universe, the structures that eventually evolved into the galaxy clusters seen today are referred to as ``protoclusters'', which consist of fewer individual galaxies contained within more complex dark matter halos that are yet to be virialized \citep[for a review of protoclusters, see][]{Overzier:2016}. 

Observations have suggested that the majority of the stellar mass in the local Universe resides in massive elliptical galaxies, which are preferentially found within galaxy clusters \citep{Dressler:1980}. Additionally, the average formation timescales for galaxies in clusters is shorter than that for analogous galaxies in the field \citep{Webb:2020}. These results suggest that the physical processes associated with extreme matter overdensities induce earlier star formation, earlier stellar mass assembly, and earlier quenching. However, massive clusters at relatively high redshift ($z = 1-2$) have also been observed to have large amounts of star formation, on par with field populations \citep{Alberts:2014, Alberts:2016, Alberts:2021}. Quantifying these effects across the protocluster to cluster boundary remains an important open problem for extragalactic astronomy \citep{Wang:2013}. 

The impact of environment on galaxy formation and evolution is best understood in the local Universe, where we can observe the effects of transformational processes such as dynamical relaxation, tidal interactions, and mergers \citep{Zabludoff:1996}. However, these processes make it difficult to ascertain important evidence (e.g., both the initial relative positions and velocities of the constituent galaxies) related to the early formation and evolution of the most massive gravitationally bound structures. For this reason, searching for protoclusters in the early Universe offers our best chance of understanding the initial formation and subsequent evolution of galaxy clusters today \citep[e.g.,][]{Li:2022, Brinch:2022, Morishita:2022}. 

In this paper, we present the discovery of an extreme galaxy overdensity at $z = 5.4$ in the GOODS-S field using data from the Near Infrared Camera \citep[NIRCam;][]{Rieke:2005, Rieke:2022} on JWST. The powerful combination of deep imaging and wide field slitless spectroscopy (WFSS) provided by JWST/NIRCam allows us to identify this overdensity, characterize the stellar populations of galaxies both inside and outside this large-scale structure, and estimate the dark matter halo mass associated with this protocluster. These observations provide important insights into the impact of environment on galaxy formation and evolution immediately after the epoch of reionization (EoR; $z > 6$) when the Universe was approximately a billion years old. 

This paper proceeds as follows. In Section~\ref{SectionTwo}, we describe the various data and observations that are used in our analysis, including the photometric redshift determination and emission line detection. In Section~\ref{SectionThree}, we present our analysis and results, including the stellar population modeling and halo mass inference. In Section~\ref{SectionFour}, we summarize our findings and their implications for galaxy evolution in the early Universe. All magnitudes are in the AB system \citep{Oke:1983}. \textcolor{black}{Uncertainties are quoted as 68\% confidence intervals.} Throughout this work, we report wavelengths in vacuum and adopt the standard flat $\Lambda$CDM cosmology from Planck18 with $H_{0} = 67.4\ \mathrm{km/s/Mpc}$ and $\Omega_{m} = 0.315$ \citep[][]{Planck:2020}. 

%% Start of section two.
\section{Data \& Observations}
\label{SectionTwo}

In this work, we use deep optical imaging from the Advanced Camera for Surveys (ACS) on HST alongside deep infrared imaging and WFSS from JWST/NIRCam in the Great Observatories Origins Deep Survey South \citep[GOODS-S;][]{Giavalisco:2004} field. The imaging data and photometry from HST/ACS and JWST/NIRCam are described in Section~\ref{SectionTwoOne}. The photometric redshifts and sample selection are described in Section~\ref{SectionTwoTwo}. The spectral data and line detection from JWST/NIRCam WFSS are described in Section~\ref{SectionTwoThree}. 

\subsection{Imaging Data \& Photometry}
\label{SectionTwoOne}

Our imaging data consist of: (1) deep optical imaging taken with HST/ACS in five photometric bands (F435W, F606W, F775W, F814W, and F850LP) and (2) deep infrared imaging taken with JWST/NIRCam in fourteen photometric bands (F090W, F115W, F150W, F182M, F200W, F210M, F277W, F335M, F356W, F410M, F430M, F444W, F460M, and F480M). 

The HST/ACS mosaics used here were produced as part of the Hubble Legacy Fields (HLF) project v2.0 and include observations covering a $25^{\prime} \times 25^{\prime}$ area over the GOODS-S field \citep[][]{Illingworth:2016, Whitaker:2019}. The JWST/NIRCam data were obtained by the JWST Advanced Deep Extragalactic Survey \citep[JADES;][PID: 1180 and 1210]{Eisenstein:2023} and the JWST Extragalactic Medium-band Survey \citep[JEMS;][PID: 1963]{Williams:2023} in September and October of 2022. The JADES observations consist of a deep mosaic covering a $4.4^{\prime} \times 6.2^{\prime}$ area with nine filters (F090W, F115W, F150W, F200W, F277W, F335M, F356W, F410M, and F444W) and a medium region covering an additional $6.1^{\prime} \times 6.5^{\prime}$ area with eight filters (F090W, F115W, F150W, F200W, F277W, F356W, F410M, and F444W). The JEMS observations consist of two $2.2^{\prime} \times 2.2^{\prime}$ regions with five filters (F182M, F210M, F430M, F460M, and F480M), all lying in JADES coverage. For all of the subsequent analysis, we do not require any of our objects to have JEMS observations, but we use these data when available. 

A detailed description of the JWST/NIRCam imaging data reduction and mosaicing will be presented in a forthcoming paper from the JADES Collaboration (Tacchella et al., in preparation). We briefly summarize here the main steps of the reduction and mosaicing process. The data are initially processed with the standard JWST calibration pipeline\footnote{\href{https://github.com/spacetelescope/jwst}{https://github.com/spacetelescope/jwst}}. Customized steps are included to aid in the removal of ``1/f'' noise, ``wisp'' artifacts, ``snowball'' artifacts, and persistence from previous observations \citep[see also][]{Rigby:2022}. The JWST Calibration Reference Data System (CRDS) context map \texttt{jwst\_1008.pmap} is used, including the flux calibration for JWST/NIRCam from Cycle 1. The background from the sky is modeled and removed using the BackGround2D class from \texttt{photutils} \citep[][]{Bradley:2022}. Finally, the image mosaics for each of the fourteen JWST/NIRCam filters are registered to the \textsc{Gaia} DR3 frame \citep[][]{GaiaDR3} and resampled onto the same world coordinate system (WCS) with a $30\,\mathrm{mas/pixel}$ grid. Assuming a circular aperture with a diameter of $0.3^{\prime\prime}$, the $5\sigma$ point-source detection limit in the F200W filter is $m \approx 30.0\,\mathrm{AB\,mag}$ and $m \approx 29.0\,\mathrm{AB\,mag}$ for the deep and medium regions, respectively. 

A detailed description of the JWST/NIRCam source detection was outlined in \citet[][]{Robertson:2022} and will be presented in detail in another forthcoming paper from the JADES Collaboration (Robertson et al., in preparation). We briefly summarize here the main steps of the source detection process. Six image mosaics (F200W, F277W, F335M, F356W, F410M, and F444W) are initially stacked using the corresponding error images and inverse-variance weighting to produce a single detection image. \textcolor{black}{These filters were chosen in order to avoid biasing our catalog against SW dropouts (e.g. dropouts in F090W, F115W, or F150W).} In this detection image, we construct a source catalog by selecting contiguous regions \textcolor{black}{of greater than five pixels} with signal-to-noise ratios $\mathrm{S/N} > 3$ and applying a standard Source Extractor \citep[\texttt{SExtractor};][]{SourceExtractor} deblending algorithm with parameters \texttt{nlevels = 32} and \texttt{contrast = 0.001} using \texttt{photutils} \citep[][]{Bradley:2022}. \textcolor{black}{Finally, we perform forced convolved photometry at the source centroids in all HST/ACS and JWST/NIRCam photometric bands, assuming elliptical Kron apertures with \texttt{parameter = 1.2} (i.e. Kron small) and \texttt{parameter = 2.5} (i.e. Kron large). To correct for potential missing light, we rescale the Kron small photometry by the flux ratio of Kron large to Kron small in the F444W filter.} Using model point spread functions (PSFs) from the \texttt{TinyTim} \citep[][]{Krist:2011} package for HST/ACS and the \texttt{WebbPSF} \citep[][]{Perrin:2014} package for JWST/NIRCam, we apply aperture corrections assuming point source morphologies. Uncertainties are estimated by placing random apertures across regions of the image mosaics to compute a flux variance \citep[e.g.,][]{Labbe:2005, Quadri:2007, Whitaker:2011}, which are summed in quadrature with the associated Poisson uncertainty for each detected source. 

\subsection{Photometric Redshifts \& Sample Selection}
\label{SectionTwoTwo}

Using the previously described photometry, we measure photometric redshifts with the template-fitting code \texttt{EAZY} \citep[][]{Brammer:2008}. A more detailed description of this procedure will be discussed in a forthcoming paper from the JADES Collaboration (Hainline et al., in preparation). We briefly summarize here the main steps of the photometric redshift process. \texttt{EAZY} uses a chi-square ($\chi^2$) minimization technique to model the broadband spectral energy distributions (SEDs) for galaxies using linear combinations of galaxy templates. It is designed to be both fast and flexible, and has been used extensively in the literature to model the photometric redshifts of galaxies \citep[e.g.,][]{Newman:2013, Skelton:2014, Bouwens:2015}. 

We fit all of the available photometry for each object with \texttt{EAZY}, \textcolor{black}{assuming the rescaled Kron small photometry described in Section~\ref{SectionTwoOne}}. For objects in the UDF, this includes photometry in nineteen filters. For objects in the JADES deep region but not in the UDF, this includes photometry in fourteen filters. For objects in the JADES medium region, this includes photometry in thirteen filters. We utilize sixteen templates in total to perform the fitting, which includes the nine \texttt{EAZY} ``v1.3'' templates, two additional templates for simple stellar populations with ages of $5\,\mathrm{Myr}$ and $25\,\mathrm{Myr}$, and five more templates with strong nebular continuum emission that were created using the Flexible Stellar Population Synthesis code \citep[\texttt{FSPS};][]{Conroy:2009, Conroy:2010}. These templates span a large range of stellar population properties and include contributions from both nebular continuum and line emission, as well as obscuration from dust. 

While the photometric calibration of JWST/NIRCam has improved significantly over the course of the last few months, there is still some uncertainty with these calibrations that needs to be taken into account. To this end, we iteratively calculate the photometric offset from the \texttt{EAZY} templates compared to the true JWST/NIRCam photometry, \textcolor{black}{using a sample of galaxies with signal-to-noise ratios between 5 and 20 in F200W}. These photometric offsets are relatively small (on the order of a few percent for both HST/ACS and JWST/NIRCam) and are subsequently applied to the entire photometric catalog. We choose not to adopt any apparent magnitude priors, but we do make use of the template error file \textsc{``template\_error.v2.0.zfourge''}. 

\textcolor{black}{The primary measurements used here are the \texttt{EAZY} ``$z_{\mathrm{a}}$'' and ``$z_{\mathrm{peak}}$'' redshifts. The former corresponds to the fit where the likelihood is maximized ($\chi^2$ is minimized), while the latter corresponds to the fit where the likelihood is maximized ($\chi^2$ is minimized) after taking into account the template error file. We allow \texttt{EAZY} to fit across the redshift range of $z = 0.2 - 22$ with a redshift step size of $\Delta z = 0.01\left(1 + z\right)$.} To test the accuracy of these photometric redshifts, we compare these predictions with existing spectroscopic redshifts in the GOODS-S field from the Multi Unit Spectroscopic Explorer \citep[MUSE;][]{Inami:2017, Urrutia:2019}. While the MUSE spectroscopic redshifts are biased toward the brightest objects detected by JWST at $z < 7$, we found catastrophic outlier fractions of only $5\%$ \citep{Rieke:2023} when comparing to the highest quality spectroscopic redshifts available with MUSE. The catastrophic outlier fraction is defined to be the fraction of objects that satisfy Equation~\ref{EquationOne}:
\begin{equation}
    \label{EquationOne}
    \frac{| z_{\,\mathrm{spec}} - z_{\,\mathrm{phot}} |}{1 + z_{\,\mathrm{spec}}} > 0.15.
\end{equation}

To perform an accurate and efficient targeted emission line search within the available spectroscopic data, we require a sample of relatively bright objects, since these are the only objects for which we expect to detect an emission line (see Section~\ref{SectionTwoThree}). We also require these objects to have tight photometric redshift constraints, which allow for spectroscopic redshift confirmation using only a single line detection. Most objects with tight photometric redshift constraints have emission lines that fall in one of the medium band filters (e.g., F410M) which allows for tight constraints when paired with the broad band filter coverage of JADES (e.g., F444W). \textcolor{black}{Our selection criteria for the final photometric catalog consist of the following: $m < 28.5\ \mathrm{AB\ mag}$ in F444W assuming \textcolor{black}{elliptical Kron apertures with \texttt{parameter = 2.5}}, $4.5 < z_{\mathrm{a}} < 9.5$, $4.5 < z_{\mathrm{peak}} < 9.5$, $\Delta z_{\,1} < 1$, and $\Delta z_{\,2} < 2$. The first EAZY confidence interval ($\Delta z_{\,1}$) is defined to be the difference between the 16th and 84th percentiles of the photometric redshift posterior distribution and is roughly twice the standard deviation. The second EAZY confidence interval ($\Delta z_{\,2}$) is defined to be the difference between the 5th and 95th percentiles of the photometric redshift posterior distribution and is roughly four times the standard deviation.} 

\subsection{Spectroscopic Data \& Emission Line Detection}
\label{SectionTwoThree}

Our spectroscopic data consist of WFSS observations taken with JWST/NIRCam in the F444W filter ($\lambda = 3.9-5.0\ \mu\mathrm{m}$). These data were obtained by the First Reionization Epoch Spectroscopic COmplete Survey (FRESCO; PI: Oesch; PID: 1895) in November of 2022. The FRESCO observations cover a $8.2^{\prime} \times 8.6^{\prime}$ area using the row-direction grisms on both modules of JWST/NIRCam (Grism R; $R \approx 1600$). The total overlapping area between the JADES and FRESCO footprints is $\approx 41$ square arcminutes. The total spectroscopic observing time for FRESCO in GOODS-S is $\approx 16$ hours with a typical on-source time of $\approx 2$ hours. The $3\sigma$ unresolved emission line detection limit around $4.2\,\mu\mathrm{m}$ in the F444W filter is $\sim 1.2 \times 10^{-18}\,\mathrm{erg/s/cm^{2}}$, which corresponds to a star formation rate (SFR) detection limit of $\sim 2.1\,M_{\odot}/\mathrm{yr}$ at $z = 5.4$ using the conversion factor from \citet{Kennicutt:2012}. 

A detailed description of the JWST/NIRCam grism data reduction can be found in \citet[][]{Sun:2022b}. We briefly summarize here the main steps of the reduction process. The data are initially processed with the standard JWST calibration pipeline\footnote{\href{https://github.com/spacetelescope/jwst}{https://github.com/spacetelescope/jwst}}. We assign WCS to the rate files, perform flat-fielding, and subtract out the sigma-clipped median sky background from each individual exposure after the ``ramp-to-slope'' fitting in the calibration pipeline. Because we are interested in conducting a targeted emission line search, and we do not expect any of our sources to have a strong continuum \textcolor{black}{due to their general faintness ($m = 27-28\ \mathrm{AB\ mag}$)}, we utilize a median-filtering technique to subtract out any remaining continuum or background on a row-by-row basis following the methodology of \citet[][]{Kashino:2022}. This produces emission line maps for each grism exposure that are void of any continuum. Although this median-filtering technique is able to properly remove continuum contamination, it sometimes over-subtracts signal in the spectral regions immediately surrounding the brightest emission lines (e.g., emission from [\ion{N}{2}] on either side of H$\alpha$). The full widths at half maximum (FWHMs) of these emission lines are relatively small and typically on the order of a few pixels, which means that the H$\alpha$ line flux is preserved by the nine pixel central gap of the median filter and is not affected by the aforementioned over-subtraction. We further remove the ``1/f'' noise using the \texttt{tshirt/roeba} algorithm\footnote{\href{https://github.com/eas342/tshirt}{https://github.com/eas342/tshirt}} in both the row and column directions. 

We extract two-dimensional (2d) grism spectra using the reduced emission-line maps for all of the objects that are part of the final photometric catalog discussed in Section~\ref{SectionTwoTwo}. Short wavelength (SW) parallel observations were conducted in two photometric bands (F182M and F210M) and are used for both astrometric and wavelength calibration of the long wavelength (LW) spectroscopic data. We use the spectral tracing and grism dispersion models \citep[][]{Sun:2022b} that were produced using the JWST/NIRCam commissioning data of the Large Magellanic Cloud (LMC; PID: 1076), which are also outlined in \citet{Wang:2023}. We additionally use the flux calibration models that were produced using JWST/NIRCam Cycle-1 absolute flux calibration observations (PID: 1536/1537/1538). 

% Distribution_Redshift
\begin{figure*}
    \centering
    \includegraphics[width=0.6\linewidth]{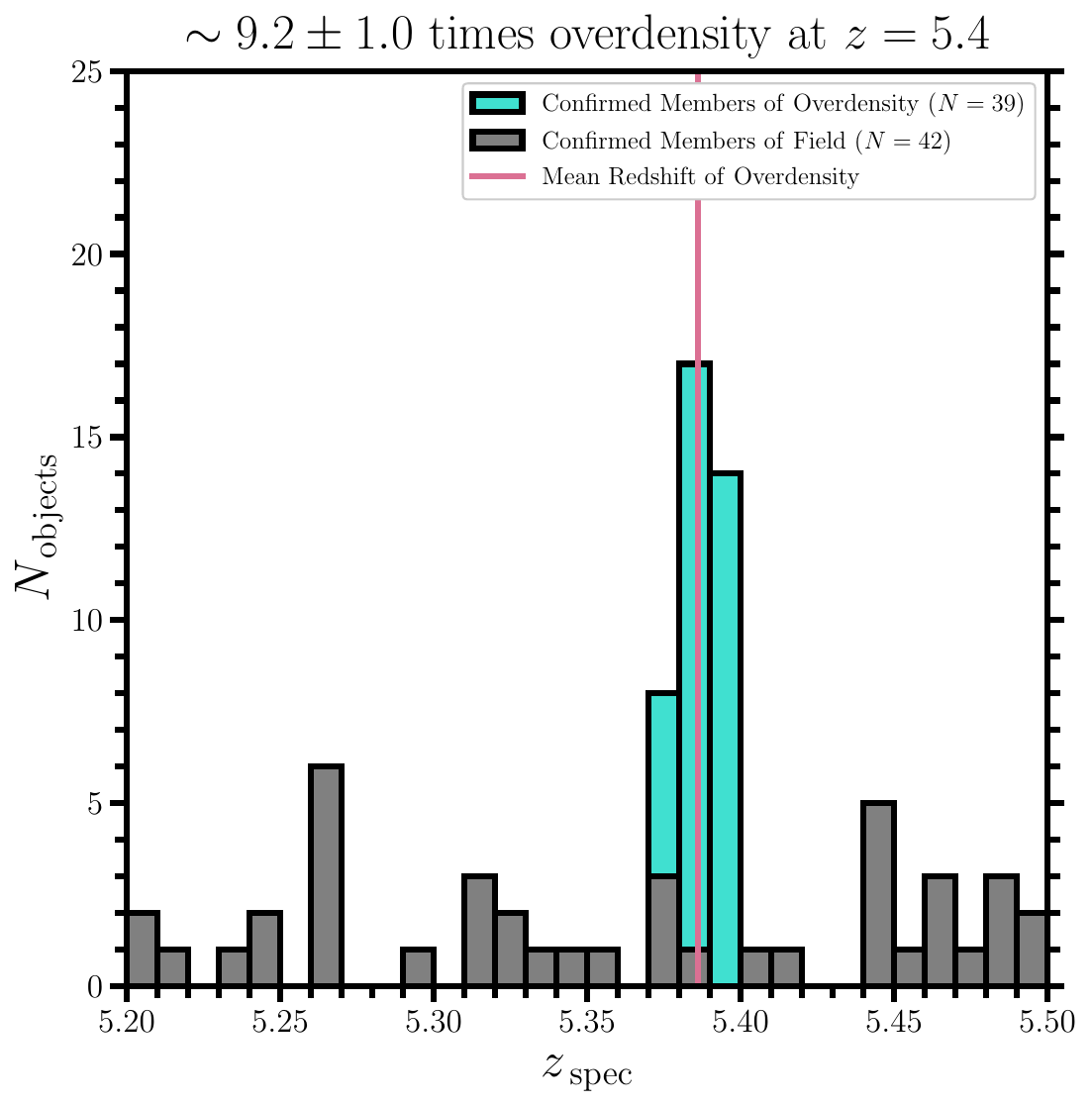}
    \caption{The distribution of spectroscopic redshifts for the $N = 81$ objects at $z = 5.2-5.5$ that are part of the final spectroscopic sample identified in Section~\ref{SectionTwoThree}. As defined in Section~\ref{SectionThreeOne}, the grey histograms represent the $N = 42$ confirmed members of the field while the turquoise histograms represent the $N = 39$ confirmed members of the overdensity. The median redshift of the overdensity is given by the solid magenta line. Compared to the field members, the overdensity members appear much more clustered, representing a $\sim 10$ times overdensity at $z = 5.4$. \label{fig:distribution_redshift}}
\end{figure*}

Using the already extracted 2d spectra, we further extract one-dimensional (1d) grism spectra using a boxcar aperture, assuming a height of five pixels ($0.31^{\prime\prime}$). We subsequently identify $>3\sigma$ peaks automatically in the 1d spectra, assuming various bin sizes (integer units of $\mathrm{nm}$ from one to eight) and fit these detected peaks with Gaussian profiles. \textcolor{black}{For each line that is detected with $\mathrm{S/N} > 3$, we tentatively assign a line identification of either $\mathrm{H}\alpha$ or $\mathrm{[OIII]}\lambda5008$, whichever one minimizes the difference between the best-fit photometric redshift and the tentative spectroscopic redshift. For example, if a line were detected at $\lambda = 4.2\,\mu\mathrm{m}$ and the best-fit photometric redshift is $z_{\mathrm{phot}} = 5.8$, then the initial line identification would be $\mathrm{H}\alpha$, since the predicted wavelength of this line would be at $\lambda = 4.5\,\mu\mathrm{m}$, which is closer to the observed wavelength than the predicted wavelength of $\mathrm{[OIII]}\lambda5008$ ($\lambda = 3.4\,\mu\mathrm{m}$). } Visual inspection is performed on each of these tentative spectroscopic redshift solutions to remove spurious detections caused by either noise or contamination. For sources that pass our visual inspection and have secure line detections, we optimally re-extract the 1d spectra using the F444W surface brightness profile \citep[][]{Horne:1986} and once again fit these detected peaks with Gaussian profiles. According to the grism wavelength calibration uncertainty, the typical absolute uncertainties of our spectroscopic redshifts are $\Delta z_{\mathrm{spec}} = 0.001$. 

% Distribution_Spatial
\begin{figure*}
    \centering
    \includegraphics[width=0.6\linewidth]{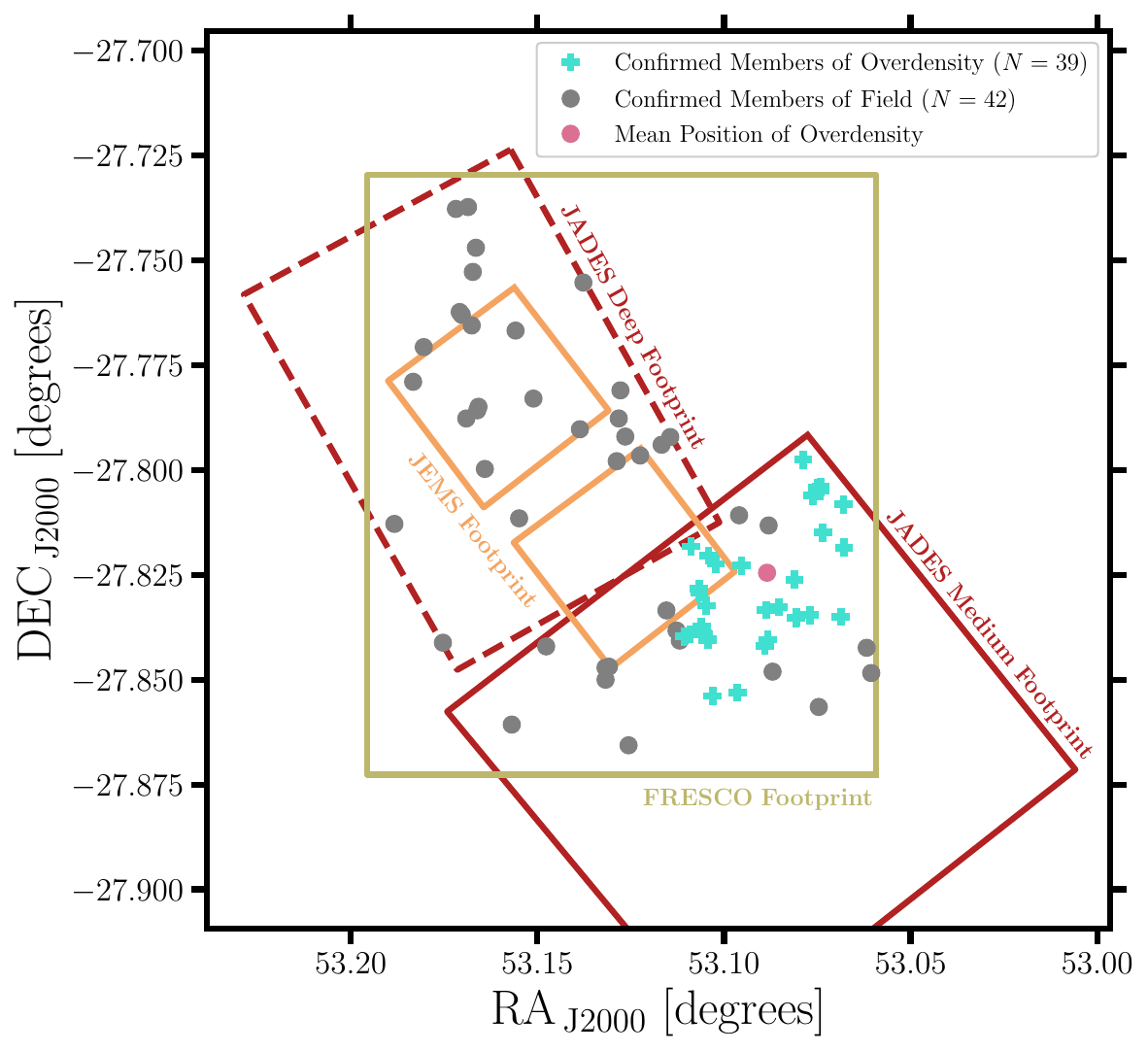}
    \caption{The on-sky distribution in angular units for the $N = 81$ objects at $z = 5.2-5.5$ that are part of the final spectroscopic sample. The grey points represent the $N = 42$ confirmed members of the field while the turquoise pluses represent the $N = 39$ confirmed members of the overdensity. The median position of the overdensity is given by the magenta point. The JADES deep (medium) footprint is illustrated by the dashed (solid) red line, the JEMS footprint by the solid orange line, and the FRESCO footprint by the solid yellow line. It is apparent that the overdensity falls near the edge of the JADES medium and FRESCO footprints, which means we cannot rule out the overdensity extending well beyond the region for which we currently have data. \label{fig:distribution_spatial}}
\end{figure*}

Our final spectroscopic sample includes $N = 81$ objects at $z = 5.2-5.5$ with $>3\sigma$ detections of H$\alpha$ from the FRESCO spectra. This redshift range was chosen to ensure that H$\alpha$ would fall in the F410M filter, which is the only medium-band filter for which we have uniform coverage, providing a sanity check for the derived emission line fluxes through a comparison with the F410M excess relative to F444W. Our final spectroscopic sample represents a subset of a larger spectroscopic sample of galaxies from both GOODS fields across a much broader redshift range (Sun et al., in preparation.). For the majority of galaxies in our final spectroscopic sample, neither of the [\ion{N}{2}] lines were detected, partially as a result of the aforementioned median-filtering technique \citep{Kashino:2022} but primarily because the line ratio [\ion{N}{2}]/H$\alpha$ is typically low at these redshifts \citep[e.g.,][]{Cameron:2023}. The NIRCam cutout images alongside the 2d and 1d extracted spectra for these objects are shown in Appendix~\ref{AppendixOne}. Figure~\ref{fig:distribution_redshift} shows the distribution of spectroscopic redshifts for these $N = 81$ objects while Figure~\ref{fig:distribution_spatial} shows the on-sky distribution in angular units. These distributions enabled us to visually identify an overdensity of galaxies around $z = 5.4$. 

%% Start of section three.
\section{Analysis \& Results}
\label{SectionThree}

Using the data and observations from Section~\ref{SectionTwo}, we perform various analyses on the $N = 81$ galaxies in our final spectroscopic sample and present the results. Identification of the extreme galaxy overdensity is described in Section~\ref{SectionThreeOne}. Detailed physical modeling of the stellar populations, presentation of the star-forming main sequence, and comparison of inferred SFRs are described in Section~\ref{SectionThreeTwo}. Determining the dynamic state, estimating the dark matter halo mass, and predicting the future evolution of the overdensity are described in Section~\ref{SectionThreeThree}. Placing this overdensity in context with previous works is described in Section~\ref{SectionThreeFour}. 

\subsection{Overdensity Identification}
\label{SectionThreeOne}

Following the technique described in \citet{Calvi:2021}, we use a Friends-of-Friends (FoF) algorithm to identify the overdensity after looking for three-dimensional (3d; two spatial, one spectral) structural groupings \citep[see also][]{Huchra:1982, Eke:2004, Berlind:2006}. This algorithm iteratively selects groups, which consist of one or more galaxies that have projected separations and line-of-sight (LOS) velocity dispersions below the adopted linking parameters (projected separation $d_{\mathrm{link}} = 500\,\mathrm{kpc}$, chosen to be roughly the virial radius of a typical galaxy cluster; LOS velocity dispersion $\sigma_{\mathrm{link}} = 500\,\mathrm{km/s}$, chosen to be roughly the velocity dispersion of such a cluster). These groupings do not depend strongly on the adopted linking parameters, producing similar results when varying either the projected separation or the LOS velocity dispersion by a factor of a few. 

\textcolor{black}{We identify one large-scale structure consisting of $N = 39$ galaxies out of the $N = 81$ galaxies that are part of our final spectroscopic sample. Throughout the rest of this work, we refer to the remaining $N = 42$ galaxies at $z = 5.2-5.5$ as field galaxies, which consist of (1) isolated galaxies and (2) those in smaller groups as determined by the FoF algorithm\footnote{The smaller groups include three groups of two, a group of three, two groups of four, and a group of five. We consider galaxies that fall into these groups as field galaxies since they are not part of the extreme galaxy overdensity, which is the primary focus of this paper. A more complete clustering analysis will be the subject of future work.}. The average spectroscopic redshift of the large-scale structure is $z = 5.386$, spanning a relatively narrow redshift range of $5.374 < z < 5.398$. The maximum on-sky separation of the clustered galaxies is roughly $3.6^{\prime}$, corresponding to a physical separation of $8.6\,\mathrm{cMpc}$.} 

\textcolor{black}{Following the methodology of \citet{Chiang:2013}, we calculate 3d galaxy overdensities ($\delta_{\mathrm{gal}} = n_{\mathrm{gal}} / \langle n_{\mathrm{gal}} \rangle - 1$), where $n_{\mathrm{gal}}$ is the number density of galaxies and $\langle n_{\mathrm{gal}} \rangle$ is the ensemble average number density of galaxies, for each of the $N = 81$ galaxies that are part of our final spectroscopic sample. To calculate these values, we assume a tophat-weighted spherical window which has a comoving volume equal to $\left( 15\,\mathrm{cMpc} \right)^{3}$. The average and standard deviation of the constituent overdensity values for the large-scale structure identified here is $\langle \delta_{\mathrm{gal}} \rangle = 9.2 \pm 1.0$. This structure is an extreme overdensity at $z = 5.4$, $\sim 10$ times more dense in 3d than the ensemble average at $z = 5.2-5.5$. For comparison, \citet{Chiang:2013} found that an average overdensity value of $\langle \delta_{\mathrm{gal}} \rangle = 3.04$ identifies structures within cosmological simulations as protocluster candidates with $80\%$ confidence. This value is for the $z = 5$ SFR-limited sample ($\mathrm{SFR} > 1\,M_{\odot}/\mathrm{yr}$), which is the sample that is most similar to our own in terms of selection. Throughout this work, we define a protocluster to be a structure that will eventually collapse into a galaxy cluster at $z = 0$.} 

% Distribution_Bimodality
\begin{figure*}
    \centering
    \includegraphics[width=1.0\linewidth]{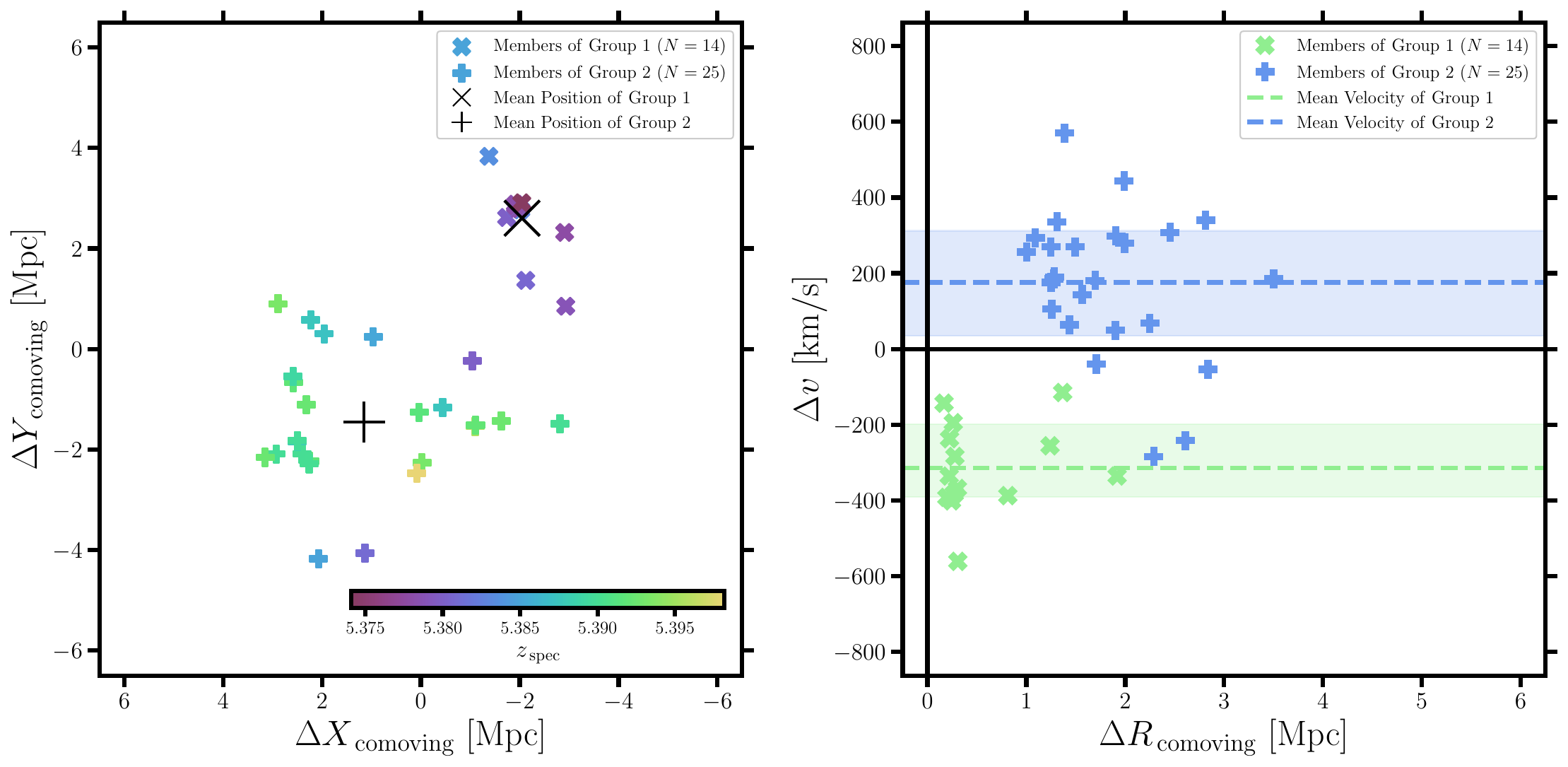}
    \caption{\textit{Left panel:} The on-sky distribution in physical units for the $N = 39$ confirmed members of the overdensity, color-coded by their spectroscopic redshift. The color-coded crosses (pluses) represent members of the first (second) component of the overdensity. The median position of the first (second) component is given by the black cross (plus). \textit{Right panel:} The observed phase-space diagram for the $N = 39$ confirmed members of the overdensity. The median velocity offset of the first (second) component of the overdensity is given by the green (blue) dashed line. The $1\sigma$ confidence interval of the velocity offsets for each of the groups are given by the shaded regions. \label{fig:distribution_bimodality}}
\end{figure*}

Throughout the rest of Section~\ref{SectionThree}, the distinction between the $N = 39$ ``confirmed members of overdensity'' and the $N = 42$ ``confirmed members of field'' will be used. The confirmed members of the overdensity are shown in Figure~\ref{fig:distribution_redshift} (Figure~\ref{fig:distribution_spatial}) by the turquoise histograms (turquoise pluses) while the confirmed members of the field are shown by the grey histograms (grey points). The median redshift or position of the overdensity is given by the solid magenta line or the magenta point. In both of these figures, the overdensity members appear much more clustered when compared to the field members. The confirmed members of the overdensity are additionally shown in the left panel of Figure~\ref{fig:distribution_bimodality}, color-coded by their spectroscopic redshift. We identify a spatial and kinematic bimodality within the overdensity at $z = 5.4$, which we return to in Section~\ref{SectionThreeThree}. To assign objects between the two components of the bimodality, we adopt an iterative process that minimizes the 3d separation within each of these components, finding $N = 14$ galaxies that are part of the first component ($\Delta v = -340^{+140}_{-50}\,\mathrm{km/s}$) and $N = 25$ galaxies that are part of the second component ($\Delta v = +190^{+120}_{-150}\,\mathrm{km/s}$). The median positions of these two groups are given by the black plus and cross in the left panel of Figure~\ref{fig:distribution_bimodality}. Velocity offsets are calculated relative to the median spectroscopic redshift of the overdensity. 

\subsection{Stellar Population Modeling}
\label{SectionThreeTwo}

Following the methodology outlined in \citet{Tacchella:2022b}, we utilize the SED fitting code \texttt{Prospector} \citep[v1.1.0;][]{Johnson:2021} to infer the stellar populations for the $N = 81$ objects that are part of our final spectroscopic sample. \textcolor{black}{Fits are performed on the rescaled Kron small photometry described in Section~\ref{SectionTwoOne}}, while the redshift is fixed at the spectroscopic redshift determined in Section~\ref{SectionTwoTwo}. \texttt{Prospector} uses a Bayesian inference framework and we choose to sample posterior distributions with the dynamic nested sampling code \texttt{dynesty} \citep[v1.2.3;][]{Speagle:2020}. 

We use the Flexible Stellar Population Synthesis code \citep[\texttt{FSPS};][]{Conroy:2009, Conroy:2010} via \texttt{python-FSPS} \citep{Foreman-Mackey:2014} with the Modules for Experiments in Stellar Astrophysics Isochrones and Stellar Tracks \citep[MIST;][]{Choi:2016, Dotter:2016}, which makes use of the Modules for Experiments in Stellar Astrophysics (MESA) stellar evolution package \citep{Paxton:2011, Paxton:2013, Paxton:2015, Paxton:2018}. We assume the MILES stellar spectral library \citep{Falcon-Barroso:2011, Vazdekis:2015} and adopt a \citet{Chabrier:2003} initial mass function (IMF). Absorption by the intergalactic medium (IGM) is modeled after \citet{Madau:1995}, where the overall scaling of the IGM attenuation curve is set to be a free parameter. Dust attenuation is modeled using a two-component dust attenuation model \citep{Charlot:2000} with a flexible attenuation curve where the slope is tied to the strength of the ultraviolet (UV) bump \citep{Kriek:2013}. Nebular emission (both from emission lines and continuum) is self-consistently modeled with the spectral synthesis code \texttt{Cloudy} \citep{Byler:2017}. 

To test the robustness of the inferred stellar populations and SFHs, we assume three different models for the SFH to see how our results depend on the assumed prior. Two of these models are non-parametric (one with the ``continuity'' prior, the other with the ``bursty continuity'' prior) while the third is parametric (with the shape of a delayed-tau function). For each of the non-parametric models, we assume that the SFH can be described by $N_{\mathrm{SFR}}$ distinct time bins of constant star-formation. The time bins are specified in units of lookback time and the number of distinct bins is fixed at $N_{\mathrm{SFR}} = 6$. The first two bins are fixed at $0-30\ \mathrm{Myr}$ and $30-100\ \mathrm{Myr}$ while the last bin is fixed between $0.15\,t_{\mathrm{univ}}$ and $t_{\mathrm{univ}}$, where $t_{\mathrm{univ}}$ is the age of the Universe at the galaxy's spectroscopic redshift, measured with respect to the formation redshift $z_{\mathrm{form}} = 20$. The rest of the bins are spaced equally in logarithmic time between $100\,\mathrm{Myr}$ and $0.85\,t_{\mathrm{univ}}$. Both the total stellar mass and the ratios of adjacent time bins are set to be free parameters. A summary of the parameters and priors associated with this \texttt{Prospector} model is presented in Table~\ref{tab:prospector_model}. \textcolor{black}{We adopt the results from the non-parametric SFH with ``continuity'' prior as fiducial, since the Bayesian evidence does not strongly favor one model over another for the vast majority of the galaxies considered here.} 

\begin{figure*}
    \centering
    \includegraphics[width=0.6\linewidth]{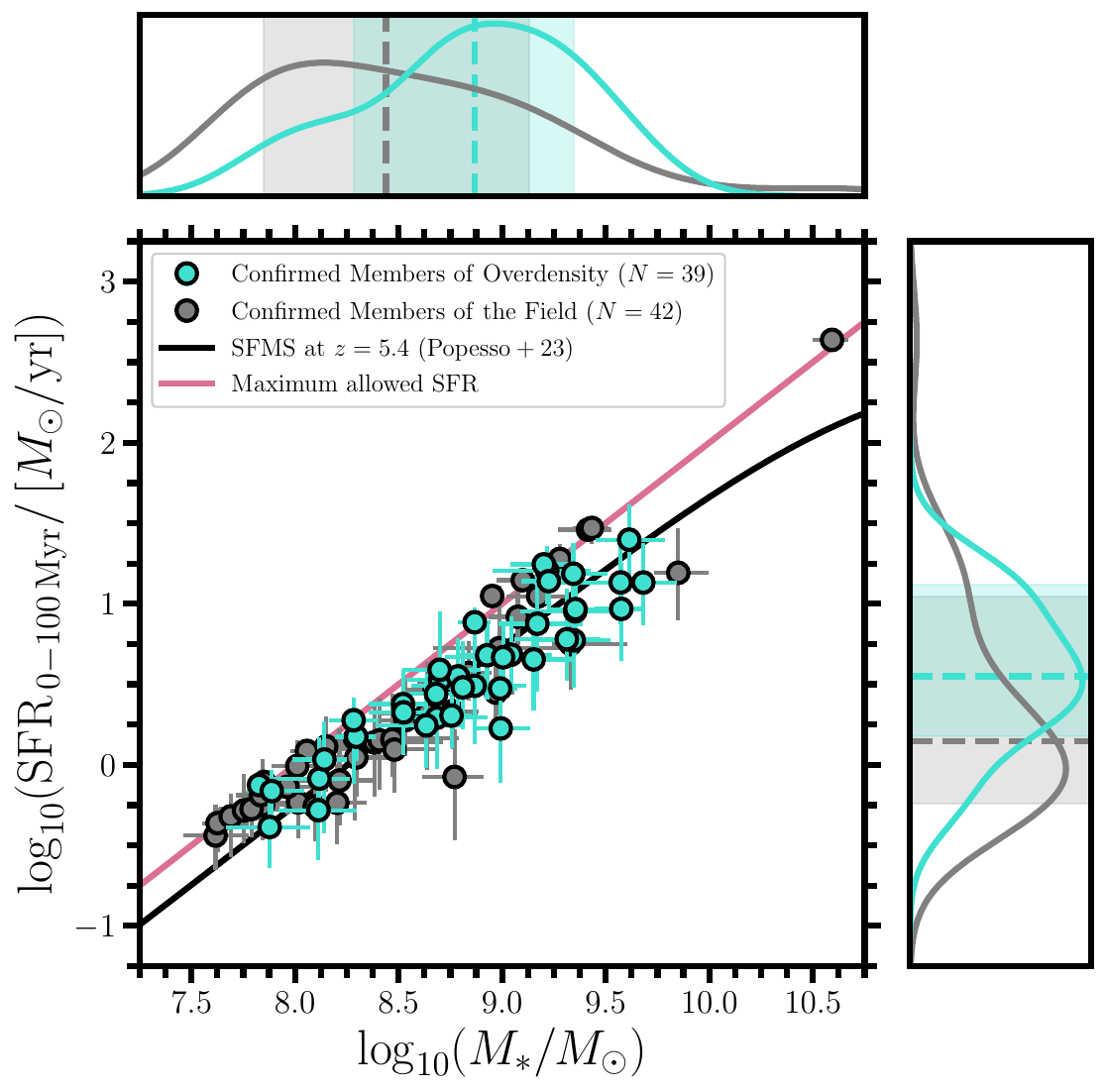}
    \caption{The star-forming main sequence for the $N = 81$ objects at $z = 5.2-5.5$ that are part of the final spectroscopic sample. The stellar masses and SFRs reported here are derived from the \texttt{Prospector} fits. The SFRs are averaged over the last $100\ \mathrm{Myr}$ of lookback time. The grey points represent the $N = 42$ confirmed members of the field while the turquoise points represent the $N = 39$ confirmed members of the overdensity. The empirical star-forming main sequence at $z = 5.4$ derived by \citet{Popesso:2023} is given by the solid black line. The maximum allowed SFR assuming all of the stellar mass was formed in the last $100\ \mathrm{Myr}$ of lookback time is given by the solid magenta line. The median values of the inferred stellar masses and SFRs for the confirmed members of the field (overdensity) are given by the grey (turquoise) dashed lines in the twin axes to the top and right. The $1\sigma$ confidence interval for these parameters are given by the shaded regions in the same axes. The overdensity members appear to have larger inferred stellar masses and SFRs when compared to the field members. Additionally, nearly all of these objects agree with the empirically derived star-forming main sequence at $z = 5.4$  within $1\sigma$. \label{fig:starforming_mainsequence}}
    
\end{figure*}

Figure~\ref{fig:starforming_mainsequence} shows the star-forming main sequence for the $N = 81$ objects at $z = 5.2-5.5$ that are part of the final spectroscopic sample identified in Section~\ref{SectionTwoThree}. The confirmed members of the field are given by the grey points while the confirmed members of the overdensity are given by the turquoise points. The reported stellar masses and SFRs are derived from the \texttt{Prospector} fits using the non-parametric SFH with ``continuity'' prior, where the SFRs are averaged over the last $100\ \mathrm{Myr}$ of lookback time. These stellar masses and SFRs are reported in Table~\ref{tab:physical_properties}, which gives a summary of the physical properties for our final spectroscopic sample. \textcolor{black}{Based on these stellar masses and the mass-size relation reported in \citet{Shibuya:2015}, we find that the minimum separation between the sources in our sample is always larger than twice the effective radius of the sources in their sample. Therefore, each of the objects that is part of our final spectroscopic sample is likely an individual star-forming galaxy rather than individual star-forming clumps within a much larger galaxy, although some are clearly merging systems in the final phase of coalescence.} To be used as a point of comparison, the empirical star-forming main sequence at $z = 5.4$ derived by \citet{Popesso:2023} is given by the solid black line. Additionally, the maximum allowed SFR assuming all of the stellar mass was formed in the last $100\ \mathrm{Myr}$ of lookback time is given by the solid magenta line. 

% StarForming_MainSequence

We find that nearly all of the $N = 81$ objects at $z = 5.2-5.5$ that are part of our final spectroscopic sample agree with the empirically derived star-forming main sequence at $z = 5.4$ given by \citet{Popesso:2023} within $1\sigma$, despite our sample being biased as a result of our requirement to detect the $\mathrm{H} \alpha$ emission line at greater than $3\sigma$. If we were to instead use $\mathrm{H}\alpha$-based SFRs derived with the conversion factor from \citet{Kennicutt:2012}, the objects within our final spectroscopic sample would be shifted upward by $\Delta \sim 0.5\,\mathrm{dex}$. However, this is likely because the canonical hydrogen ionizing photon production efficiency ($\xi_\mathrm{ion} \propto L_{\mathrm{H}\alpha}/L_{\mathrm{UV}}$) used in \citet{Kennicutt:2012} is only $\xi_\mathrm{ion} \sim 10^{25.1}\,\mathrm{erg}^{-1}\mathrm{Hz}$, lower than measurements at $z = 5-6$ by $\Delta \xi_\mathrm{ion} \sim 0.5\,\mathrm{dex}$ \citep[e.g.,][]{Bouwens:2016, Ning:2022, Sun:2022b}. \textcolor{black}{Furthermore, if we were to instead use UV-based SFRs derived with the conversion factor from \citet{Kennicutt:2012} alongside measurements of the rest-UV magnitudes derived as $\nu L_{\nu}$ at $\lambda_{\mathrm{rest}} = 1500$\AA, it would not change the distribution of our final spectroscopic sample in Figure~\ref{fig:starforming_mainsequence}. This is expected, since the \texttt{Prospector}-based SFRs largely rely on the available rest-UV photometry.} 

The most notable examples of galaxies disagreeing with the empirical relation include one well above the main sequence at high stellar mass (JADES$-$GS$+$53.13859$-$27.79025) and two somewhat below the main sequence at intermediate stellar mass (JADES$-$GS$+$53.06799$-$27.80816 and JADES$-$GS$+$53.18328$-$27.77894) despite large uncertainties in the inferred SFRs. We note that the galaxy well above the main sequence at high stellar mass is an active galactic nucleus (AGN), \textcolor{black}{originally identified as a broad line $\mathrm{H} \alpha$ emitter in \citet{Matthee:2023} with a line width of $2200 \pm 500\,\mathrm{km/s}$}. Combined with our relatively low SFR detection limit ($\sim 2.1\,M_{\odot}/\mathrm{yr}$, see Section~\ref{SectionTwoThree}), the fact that our sample agrees with the empirical relation suggests that we are sampling the bulk of the star-forming population at these redshifts despite our selection criteria. However, we should mention that we are likely missing some amount of dusty star-forming galaxies (DSFGs), which historically have been used as tracers to identify protocluster candidates at these redshifts \citep[for a review of environmental galaxy evolution, see][]{Alberts:2022}. 

To test the impact of the assumed SFH, we compare the \texttt{Prospector} derived stellar masses and SFRs for the three different SFH models. We remind the reader that for Figure~\ref{fig:starforming_mainsequence} and Table~\ref{tab:physical_properties}, the assumed SFH is non-parametric with the ``continuity'' prior. Compared to the non-parametric SFH with the ``bursty continuity'' prior, the derived stellar masses are a bit larger ($\Delta \approx 0.1\,\mathrm{dex}$) with large scatter ($\sigma \approx 0.7\,\mathrm{dex}$) while the derived SFRs are broadly consistent ($\Delta \approx 0.0\,\mathrm{dex}$) with large scatter ($\sigma \approx 0.8\,\mathrm{dex}$). Compared to the parametric SFH, both the derived stellar masses and SFRs are broadly consistent ($\Delta \approx 0.0\,\mathrm{dex}$) with large scatter ($\sigma \approx 0.8\,\mathrm{dex}$ and $\sigma \approx 1.0\,\mathrm{dex}$, respectively). We find that the inferred \texttt{Prospector} parameters considered here do not depend strongly on the assumed SFH. 

\begin{figure*}
    \centering
    \includegraphics[width=0.6\linewidth]{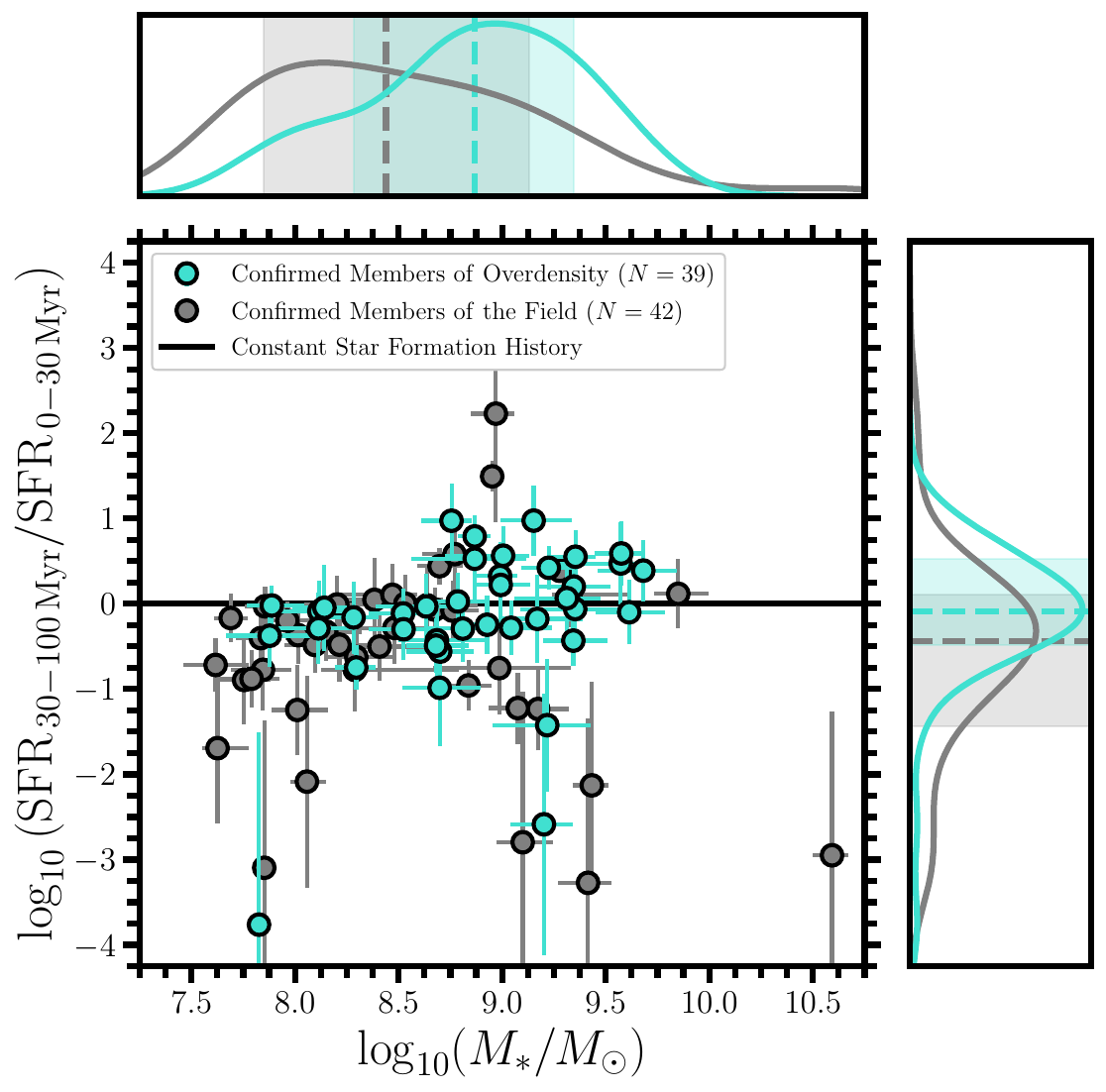}
    \caption{The comparison of SFRs in the two most recent time bins for the $N = 81$ objects at $z = 5.2-5.5$ that are part of the final spectroscopic sample. The stellar masses and SFRs reported here are derived from the \texttt{Prospector} fits. The grey points represent the $N = 42$ confirmed members of the field while the turquoise points represent the $N = 39$ confirmed members of the overdensity. A constant SFH is given by the solid black line. Values above (below) this line represent a falling (rising) SFH over the last $100\ \mathrm{Myr}$ of lookback time. The twin axes to the top and right are similar to those shown in Figure~\ref{fig:starforming_mainsequence}. The overdensity (field) members appear to have non-parametric SFHs consistent with a constant (rising) SFH with relatively small (large) scatter. \label{fig:Prospector_comparison_quenching}}
\end{figure*}

Figure~\ref{fig:Prospector_comparison_quenching} shows the comparison of SFRs in the two most recent time bins (corresponding to lookback times of $0-30\ \mathrm{Myr}$ and $30-100\ \mathrm{Myr}$, respectively) for the $N = 81$ objects at $z = 5.2-5.5$. Once again, the confirmed members of the field are given by the grey points while the confirmed members of the overdensity are given by the turquoise points. A constant SFH is given by the solid black line, which assumes that the average SFRs in the two most recent time bins are equal. Values above this line represent a falling SFH over the last $100\ \mathrm{Myr}$, while values below this line represent a rising SFH. The twin axes to the top and right are similar to those shown in Figure~\ref{fig:starforming_mainsequence}, \textcolor{black}{where the median values for the confirmed members of the field (overdensity) are given by the grey (turquoise) dashed lines}. 

% Comparison_Quenching

By comparing members of the overdensity with members of the field in Figures~\ref{fig:starforming_mainsequence} and \ref{fig:Prospector_comparison_quenching}, we can begin to explore the impact of environment on galaxy formation and evolution at $z = 5.2-5.5$. Table~\ref{tab:percentiles} gives a summary of percentiles for some of the \texttt{Prospector} inferred physical parameters for both the members of the field and members of the overdensity. In the star-forming main sequence \textcolor{black}{and the summary of percentiles}, overdensity members appear to have larger inferred stellar masses and SFRs when compared to the field members. \textcolor{black}{We further compare the distributions of these inferred parameters by performing Kolmogorov-Smirnov (KS) and Anderson-Darling (AD) tests, which are two-sided tests for the null hypothesis that two independent samples are drawn from the same continuous distribution. The results of these tests indicate that the stellar masses and SFRs of the field and overdensity samples exhibit statistically significant differences at roughly the $2-3\sigma$ level (corresponding to $p$-values of $0.003 < p < 0.05$).} However, to fairly compare these two samples, the stellar mass distributions must be similar such that any observed differences can be attributed to external processes (e.g., environment) rather than internal processes (e.g., feedback). To create a mass-matched sample of field galaxies, we select the field galaxy that is closest in stellar mass to each overdensity galaxy, \textcolor{black}{which produces a sample of galaxies with stellar masses that are consistent with being drawn from the same parent distribution as the overdensity members based on the KS- and AD-tests}. The percentiles for the mass-matched members of the field also appear in Table~\ref{tab:percentiles}. 

\begin{table*}
	\caption{A summary of percentiles for some of the \texttt{Prospector} inferred physical parameters.}
	\label{tab:percentiles}
	\hspace*{-12mm}
        \makebox[\textwidth]{
	\begin{threeparttable}
	\begin{tabular}{lrrrrrrrrr} 
		\hline
		\hline
		Parameter & \multicolumn{3}{c}{Field} & \multicolumn{3}{c}{Overdensity} & \multicolumn{3}{c}{Mass-Matched Field} \\
		& 16th & 50th & 84th & 16th & 50th & 84th & 16th & 50th & 84th \\
		\hline
            $A_{V}/\mathrm{mag}$ & $0.06$ & $0.16$ & $0.53$ & $0.08$ & $0.23$ & $0.48$ & $0.07$ & $0.20$ & $0.74$ \\
            $\mathrm{log}_{10}(M_{\ast}/M_{\odot})$ & $7.85$ & $8.44$ & $9.13$ & $8.28$ & $8.87$ & $9.35$ & $8.29$ & $8.84$ & $9.33$ \\
            $\mathrm{log}_{10}(\mathrm{SFR_{\,0-100\,Myr}}/\left[M_{\odot}/\mathrm{yr}\right])$ & $-0.24$ & $+0.15$ & $+1.05$ & $+0.18$ & $+0.55$ & $+1.12$ & $+0.05$ & $+0.57$ & $+1.05$ \\
            $\mathrm{log}_{10}\left(\mathrm{SFR_{\,30-100\,Myr}}/\mathrm{SFR_{\,0-30\,Myr}}\right)$ & $-1.44$ & $-0.44$ & $+0.11$ & $-0.49$ & $-0.09$ & $+0.52$ & $-1.23$ & $-0.40$ & $+0.11$ \\
            \hline
	\end{tabular}
	\end{threeparttable}
	\hspace*{+4mm}
	}
\end{table*}

When compared to the mass-matched field members, the overdensity members have \textcolor{black}{similar} inferred SFRs, \textcolor{black}{consistent with being drawn from the same parent distribution based on the KS- and AD-tests}. The ratio of SFRs in the two most recent time bins (corresponding to lookback times of $0-30\ \mathrm{Myr}$ and $30-100\ \mathrm{Myr}$, respectively) also show an interesting trend, where overdensity members have SFHs consistent with constant while mass-matched field members have SFHs consistent with rising. \textcolor{black}{Furthermore, KS- and AD-tests suggest that the SFHs of the mass-matched field and overdensity sample exhibit statistically significant differences at roughly the $2-3\sigma$ level (corresponding to $p$-values of $0.003 < p < 0.05$).} These results suggest that the physical processes associated with this extreme galaxy overdensity at $z = 5.4$ have induced earlier star formation and earlier stellar mass assembly relative to the field, although there are large uncertainties associated with all of these parameters, \textcolor{black}{and the distributions of these parameters appear consistent within the associated uncertainties}. 

\subsection{Dark Matter Halo Mass Estimates}
\label{SectionThreeThree}

To further understand the dynamical state of the overdensity identified in Section~\ref{SectionThreeOne}, the right panel of Figure~\ref{fig:distribution_bimodality} shows the phase-space diagram for the $N = 39$ confirmed members of the overdensity. Velocity offsets are calculated from the median redshift of the overdensity (see Figure~\ref{fig:distribution_redshift}) while spatial offsets are calculated from the median on-sky positions of the two groups that make up the bimodality (see the left panel of Figure~\ref{fig:distribution_bimodality}). The median velocity offset of the overdensity is given by the grey dashed line while the median velocity offsets of the first and second groups are given by the green and blue dotted lines, respectively. For each of the groups, the $1\sigma$ confidence interval of the velocity offsets are given by the shaded regions (for group one, $\Delta v = -340^{+140}_{-50}\,\mathrm{km/s}$; for group two, $\Delta v = +190^{+120}_{-150}\,\mathrm{km/s}$). 

Following the methodology described in \citet{Long:2020}, we derive two different estimates of the dark matter halo mass for the overdensity at $z = 5.4$. For both of these methods, we use the stellar-to-halo abundance matching relation described in \citet{Behroozi:2013} to convert stellar masses into halo masses. Uncertainties on our estimates are calculated by adopting the mean stellar-to-halo abundance matching relation from \citet{Behroozi:2013} and propagating the uncertainties on the stellar masses for each individual galaxy. This is an underestimate of the true uncertainties since there is scatter in the stellar-to-halo abundance matching relation which we do not account for. 

The first halo mass estimate is calculated by summing the halo masses for each individual galaxy that is part of the overdensity. This estimate assumes that (1) each galaxy is formed in a separate dark matter halo and (2) the individual galaxies are closer to virialization than the large-scale structure associated with the overdensity. We find that the first method gives a total halo mass of $\mathrm{log}_{10} \left( M_{\mathrm{halo}}/M_{\odot} \right) = 12.8 \pm 0.1$. The second halo mass estimate is determined by summing the stellar masses for each individual galaxy that is part of the overdensity and converting to a halo mass. This estimate instead assumes that each galaxy is formed in the same dark matter halo. We find that the second method gives a total halo mass of $\mathrm{log}_{10} \left( M_{\mathrm{halo}}/M_{\odot} \right) = 12.6 \pm 0.3$. 

Weighing this kind of large-scale structure in the early Universe is challenging and requires a variety of assumptions. Typical methods for weighing galaxy clusters include (1) gravitational lensing, (2) the Sunyaev-Zel’dovich effect, and (3) using X-ray observations of the hot gas in the intracluster medium (ICM). However, current X-ray and sub-millimeter observations are not sensitive enough to properly weigh this extreme overdensity at $z = 5.4$. For this reason, we use the above dark matter halo mass estimates to infer a probable halo mass range of $12.6 \lesssim \mathrm{log}_{10} \left( M_{\mathrm{halo}}/M_{\odot} \right) \lesssim 12.8$ for the overdensity identified in Section~\ref{SectionThreeOne}. 

We note that none of these methods account for additional members of the overdensity that were not identified and included in the final spectroscopic sample, including objects that fall outside either the JADES or the FRESCO footprints. This is a non-negligible effect since the first component of the overdensity (see Figure~\ref{fig:distribution_bimodality}) falls right at the edge of the JADES footprint ($\theta \ll 1^{\prime}$). Additionally, since our final spectroscopic sample only includes galaxies with narrow photometric redshift constraints and secure $\mathrm{H}\alpha$ line detections, we are likely missing some additional subset of objects with relatively unconstrained photometric redshifts  and/or low levels of star formation (e.g., DSFGs and/or obscured AGNs). \textcolor{black}{There are zero known DSFGs at $z = 5.4$ in GOODS-S in the literature \citep[e.g.,][]{Franco:2018, Hatsukade:2018, Gonzalez-Lopez:2020, Gomez-Guijarro:2022}. However, wide and shallow surveys like GOODS-ALMA would miss DSFGs with infrared luminosities of $L_{\mathrm{IR}} \lesssim 3 \times 10^{11}\,L_{\odot}$ at $z = 5.4$.} Given that clusters induce earlier quenching for their constituent members, we cannot rule out the existence of a significant population of these kinds of objects. For these reasons, the halo mass range quoted above is likely an underestimate of the true halo mass associated with this extreme galaxy overdensity at $z = 5.4$. 

% Protocluster_Evolution
\begin{figure*}
    \centering
    \includegraphics[width=0.6\linewidth]{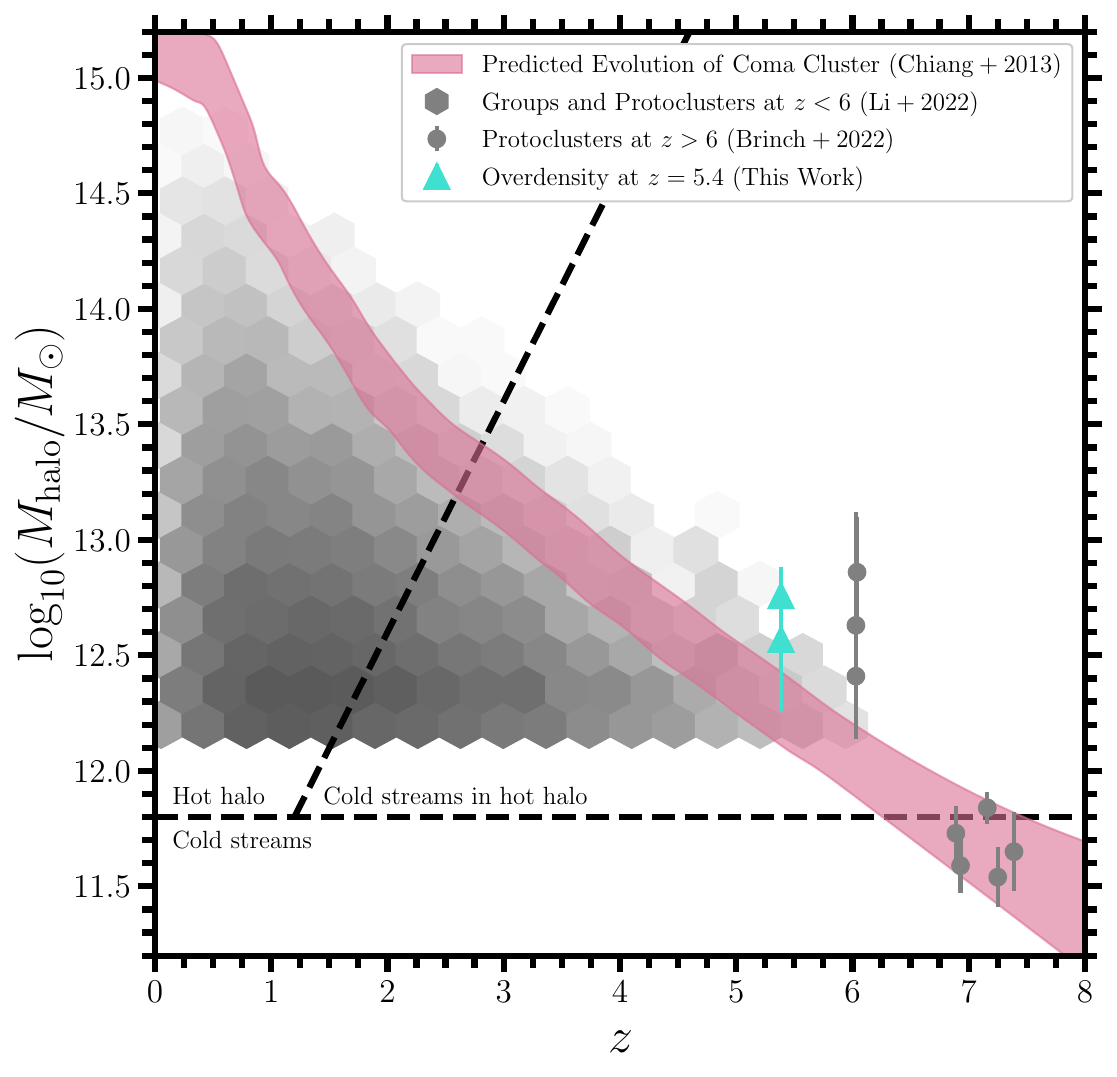}
    \caption{The dark matter halo mass distribution for groups and protoclusters at $z < 8$. The two halo mass estimates derived in Section~\ref{SectionThreeThree} are given by the turquoise triangles. For comparison, groups and protoclusters at $z < 6$ from \citet{Li:2022} are given in grayscale while protoclusters at $z > 6$ from \citet{Brinch:2022} are given by the grey points. The magenta shaded region shows the expected halo mass evolution of a Coma-like cluster \citep{Chiang:2013}. The black horizontal dashed line represents the typical threshold for shock stability assuming a spherical infall, below which the flows are predominantly cold and above which a shock-heated ICM is expected \citep{Dekel:2006}. The black diagonal dashed line represents the typical threshold for penetrating cold gas flows. The overdensity at $z = 5.4$ is expected to evolve into a Coma-like cluster with $\mathrm{log}_{10}(M_{\mathrm{halo}}/M_{\odot}) > 15$ by $z = 0$.
    \label{fig:protocluster_evolution}}
\end{figure*}

Figure~\ref{fig:protocluster_evolution} shows the dark matter halo mass distribution for some representative groups and protoclusters at $z < 8$. The two halo mass estimates derived here in Section~\ref{SectionThree} are given by the turquoise triangles, where the triangles indicate that these are likely underestimates of the true halo mass. Groups and protoclusters at $z < 6$ from \citet{Li:2022} are given in grayscale while protocluster candidates in the COSMOS field at $z > 6$ from \citet{Brinch:2022} are given by the grey points for comparison, both selected based on photometric redshifts, with dark matter halo mass estimates that assume the same stellar-to-halo abundance matching relation used here. The magenta shaded region shows the expected halo mass evolution of a Coma-like cluster \citep{Chiang:2013} assuming a smooth evolution at $z > 6$. The \textcolor{black}{black} dashed line represents the typical threshold for shock stability assuming a spherical infall, below which the flows are predominantly cold and above which a shock-heated ICM is expected \citep{Dekel:2006}. The black diagonal dashed line represents the typical threshold for penetrating cold gas flows. The overdensity identified in Section~\ref{SectionThreeOne} is expected to eventually evolve into a Coma-like cluster with $\mathrm{log}_{10}(M_{\mathrm{halo}}/M_{\odot}) > 15$ by $z = 0$, rivaling some of the most massive galaxy clusters found in the local Universe \citep{Ruel:2014, Buddendiek:2015}. For a Coma-like cluster at $z = 0$, the effective radius at $z = 5.4$ would be $R_{e} \approx 10\,\mathrm{cMpc}$ \citep{Chiang:2013}. This value is much larger than the physical size of this extreme galaxy overdensity (see Figure~\ref{fig:distribution_bimodality}). Thus, we conclude that the classification of this large-scale structure as a protocluster is justified. 

\subsection{Comparison with Previous Works}
\label{SectionThreeFour}

A number of previous works have found overdensities at similar redshifts to the one reported here. The earliest one identified at $z > 5$ is from \citet{Ouchi:2005}, who used narrow-band imaging with Subaru-Suprime Cam to identify galaxies with strong $\mathrm{Ly}\alpha$ emission lines at $z = 5.7$ ($\Delta z = 0.05$). The distribution of sources was described as ``clumpy'' with one prominent overdensity significant at the $4.8\sigma$ level, and a second one at the $2.2\sigma$ level. Follow-up spectroscopy confirmed the narrow redshift range for these clumps ($\Delta z = 0.03$), each being about $2^{\prime}$ in diameter with a separation of about $9^{\prime}$. The narrow redshift range combined with the small on-sky separations suggests that these two overdensities may be clumps within a large-scale structure analogous to the one reported here. 

Similar structures have been reported more recently \citep[e.g.,][]{ Jiang:2018, Chan:2019, Harikane:2019, Wang:2021}, with similar techniques in the first three cases, and based on overdensities in the sub-millimeter found with the South Pole Telescope in the fourth. \citet{Jiang:2018} found two overdensities near that which was identified by \citet{Ouchi:2005}, one at $z = 5.68$ with a diameter of $\sim 15^{\prime}$ and a smaller one at $z = 5.75$. These both are presumably related to the overdensities reported by \citet{Ouchi:2005}, given their proximity in redshift and on-sky. \citet{Chan:2019} and \citet{Harikane:2019} identified two similar structures at $z = 6.5$. Since the search method for these works starts with narrow-band imaging, the samples are from a narrow redshift range and the actual density of such overdensities on the sky must be significantly higher than implied by the existing detections. Thus, massive clusters of galaxies must be well on the way to formation by $z = 6$, which is something that we expect based on simulations \citep{Chiang:2013, Chiang:2017} and more recent observations \citep{Laporte:2022, Morishita:2022}. 

One of the key breakthroughs illustrated here with respect to these previous works is the efficient spectroscopic confirmation of such a large sample of galaxies at $z = 5.2-5.5$, made possible by the powerful combination of deep imaging and WFSS provided by JWST/NIRCam. As also presented in \citealt{Kashino:2022}, JWST/NIRCam WFSS enabled the discovery of three galaxy overdensities along the sight line of quasar $\mathrm{J}0100+2802$ at $z = 6.19$, $z_{\mathrm{quasar}} = 6.33$, and $z = 6.78$ through the blind detections of [\ion{O}{3}]-emitting galaxies. It is also worth mentioning that a similar $z = 5.2$ galaxy overdensity was identified in the GOODS-N field with a probable halo mass range of $12.3 < \mathrm{log}_{10} \left( M_{\mathrm{halo}}/M_{\odot} \right) < 12.9$ through $\mathrm{Ly}\alpha$ and sub-millimeter spectroscopy \citep{Walter:2012, Calvi:2021}, which was also partially observed by FRESCO in February 2023 \citep[see recent papers by][]{Herard-Demanche:2023, Sun:2023}. Future observations with JWST will certainly find more structures similar to those highlighted here, allowing a more complete look at the progenitors of the most massive gravitationally bound structures in the local Universe: galaxy clusters. 

%% Start of section four.
\section{Summary \& Conclusions}
\label{SectionFour}

We have presented the discovery of an extreme galaxy overdensity at $z = 5.4$ in the GOODS-S field using data from the Near Infrared Camera (NIRCam) on JWST. These data consist of JWST/NIRCam imaging from the JWST Advanced Deep Extragalactic Survey (JADES) and JWST/NIRCam wide field slitless spectroscopy from the First Reionization Epoch Spectroscopic COmplete Survey (FRESCO). Our findings can be summarized as follows.
\begin{enumerate}
    \item Galaxies were initially selected using HST+JWST photometry spanning $\lambda = 0.4-5.0\ \mu\mathrm{m}$. These data provide well-constrained photometric redshifts down to $m \approx 29-30\,\mathrm{mag}$, particularly at $z = 5.2-5.5$, where $\mathrm{H} \alpha$ excess can be traced by comparing photometry in the F410M and F444W filters. Galaxies were subsequently selected using slitless spectroscopy over $\lambda = 3.9-5.0\ \mu\mathrm{m}$ via a targeted emission line search for $\mathrm{H} \alpha$ around the best-fit photometric redshift. The final spectroscopic sample of galaxies includes $N = 81$ objects at $z = 5.2-5.5$.
    \item A Friends-of-Friends (FoF) algorithm was used to identify this extreme galaxy overdensity by iteratively looking for three-dimensional (3d) structural groupings within the final spectroscopic sample. One large-scale structure consisting of $N = 39$ galaxies was discovered, which is $\sim 10$ times more dense in one-dimension (1d) and $\sim 12$ times more dense in 3d than the $N = 42$ analogous field galaxies at $z = 5.2-5.5$.
    \item The stellar populations for these $N = 81$ objects at $z = 5.2-5.5$ were inferred using the HST+JWST photometry spanning $\lambda = 0.4-5.0\ \mu\mathrm{m}$, the spectroscopic redshifts determined by the targeted line search, and the SED fitting code \texttt{Prospector} \citep{Johnson:2021}. We constructed the star-forming main sequence at $z = 5.2-5.5$ and found that nearly all the galaxies in our sample agree with the empirically derived star-forming main sequence at $z = 5.4$ derived by \citet{Popesso:2023}. Combined with our relatively low SFR detection limit, this suggests that we are sampling the bulk of the star-forming population at these redshifts despite our $\mathrm{H}\alpha$ selection criteria. By comparing members of the overdensity with a mass-matched sample of members of the field, we find evidence suggesting that environment has induced earlier star formation and earlier stellar mass assembly within the overdensity relative to the field.
    \item Using two different methods, we estimated the total dark matter halo mass associated with this extreme galaxy overdensity at $z = 5.4$ to be within $12.6 \lesssim \mathrm{log}_{10} \left( M_{\mathrm{halo}}/M_{\odot} \right) \lesssim 12.8$. As a result of our selection criteria, we are potentially missing objects that fall outside either the JADES or the FRESCO footprints, as well as some subset of objects with relatively unconstrained photometric redshifts and/or low levels of star formation. This means the total dark matter halo mass range quoted above is likely an underestimate of the true halo mass. This massive large-scale structure is expected to evolve into a Coma-like cluster with $\mathrm{log}_{10}(M_{\mathrm{halo}}/M_{\odot}) > 15$ by $z = 0$.
\end{enumerate}

In this work, we have demonstrated the powerful combination of JWST/NIRCam imaging and slitless spectroscopy by efficiently confirming the redshifts for $N = 81$ galaxies at $z = 5.2-5.5$, inferring the physical properties of these galaxies, and assessing the large-scale structure in which these galaxies reside. Follow-up spectroscopic observations using JWST and/or the Atacama Large Millimeter/sub-millimeter Array (ALMA) will (1) inform us about the chemical compositions of these (and similar) galaxies, (2) provide insight into the formation and evolution of extreme galaxy overdensities in the early Universe, and (3) constrain the total number of these kinds of large-scale structures immediately after the epoch of reionization (EoR; $z > 6$) when the Universe was less than a billion years old.

%% Start of section five.
\section*{Acknowledgements}
\label{SectionFive}

This work is based on observations made with the NASA/ESA/CSA James Webb Space Telescope. The data were obtained from the Mikulski Archive for Space Telescopes \textcolor{black}{(MAST)} at the Space Telescope Science Institute, which is operated by the Association of Universities for Research in Astronomy, Inc., under NASA contract NAS 5-03127 for JWST. These observations are associated with program \#1180, 1210, 1895 and 1963. \textcolor{black}{The specific observations analyzed here can be accessed via \dataset[DOI: 10.17909/4k15-5x09]{https://doi.org/10.17909/4k15-5x09} and \dataset[DOI: 10.17909/T91019]{https://doi.org/10.17909/T91019}.} The authors sincerely thank the FRESCO team (PI: Pascal Oesch) for developing their observing program with a zero-exclusive-access period. Additionally, this work made use of the {\it lux} supercomputer at UC Santa Cruz which is funded by NSF MRI grant AST 1828315, as well as the High Performance Computing (HPC) resources at the University of Arizona which is funded by the Office of Research Discovery and Innovation (ORDI), Chief Information Officer (CIO), and University Information Technology Services (UITS).

We respectfully acknowledge the University of Arizona is on the land and territories of Indigenous peoples. Today, Arizona is home to 22 federally recognized tribes, with Tucson being home to the Oodham and the Yaqui. Committed to diversity and inclusion, the University strives to build sustainable relationships with sovereign Native Nations and Indigenous communities through education offerings, partnerships, and
community service.

FS, CNAW, GHR, MJR, BR, BDJ, DJE, and EE acknowledge support from the JWST/NIRCam contract to the University of Arizona, NAS5-02015. DJE is supported as a Simons Investigator. RH acknowledges funding from by the Johns Hopkins University, Institute for Data Intensive Engineering and Science (IDIES). NB acknowledges support from the Cosmic Dawn Center (DAWN), which is funded by the Danish National Research Foundation under grant No. 140. AB acknowledges funding from the “FirstGalaxies” Advanced Grant from the European Research Council (ERC) under the European Union’s Horizon 2020 research and innovation programme (Grant agreement No. 789056). ECL acknowledges support of an STFC Webb Fellowship (ST/W001438/1). TJL, RM, and JW acknowledge support from the ERC Advanced Grant 695671, ``QUENCH''. TJL and RM acknowledge support by the Science and Technology Facilities Council (STFC) and funding from a research professorship from the Royal Society. JW acknowledges support from the Fondation MERAC. KB acknowledges support in part by the Australian Research Council Centre of Excellence for All Sky Astrophysics in 3 Dimensions (ASTRO 3D), through project number CE170100013. REH acknowledges support from the National Science Foundation Graduate Research Fellowship Program under Grant No. DGE-1746060. 

\facilities{HST (ACS), JWST (NIRCam)}

\software{\texttt{AstroPy} \citep[][]{Astropy:2013, Astropy:2020}, \texttt{Cloudy} \citep{Byler:2017}, \texttt{dynesty} \citep[][]{Speagle:2020}, \texttt{FSPS} \citep[][]{Conroy:2009, Conroy:2010}, \texttt{Matplotlib} \citep[][]{Matplotlib:2007}, \texttt{NumPy} \citep[][]{NumPy:2011, NumPy:2020}, \texttt{Pandas} \citep[][]{Pandas:2022}, \texttt{photutils} \citep[][]{Bradley:2022}, \texttt{Prospector} \citep[][]{Johnson:2021}, \texttt{python-FSPS} \citep[][]{Foreman-Mackey:2014}, \texttt{SciPy} \citep[][]{SciPy:2020}, \texttt{seaborn} \citep[][]{Waskom:2021}, \texttt{TinyTim} \citep[][]{Krist:2011}, \texttt{WebbPSF} \citep[][]{Perrin:2014}}

% \newpage
\bibliographystyle{aasjournal}
\bibliography{main}{}

\begin{thebibliography}{}
\expandafter\ifx\csname natexlab\endcsname\relax\def\natexlab#1{#1}\fi
\providecommand{\url}[1]{\href{#1}{#1}}
\providecommand{\dodoi}[1]{doi:~\href{http://doi.org/#1}{\nolinkurl{#1}}}
\providecommand{\doeprint}[1]{\href{http://ascl.net/#1}{\nolinkurl{http://ascl.net/#1}}}
\providecommand{\doarXiv}[1]{\href{https://arxiv.org/abs/#1}{\nolinkurl{https://arxiv.org/abs/#1}}}

\bibitem[{{Alberts} \& {Noble}(2022)}]{Alberts:2022}
{Alberts}, S., \& {Noble}, A. 2022, Universe, 8, 554,
  \dodoi{10.3390/universe8110554}

\bibitem[{{Alberts} {et~al.}(2014){Alberts}, {Pope}, {Brodwin}, {Atlee}, {Lin},
  {Dey}, {Eisenhardt}, {Gettings}, {Gonzalez}, {Jannuzi}, {Mancone},
  {Moustakas}, {Snyder}, {Stanford}, {Stern}, {Weiner}, \&
  {Zeimann}}]{Alberts:2014}
{Alberts}, S., {Pope}, A., {Brodwin}, M., {et~al.} 2014, \mnras, 437, 437,
  \dodoi{10.1093/mnras/stt1897}

\bibitem[{{Alberts} {et~al.}(2016){Alberts}, {Pope}, {Brodwin}, {Chung},
  {Cybulski}, {Dey}, {Eisenhardt}, {Galametz}, {Gonzalez}, {Jannuzi},
  {Stanford}, {Snyder}, {Stern}, \& {Zeimann}}]{Alberts:2016}
---. 2016, \apj, 825, 72, \dodoi{10.3847/0004-637X/825/1/72}

\bibitem[{{Alberts} {et~al.}(2021){Alberts}, {Lee}, {Pope}, {Brodwin},
  {Chiang}, {McKinney}, {Xue}, {Huang}, {Brown}, {Dey}, {Eisenhardt},
  {Jannuzi}, {Popescu}, {Ramakrishnan}, {Stanford}, \& {Weiner}}]{Alberts:2021}
{Alberts}, S., {Lee}, K.-S., {Pope}, A., {et~al.} 2021, \mnras, 501, 1970,
  \dodoi{10.1093/mnras/staa3357}

\bibitem[{{Astropy Collaboration} {et~al.}(2013){Astropy Collaboration},
  {Robitaille}, {Tollerud}, {Greenfield}, {Droettboom}, {Bray}, {Aldcroft},
  {Davis}, {Ginsburg}, {Price-Whelan}, {Kerzendorf}, {Conley}, {Crighton},
  {Barbary}, {Muna}, {Ferguson}, {Grollier}, {Parikh}, {Nair}, {Unther},
  {Deil}, {Woillez}, {Conseil}, {Kramer}, {Turner}, {Singer}, {Fox}, {Weaver},
  {Zabalza}, {Edwards}, {Azalee Bostroem}, {Burke}, {Casey}, {Crawford},
  {Dencheva}, {Ely}, {Jenness}, {Labrie}, {Lim}, {Pierfederici}, {Pontzen},
  {Ptak}, {Refsdal}, {Servillat}, \& {Streicher}}]{Astropy:2013}
{Astropy Collaboration}, {Robitaille}, T.~P., {Tollerud}, E.~J., {et~al.} 2013,
  \aap, 558, A33, \dodoi{10.1051/0004-6361/201322068}

\bibitem[{{Astropy Collaboration} {et~al.}(2018){Astropy Collaboration},
  {Price-Whelan}, {Sip{\H{o}}cz}, {G{\"u}nther}, {Lim}, {Crawford}, {Conseil},
  {Shupe}, {Craig}, {Dencheva}, {Ginsburg}, {VanderPlas}, {Bradley},
  {P{\'e}rez-Su{\'a}rez}, {de Val-Borro}, {Aldcroft}, {Cruz}, {Robitaille},
  {Tollerud}, {Ardelean}, {Babej}, {Bach}, {Bachetti}, {Bakanov}, {Bamford},
  {Barentsen}, {Barmby}, {Baumbach}, {Berry}, {Biscani}, {Boquien}, {Bostroem},
  {Bouma}, {Brammer}, {Bray}, {Breytenbach}, {Buddelmeijer}, {Burke},
  {Calderone}, {Cano Rodr{\'\i}guez}, {Cara}, {Cardoso}, {Cheedella}, {Copin},
  {Corrales}, {Crichton}, {D'Avella}, {Deil}, {Depagne}, {Dietrich}, {Donath},
  {Droettboom}, {Earl}, {Erben}, {Fabbro}, {Ferreira}, {Finethy}, {Fox},
  {Garrison}, {Gibbons}, {Goldstein}, {Gommers}, {Greco}, {Greenfield},
  {Groener}, {Grollier}, {Hagen}, {Hirst}, {Homeier}, {Horton}, {Hosseinzadeh},
  {Hu}, {Hunkeler}, {Ivezi{\'c}}, {Jain}, {Jenness}, {Kanarek}, {Kendrew},
  {Kern}, {Kerzendorf}, {Khvalko}, {King}, {Kirkby}, {Kulkarni}, {Kumar},
  {Lee}, {Lenz}, {Littlefair}, {Ma}, {Macleod}, {Mastropietro}, {McCully},
  {Montagnac}, {Morris}, {Mueller}, {Mumford}, {Muna}, {Murphy}, {Nelson},
  {Nguyen}, {Ninan}, {N{\"o}the}, {Ogaz}, {Oh}, {Parejko}, {Parley}, {Pascual},
  {Patil}, {Patil}, {Plunkett}, {Prochaska}, {Rastogi}, {Reddy Janga},
  {Sabater}, {Sakurikar}, {Seifert}, {Sherbert}, {Sherwood-Taylor}, {Shih},
  {Sick}, {Silbiger}, {Singanamalla}, {Singer}, {Sladen}, {Sooley},
  {Sornarajah}, {Streicher}, {Teuben}, {Thomas}, {Tremblay}, {Turner},
  {Terr{\'o}n}, {van Kerkwijk}, {de la Vega}, {Watkins}, {Weaver}, {Whitmore},
  {Woillez}, {Zabalza}, \& {Astropy Contributors}}]{Astropy:2020}
{Astropy Collaboration}, {Price-Whelan}, A.~M., {Sip{\H{o}}cz}, B.~M., {et~al.}
  2018, \aj, 156, 123, \dodoi{10.3847/1538-3881/aabc4f}

\bibitem[{{Behroozi} {et~al.}(2013){Behroozi}, {Wechsler}, \&
  {Conroy}}]{Behroozi:2013}
{Behroozi}, P.~S., {Wechsler}, R.~H., \& {Conroy}, C. 2013, \apj, 770, 57,
  \dodoi{10.1088/0004-637X/770/1/57}

\bibitem[{{Berlind} {et~al.}(2006){Berlind}, {Frieman}, {Weinberg}, {Blanton},
  {Warren}, {Abazajian}, {Scranton}, {Hogg}, {Scoccimarro}, {Bahcall},
  {Brinkmann}, {Gott}, {Kleinman}, {Krzesinski}, {Lee}, {Miller}, {Nitta},
  {Schneider}, {Tucker}, {Zehavi}, \& {SDSS Collaboration}}]{Berlind:2006}
{Berlind}, A.~A., {Frieman}, J., {Weinberg}, D.~H., {et~al.} 2006, \apjs, 167,
  1, \dodoi{10.1086/508170}

\bibitem[{{Bertin} \& {Arnouts}(1996)}]{SourceExtractor}
{Bertin}, E., \& {Arnouts}, S. 1996, \aaps, 117, 393,
  \dodoi{10.1051/aas:1996164}

\bibitem[{{Bouwens} {et~al.}(2016){Bouwens}, {Smit}, {Labb{\'e}}, {Franx},
  {Caruana}, {Oesch}, {Stefanon}, \& {Rasappu}}]{Bouwens:2016}
{Bouwens}, R.~J., {Smit}, R., {Labb{\'e}}, I., {et~al.} 2016, \apj, 831, 176,
  \dodoi{10.3847/0004-637X/831/2/176}

\bibitem[{{Bouwens} {et~al.}(2015){Bouwens}, {Illingworth}, {Oesch}, {Trenti},
  {Labb{\'e}}, {Bradley}, {Carollo}, {van Dokkum}, {Gonzalez}, {Holwerda},
  {Franx}, {Spitler}, {Smit}, \& {Magee}}]{Bouwens:2015}
{Bouwens}, R.~J., {Illingworth}, G.~D., {Oesch}, P.~A., {et~al.} 2015, \apj,
  803, 34, \dodoi{10.1088/0004-637X/803/1/34}

\bibitem[{{Bradley} {et~al.}(2022){Bradley}, {Sip{\H{o}}cz}, {Robitaille},
  {Tollerud}, {Vin{\'\i}cius}, {Deil}, {Barbary}, {Wilson}, {Busko}, {Donath},
  {G{\"u}nther}, {Cara}, {Lim}, {Me{\ss}linger}, {Conseil}, {Bostroem},
  {Droettboom}, {Bray}, {Andersen Bratholm}, {Barentsen}, {Craig}, {Rathi},
  {Pascual}, {Perren}, {Georgiev}, {De Val-Borro}, {Kerzendorf}, {Bach},
  {Quint}, \& {Souchereau}}]{Bradley:2022}
{Bradley}, L., {Sip{\H{o}}cz}, B., {Robitaille}, T., {et~al.} 2022,
  {astropy/photutils: 1.5.0}, 1.5.0, Zenodo,  Zenodo,
  \dodoi{10.5281/zenodo.6825092}

\bibitem[{{Brammer} {et~al.}(2008){Brammer}, {van Dokkum}, \&
  {Coppi}}]{Brammer:2008}
{Brammer}, G.~B., {van Dokkum}, P.~G., \& {Coppi}, P. 2008, \apj, 686, 1503,
  \dodoi{10.1086/591786}

\bibitem[{{Brinch} {et~al.}(2023){Brinch}, {Greve}, {Weaver}, {Brammer},
  {Ilbert}, {Shuntov}, {Jin}, {Liu}, {Gim{\'e}nez-Arteaga}, {Casey},
  {Davidson}, {Fujimoto}, {Koekemoer}, {Kokorev}, {Magdis}, {McCracken},
  {McPartland}, {Mobasher}, {Sanders}, {Toft}, {Valentino}, {Zamorani},
  {Zavala}, \& {Cosmos Team}}]{Brinch:2022}
{Brinch}, M., {Greve}, T.~R., {Weaver}, J.~R., {et~al.} 2023, \apj, 943, 153,
  \dodoi{10.3847/1538-4357/ac9d96}

\bibitem[{{Buddendiek} {et~al.}(2015){Buddendiek}, {Schrabback}, {Greer},
  {Hoekstra}, {Sommer}, {Eifler}, {Erben}, {Erler}, {Hicks}, {High},
  {Hildebrandt}, {Marrone}, {Morris}, {Muzzin}, {Reiprich}, {Schirmer},
  {Schneider}, \& {von der Linden}}]{Buddendiek:2015}
{Buddendiek}, A., {Schrabback}, T., {Greer}, C.~H., {et~al.} 2015, \mnras, 450,
  4248, \dodoi{10.1093/mnras/stv783}

\bibitem[{{Byler} {et~al.}(2017){Byler}, {Dalcanton}, {Conroy}, \&
  {Johnson}}]{Byler:2017}
{Byler}, N., {Dalcanton}, J.~J., {Conroy}, C., \& {Johnson}, B.~D. 2017, \apj,
  840, 44, \dodoi{10.3847/1538-4357/aa6c66}

\bibitem[{{Calvi} {et~al.}(2021){Calvi}, {Dannerbauer}, {Arrabal Haro},
  {Rodr{\'\i}guez Espinosa}, {Mu{\~n}oz-Tu{\~n}{\'o}n}, {P{\'e}rez
  Gonz{\'a}lez}, \& {Geier}}]{Calvi:2021}
{Calvi}, R., {Dannerbauer}, H., {Arrabal Haro}, P., {et~al.} 2021, \mnras, 502,
  4558, \dodoi{10.1093/mnras/staa4037}

\bibitem[{{Calzetti} {et~al.}(2000){Calzetti}, {Armus}, {Bohlin}, {Kinney},
  {Koornneef}, \& {Storchi-Bergmann}}]{Calzetti:2000}
{Calzetti}, D., {Armus}, L., {Bohlin}, R.~C., {et~al.} 2000, \apj, 533, 682,
  \dodoi{10.1086/308692}

\bibitem[{{Cameron} {et~al.}(2023){Cameron}, {Saxena}, {Bunker}, {D'Eugenio},
  {Carniani}, {Maiolino}, {Curtis-Lake}, {Ferruit}, {Jakobsen}, {Arribas},
  {Bonaventura}, {Charlot}, {Chevallard}, {Curti}, {Looser}, {Maseda}, {Rawle},
  {Rodr{\'\i}guez Del Pino}, {Smit}, {{\"U}bler}, {Willott}, {Witstok},
  {Egami}, {Eisenstein}, {Johnson}, {Hainline}, {Rieke}, {Robertson}, {Stark},
  {Tacchella}, {Williams}, {Willmer}, {Bhatawdekar}, {Bowler}, {Boyett},
  {Circosta}, {Helton}, {Jones}, {Kumari}, {Ji}, {Nelson}, {Parlanti},
  {Sandles}, {Scholtz}, \& {Sun}}]{Cameron:2023}
{Cameron}, A.~J., {Saxena}, A., {Bunker}, A.~J., {et~al.} 2023, \aap, 677,
  A115, \dodoi{10.1051/0004-6361/202346107}

\bibitem[{{Chabrier}(2003)}]{Chabrier:2003}
{Chabrier}, G. 2003, \pasp, 115, 763, \dodoi{10.1086/376392}

\bibitem[{{Chanchaiworawit} {et~al.}(2019){Chanchaiworawit}, {Guzm{\'a}n},
  {Salvador-Sol{\'e}}, {Rodr{\'\i}guez Espinosa}, {Calvi}, {Manrique},
  {Gallego}, {Herrero}, {Mar{\'\i}n-Franch}, \& {Mas-Hesse}}]{Chan:2019}
{Chanchaiworawit}, K., {Guzm{\'a}n}, R., {Salvador-Sol{\'e}}, E., {et~al.}
  2019, \apj, 877, 51, \dodoi{10.3847/1538-4357/ab1a34}

\bibitem[{{Charlot} \& {Fall}(2000)}]{Charlot:2000}
{Charlot}, S., \& {Fall}, S.~M. 2000, \apj, 539, 718, \dodoi{10.1086/309250}

\bibitem[{{Chiang} {et~al.}(2013){Chiang}, {Overzier}, \&
  {Gebhardt}}]{Chiang:2013}
{Chiang}, Y.-K., {Overzier}, R., \& {Gebhardt}, K. 2013, \apj, 779, 127,
  \dodoi{10.1088/0004-637X/779/2/127}

\bibitem[{{Chiang} {et~al.}(2017){Chiang}, {Overzier}, {Gebhardt}, \&
  {Henriques}}]{Chiang:2017}
{Chiang}, Y.-K., {Overzier}, R.~A., {Gebhardt}, K., \& {Henriques}, B. 2017,
  \apjl, 844, L23, \dodoi{10.3847/2041-8213/aa7e7b}

\bibitem[{{Choi} {et~al.}(2016){Choi}, {Dotter}, {Conroy}, {Cantiello},
  {Paxton}, \& {Johnson}}]{Choi:2016}
{Choi}, J., {Dotter}, A., {Conroy}, C., {et~al.} 2016, \apj, 823, 102,
  \dodoi{10.3847/0004-637X/823/2/102}

\bibitem[{{Conroy} \& {Gunn}(2010)}]{Conroy:2010}
{Conroy}, C., \& {Gunn}, J.~E. 2010, \apj, 712, 833,
  \dodoi{10.1088/0004-637X/712/2/833}

\bibitem[{{Conroy} {et~al.}(2009){Conroy}, {Gunn}, \& {White}}]{Conroy:2009}
{Conroy}, C., {Gunn}, J.~E., \& {White}, M. 2009, \apj, 699, 486,
  \dodoi{10.1088/0004-637X/699/1/486}

\bibitem[{{Dekel} \& {Birnboim}(2006)}]{Dekel:2006}
{Dekel}, A., \& {Birnboim}, Y. 2006, \mnras, 368, 2,
  \dodoi{10.1111/j.1365-2966.2006.10145.x}

\bibitem[{{Dotter}(2016)}]{Dotter:2016}
{Dotter}, A. 2016, \apjs, 222, 8, \dodoi{10.3847/0067-0049/222/1/8}

\bibitem[{{Dressler}(1980)}]{Dressler:1980}
{Dressler}, A. 1980, \apj, 236, 351, \dodoi{10.1086/157753}

\bibitem[{{Eisenstein} {et~al.}(2023){Eisenstein}, {Willott}, {Alberts},
  {Arribas}, {Bonaventura}, {Bunker}, {Cameron}, {Carniani}, {Charlot},
  {Curtis-Lake}, {D'Eugenio}, {Endsley}, {Ferruit}, {Giardino}, {Hainline},
  {Hausen}, {Jakobsen}, {Johnson}, {Maiolino}, {Rieke}, {Rieke}, {Rix},
  {Robertson}, {Stark}, {Tacchella}, {Williams}, {Willmer}, {Baker}, {Baum},
  {Bhatawdekar}, {Boyett}, {Chen}, {Chevallard}, {Circosta}, {Curti},
  {Danhaive}, {DeCoursey}, {de Graaff}, {Dressler}, {Egami}, {Helton},
  {Hviding}, {Ji}, {Jones}, {Kumari}, {L{\"u}tzgendorf}, {Laseter}, {Looser},
  {Lyu}, {Maseda}, {Nelson}, {Parlanti}, {Perna}, {Pusk{\'a}s}, {Rawle},
  {Rodr{\'\i}guez Del Pino}, {Sandles}, {Saxena}, {Scholtz}, {Sharpe},
  {Shivaei}, {Silcock}, {Simmonds}, {Skarbinski}, {Smit}, {Stone}, {Suess},
  {Sun}, {Tang}, {Topping}, {{\"U}bler}, {Villanueva}, {Wallace}, {Whitler},
  {Witstok}, \& {Woodrum}}]{Eisenstein:2023}
{Eisenstein}, D.~J., {Willott}, C., {Alberts}, S., {et~al.} 2023, arXiv
  e-prints, arXiv:2306.02465, \dodoi{10.48550/arXiv.2306.02465}

\bibitem[{{Eke} {et~al.}(2004){Eke}, {Frenk}, {Baugh}, {Cole}, {Norberg},
  {Peacock}, {Baldry}, {Bland-Hawthorn}, {Bridges}, {Cannon}, {Colless},
  {Collins}, {Couch}, {Dalton}, {de Propris}, {Driver}, {Efstathiou}, {Ellis},
  {Glazebrook}, {Jackson}, {Lahav}, {Lewis}, {Lumsden}, {Maddox}, {Madgwick},
  {Peterson}, {Sutherland}, \& {Taylor}}]{Eke:2004}
{Eke}, V.~R., {Frenk}, C.~S., {Baugh}, C.~M., {et~al.} 2004, \mnras, 355, 769,
  \dodoi{10.1111/j.1365-2966.2004.08354.x}

\bibitem[{{Falc{\'o}n-Barroso} {et~al.}(2011){Falc{\'o}n-Barroso},
  {S{\'a}nchez-Bl{\'a}zquez}, {Vazdekis}, {Ricciardelli}, {Cardiel}, {Cenarro},
  {Gorgas}, \& {Peletier}}]{Falcon-Barroso:2011}
{Falc{\'o}n-Barroso}, J., {S{\'a}nchez-Bl{\'a}zquez}, P., {Vazdekis}, A.,
  {et~al.} 2011, \aap, 532, A95, \dodoi{10.1051/0004-6361/201116842}

\bibitem[{{Foreman-Mackey} {et~al.}(2014){Foreman-Mackey}, {Sick}, \&
  {Johnson}}]{Foreman-Mackey:2014}
{Foreman-Mackey}, D., {Sick}, J., \& {Johnson}, B. 2014, {python-fsps: Python
  bindings to FSPS (v0.1.1)}, v0.1.1, Zenodo,  Zenodo,
  \dodoi{10.5281/zenodo.12157}

\bibitem[{{Franco} {et~al.}(2018){Franco}, {Elbaz}, {B{\'e}thermin},
  {Magnelli}, {Schreiber}, {Ciesla}, {Dickinson}, {Nagar}, {Silverman},
  {Daddi}, {Alexander}, {Wang}, {Pannella}, {Le Floc'h}, {Pope}, {Giavalisco},
  {Maury}, {Bournaud}, {Chary}, {Demarco}, {Ferguson}, {Finkelstein}, {Inami},
  {Iono}, {Juneau}, {Lagache}, {Leiton}, {Lin}, {Magdis}, {Messias},
  {Motohara}, {Mullaney}, {Okumura}, {Papovich}, {Pforr}, {Rujopakarn},
  {Sargent}, {Shu}, \& {Zhou}}]{Franco:2018}
{Franco}, M., {Elbaz}, D., {B{\'e}thermin}, M., {et~al.} 2018, \aap, 620, A152,
  \dodoi{10.1051/0004-6361/201832928}

\bibitem[{{Gaia Collaboration} {et~al.}(2023){Gaia Collaboration}, {Vallenari},
  {Brown}, {Prusti}, {de Bruijne}, {Arenou}, {Babusiaux}, {Biermann},
  {Creevey}, {Ducourant}, \& et~al.}]{GaiaDR3}
{Gaia Collaboration}, {Vallenari}, A., {Brown}, A.~G.~A., {et~al.} 2023, \aap,
  674, A1, \dodoi{10.1051/0004-6361/202243940}

\bibitem[{{Giavalisco} {et~al.}(2004){Giavalisco}, {Ferguson}, {Koekemoer},
  {Dickinson}, {Alexander}, {Bauer}, {Bergeron}, {Biagetti}, {Brandt},
  {Casertano}, {Cesarsky}, {Chatzichristou}, {Conselice}, {Cristiani}, {Da
  Costa}, {Dahlen}, {de Mello}, {Eisenhardt}, {Erben}, {Fall}, {Fassnacht},
  {Fosbury}, {Fruchter}, {Gardner}, {Grogin}, {Hook}, {Hornschemeier}, {Idzi},
  {Jogee}, {Kretchmer}, {Laidler}, {Lee}, {Livio}, {Lucas}, {Madau},
  {Mobasher}, {Moustakas}, {Nonino}, {Padovani}, {Papovich}, {Park},
  {Ravindranath}, {Renzini}, {Richardson}, {Riess}, {Rosati}, {Schirmer},
  {Schreier}, {Somerville}, {Spinrad}, {Stern}, {Stiavelli}, {Strolger},
  {Urry}, {Vandame}, {Williams}, \& {Wolf}}]{Giavalisco:2004}
{Giavalisco}, M., {Ferguson}, H.~C., {Koekemoer}, A.~M., {et~al.} 2004, \apjl,
  600, L93, \dodoi{10.1086/379232}

\bibitem[{{G{\'o}mez-Guijarro} {et~al.}(2022){G{\'o}mez-Guijarro}, {Elbaz},
  {Xiao}, {B{\'e}thermin}, {Franco}, {Magnelli}, {Daddi}, {Dickinson},
  {Demarco}, {Inami}, {Rujopakarn}, {Magdis}, {Shu}, {Chary}, {Zhou},
  {Alexander}, {Bournaud}, {Ciesla}, {Ferguson}, {Finkelstein}, {Giavalisco},
  {Iono}, {Juneau}, {Kartaltepe}, {Lagache}, {Le Floc'h}, {Leiton}, {Lin},
  {Motohara}, {Mullaney}, {Okumura}, {Pannella}, {Papovich}, {Pope}, {Sargent},
  {Silverman}, {Treister}, \& {Wang}}]{Gomez-Guijarro:2022}
{G{\'o}mez-Guijarro}, C., {Elbaz}, D., {Xiao}, M., {et~al.} 2022, \aap, 658,
  A43, \dodoi{10.1051/0004-6361/202141615}

\bibitem[{{Gonz{\'a}lez-L{\'o}pez} {et~al.}(2020){Gonz{\'a}lez-L{\'o}pez},
  {Novak}, {Decarli}, {Walter}, {Aravena}, {Carilli}, {Boogaard}, {Popping},
  {Weiss}, {Assef}, {Bauer}, {Bouwens}, {Cortes}, {Cox}, {Daddi}, {Cunha},
  {D{\'\i}az-Santos}, {Ivison}, {Magnelli}, {Riechers}, {Smail}, {van der
  Werf}, \& {Wagg}}]{Gonzalez-Lopez:2020}
{Gonz{\'a}lez-L{\'o}pez}, J., {Novak}, M., {Decarli}, R., {et~al.} 2020, \apj,
  897, 91, \dodoi{10.3847/1538-4357/ab765b}

\bibitem[{{Harikane} {et~al.}(2019){Harikane}, {Ouchi}, {Ono}, {Fujimoto},
  {Donevski}, {Shibuya}, {Faisst}, {Goto}, {Hatsukade}, {Kashikawa}, {Kohno},
  {Hashimoto}, {Higuchi}, {Inoue}, {Lin}, {Martin}, {Overzier}, {Smail},
  {Toshikawa}, {Umehata}, {Ao}, {Chapman}, {Clements}, {Im}, {Jing},
  {Kawaguchi}, {Lee}, {Lee}, {Lin}, {Matsuoka}, {Marinello}, {Nagao},
  {Onodera}, {Toft}, \& {Wang}}]{Harikane:2019}
{Harikane}, Y., {Ouchi}, M., {Ono}, Y., {et~al.} 2019, \apj, 883, 142,
  \dodoi{10.3847/1538-4357/ab2cd5}

\bibitem[{{Harris} {et~al.}(2020){Harris}, {Millman}, {van der Walt},
  {Gommers}, {Virtanen}, {Cournapeau}, {Wieser}, {Taylor}, {Berg}, {Smith},
  {Kern}, {Picus}, {Hoyer}, {van Kerkwijk}, {Brett}, {Haldane}, {del R{\'\i}o},
  {Wiebe}, {Peterson}, {G{\'e}rard-Marchant}, {Sheppard}, {Reddy}, {Weckesser},
  {Abbasi}, {Gohlke}, \& {Oliphant}}]{NumPy:2020}
{Harris}, C.~R., {Millman}, K.~J., {van der Walt}, S.~J., {et~al.} 2020, \nat,
  585, 357, \dodoi{10.1038/s41586-020-2649-2}

\bibitem[{{Hatsukade} {et~al.}(2018){Hatsukade}, {Kohno}, {Yamaguchi},
  {Umehata}, {Ao}, {Aretxaga}, {Caputi}, {Dunlop}, {Egami}, {Espada},
  {Fujimoto}, {Hayatsu}, {Hughes}, {Ikarashi}, {Iono}, {Ivison}, {Kawabe},
  {Kodama}, {Lee}, {Matsuda}, {Nakanishi}, {Ohta}, {Ouchi}, {Rujopakarn},
  {Suzuki}, {Tamura}, {Ueda}, {Wang}, {Wang}, {Wilson}, {Yoshimura}, \&
  {Yun}}]{Hatsukade:2018}
{Hatsukade}, B., {Kohno}, K., {Yamaguchi}, Y., {et~al.} 2018, \pasj, 70, 105,
  \dodoi{10.1093/pasj/psy104}

\bibitem[{{Herard-Demanche} {et~al.}(2023){Herard-Demanche}, {Bouwens},
  {Oesch}, {Naidu}, {Decarli}, {Nelson}, {Brammer}, {Weibel}, {Xiao},
  {Stefanon}, {Walter}, {Matthee}, {Meyer}, {Wuyts}, {Reddy}, {Arrabal Haro},
  {Dannerbauer}, {Shapley}, {Chisholm}, {van Dokkum}, {Labbe}, {Illingworth},
  {Schaerer}, \& {Shivaei}}]{Herard-Demanche:2023}
{Herard-Demanche}, T., {Bouwens}, R.~J., {Oesch}, P.~A., {et~al.} 2023, arXiv
  e-prints, arXiv:2309.04525, \dodoi{10.48550/arXiv.2309.04525}

\bibitem[{{Horne}(1986)}]{Horne:1986}
{Horne}, K. 1986, \pasp, 98, 609, \dodoi{10.1086/131801}

\bibitem[{{Huchra} \& {Geller}(1982)}]{Huchra:1982}
{Huchra}, J.~P., \& {Geller}, M.~J. 1982, \apj, 257, 423,
  \dodoi{10.1086/160000}

\bibitem[{{Hunter}(2007)}]{Matplotlib:2007}
{Hunter}, J.~D. 2007, Computing in Science and Engineering, 9, 90,
  \dodoi{10.1109/MCSE.2007.55}

\bibitem[{{Illingworth} {et~al.}(2016){Illingworth}, {Magee}, {Bouwens},
  {Oesch}, {Labbe}, {van Dokkum}, {Whitaker}, {Holden}, {Franx}, \&
  {Gonzalez}}]{Illingworth:2016}
{Illingworth}, G., {Magee}, D., {Bouwens}, R., {et~al.} 2016, arXiv e-prints,
  arXiv:1606.00841, \dodoi{10.48550/arXiv.1606.00841}

\bibitem[{{Inami} {et~al.}(2017){Inami}, {Bacon}, {Brinchmann}, {Richard},
  {Contini}, {Conseil}, {Hamer}, {Akhlaghi}, {Bouch{\'e}}, {Cl{\'e}ment},
  {Desprez}, {Drake}, {Hashimoto}, {Leclercq}, {Maseda}, {Michel-Dansac},
  {Paalvast}, {Tresse}, {Ventou}, {Kollatschny}, {Boogaard}, {Finley},
  {Marino}, {Schaye}, \& {Wisotzki}}]{Inami:2017}
{Inami}, H., {Bacon}, R., {Brinchmann}, J., {et~al.} 2017, \aap, 608, A2,
  \dodoi{10.1051/0004-6361/201731195}

\bibitem[{{Jiang} {et~al.}(2018){Jiang}, {Wu}, {Bian}, {Chiang}, {Ho}, {Shen},
  {Zheng}, {Bailey}, {Blanc}, {Crane}, {Fan}, {Mateo}, {Olszewski},
  {Oyarz{\'u}n}, {Wang}, \& {Wu}}]{Jiang:2018}
{Jiang}, L., {Wu}, J., {Bian}, F., {et~al.} 2018, Nature Astronomy, 2, 962,
  \dodoi{10.1038/s41550-018-0587-9}

\bibitem[{{Johnson} {et~al.}(2021){Johnson}, {Leja}, {Conroy}, \&
  {Speagle}}]{Johnson:2021}
{Johnson}, B.~D., {Leja}, J., {Conroy}, C., \& {Speagle}, J.~S. 2021, \apjs,
  254, 22, \dodoi{10.3847/1538-4365/abef67}

\bibitem[{{Kashino} {et~al.}(2023){Kashino}, {Lilly}, {Matthee}, {Eilers},
  {Mackenzie}, {Bordoloi}, \& {Simcoe}}]{Kashino:2022}
{Kashino}, D., {Lilly}, S.~J., {Matthee}, J., {et~al.} 2023, \apj, 950, 66,
  \dodoi{10.3847/1538-4357/acc588}

\bibitem[{{Kennicutt} \& {Evans}(2012)}]{Kennicutt:2012}
{Kennicutt}, R.~C., \& {Evans}, N.~J. 2012, \araa, 50, 531,
  \dodoi{10.1146/annurev-astro-081811-125610}

\bibitem[{{Kriek} \& {Conroy}(2013)}]{Kriek:2013}
{Kriek}, M., \& {Conroy}, C. 2013, \apjl, 775, L16,
  \dodoi{10.1088/2041-8205/775/1/L16}

\bibitem[{{Krist} {et~al.}(2011){Krist}, {Hook}, \& {Stoehr}}]{Krist:2011}
{Krist}, J.~E., {Hook}, R.~N., \& {Stoehr}, F. 2011, in Society of
  Photo-Optical Instrumentation Engineers (SPIE) Conference Series, Vol. 8127,
  Optical Modeling and Performance Predictions V, ed. M.~A. {Kahan}, 81270J,
  \dodoi{10.1117/12.892762}

\bibitem[{{Labb{\'e}} {et~al.}(2005){Labb{\'e}}, {Huang}, {Franx}, {Rudnick},
  {Barmby}, {Daddi}, {van Dokkum}, {Fazio}, {F{\"o}rster Schreiber},
  {Moorwood}, {Rix}, {R{\"o}ttgering}, {Trujillo}, \& {van der
  Werf}}]{Labbe:2005}
{Labb{\'e}}, I., {Huang}, J., {Franx}, M., {et~al.} 2005, \apjl, 624, L81,
  \dodoi{10.1086/430700}

\bibitem[{{Laporte} {et~al.}(2022){Laporte}, {Zitrin}, {Dole},
  {Roberts-Borsani}, {Furtak}, \& {Witten}}]{Laporte:2022}
{Laporte}, N., {Zitrin}, A., {Dole}, H., {et~al.} 2022, \aap, 667, L3,
  \dodoi{10.1051/0004-6361/202244719}

\bibitem[{{Li} {et~al.}(2022){Li}, {Yang}, {Liu}, {Jing}, {He}, {Huang}, {Dai},
  {Sawicki}, {Arnouts}, {Gwyn}, {Moutard}, {Mo}, {Wang}, {Katsianis}, {Cui},
  {Han}, {Chiu}, {Gu}, \& {Xu}}]{Li:2022}
{Li}, Q., {Yang}, X., {Liu}, C., {et~al.} 2022, \apj, 933, 9,
  \dodoi{10.3847/1538-4357/ac6e69}

\bibitem[{{Long} {et~al.}(2020){Long}, {Cooray}, {Ma}, {Casey}, {Wardlow},
  {Nayyeri}, {Ivison}, {Farrah}, \& {Dannerbauer}}]{Long:2020}
{Long}, A.~S., {Cooray}, A., {Ma}, J., {et~al.} 2020, \apj, 898, 133,
  \dodoi{10.3847/1538-4357/ab9d1f}

\bibitem[{{Madau}(1995)}]{Madau:1995}
{Madau}, P. 1995, \apj, 441, 18, \dodoi{10.1086/175332}

\bibitem[{{Matthee} {et~al.}(2023){Matthee}, {Naidu}, {Brammer}, {Chisholm},
  {Eilers}, {Goulding}, {Greene}, {Kashino}, {Labbe}, {Lilly}, {Mackenzie},
  {Oesch}, {Weibel}, {Wuyts}, {Xiao}, {Bordoloi}, {Bouwens}, {van Dokkum},
  {Illingworth}, {Kramarenko}, {Maseda}, {Mason}, {Meyer}, {Nelson}, {Reddy},
  {Shivaei}, {Simcoe}, \& {Yue}}]{Matthee:2023}
{Matthee}, J., {Naidu}, R.~P., {Brammer}, G., {et~al.} 2023, arXiv e-prints,
  arXiv:2306.05448, \dodoi{10.48550/arXiv.2306.05448}

\bibitem[{{Morishita} {et~al.}(2023){Morishita}, {Roberts-Borsani}, {Treu},
  {Brammer}, {Mason}, {Trenti}, {Vulcani}, {Wang}, {Acebron}, {Bah{\'e}},
  {Bergamini}, {Boyett}, {Bradac}, {Calabr{\`o}}, {Castellano}, {Chen}, {De
  Lucia}, {Filippenko}, {Fontana}, {Glazebrook}, {Grillo}, {Henry}, {Jones},
  {Kelly}, {Koekemoer}, {Leethochawalit}, {Lu}, {Marchesini}, {Mascia},
  {Mercurio}, {Merlin}, {Metha}, {Nanayakkara}, {Nonino}, {Paris},
  {Pentericci}, {Rosati}, {Santini}, {Strait}, {Vanzella}, {Windhorst}, \&
  {Xie}}]{Morishita:2022}
{Morishita}, T., {Roberts-Borsani}, G., {Treu}, T., {et~al.} 2023, \apjl, 947,
  L24, \dodoi{10.3847/2041-8213/acb99e}

\bibitem[{{Newman} {et~al.}(2013){Newman}, {Cooper}, {Davis}, {Faber}, {Coil},
  {Guhathakurta}, {Koo}, {Phillips}, {Conroy}, {Dutton}, {Finkbeiner}, {Gerke},
  {Rosario}, {Weiner}, {Willmer}, {Yan}, {Harker}, {Kassin}, {Konidaris},
  {Lai}, {Madgwick}, {Noeske}, {Wirth}, {Connolly}, {Kaiser}, {Kirby},
  {Lemaux}, {Lin}, {Lotz}, {Luppino}, {Marinoni}, {Matthews}, {Metevier}, \&
  {Schiavon}}]{Newman:2013}
{Newman}, J.~A., {Cooper}, M.~C., {Davis}, M., {et~al.} 2013, \apjs, 208, 5,
  \dodoi{10.1088/0067-0049/208/1/5}

\bibitem[{{Ning} {et~al.}(2023){Ning}, {Cai}, {Jiang}, {Lin}, {Fu}, \&
  {Spinoso}}]{Ning:2022}
{Ning}, Y., {Cai}, Z., {Jiang}, L., {et~al.} 2023, \apjl, 944, L1,
  \dodoi{10.3847/2041-8213/acb26b}

\bibitem[{{Oke} \& {Gunn}(1983)}]{Oke:1983}
{Oke}, J.~B., \& {Gunn}, J.~E. 1983, \apj, 266, 713, \dodoi{10.1086/160817}

\bibitem[{{Ouchi} {et~al.}(2005){Ouchi}, {Shimasaku}, {Akiyama}, {Sekiguchi},
  {Furusawa}, {Okamura}, {Kashikawa}, {Iye}, {Kodama}, {Saito}, {Sasaki},
  {Simpson}, {Takata}, {Yamada}, {Yamanoi}, {Yoshida}, \&
  {Yoshida}}]{Ouchi:2005}
{Ouchi}, M., {Shimasaku}, K., {Akiyama}, M., {et~al.} 2005, \apjl, 620, L1,
  \dodoi{10.1086/428499}

\bibitem[{{Overzier}(2016)}]{Overzier:2016}
{Overzier}, R.~A. 2016, \aapr, 24, 14, \dodoi{10.1007/s00159-016-0100-3}

\bibitem[{{Paxton} {et~al.}(2011){Paxton}, {Bildsten}, {Dotter}, {Herwig},
  {Lesaffre}, \& {Timmes}}]{Paxton:2011}
{Paxton}, B., {Bildsten}, L., {Dotter}, A., {et~al.} 2011, \apjs, 192, 3,
  \dodoi{10.1088/0067-0049/192/1/3}

\bibitem[{{Paxton} {et~al.}(2013){Paxton}, {Cantiello}, {Arras}, {Bildsten},
  {Brown}, {Dotter}, {Mankovich}, {Montgomery}, {Stello}, {Timmes}, \&
  {Townsend}}]{Paxton:2013}
{Paxton}, B., {Cantiello}, M., {Arras}, P., {et~al.} 2013, \apjs, 208, 4,
  \dodoi{10.1088/0067-0049/208/1/4}

\bibitem[{{Paxton} {et~al.}(2015){Paxton}, {Marchant}, {Schwab}, {Bauer},
  {Bildsten}, {Cantiello}, {Dessart}, {Farmer}, {Hu}, {Langer}, {Townsend},
  {Townsley}, \& {Timmes}}]{Paxton:2015}
{Paxton}, B., {Marchant}, P., {Schwab}, J., {et~al.} 2015, \apjs, 220, 15,
  \dodoi{10.1088/0067-0049/220/1/15}

\bibitem[{{Paxton} {et~al.}(2018){Paxton}, {Schwab}, {Bauer}, {Bildsten},
  {Blinnikov}, {Duffell}, {Farmer}, {Goldberg}, {Marchant}, {Sorokina},
  {Thoul}, {Townsend}, \& {Timmes}}]{Paxton:2018}
{Paxton}, B., {Schwab}, J., {Bauer}, E.~B., {et~al.} 2018, \apjs, 234, 34,
  \dodoi{10.3847/1538-4365/aaa5a8}

\bibitem[{{Perrin} {et~al.}(2014){Perrin}, {Sivaramakrishnan}, {Lajoie},
  {Elliott}, {Pueyo}, {Ravindranath}, \& {Albert}}]{Perrin:2014}
{Perrin}, M.~D., {Sivaramakrishnan}, A., {Lajoie}, C.-P., {et~al.} 2014, in
  Society of Photo-Optical Instrumentation Engineers (SPIE) Conference Series,
  Vol. 9143, Space Telescopes and Instrumentation 2014: Optical, Infrared, and
  Millimeter Wave, ed. J.~{Oschmann}, Jacobus~M., M.~{Clampin}, G.~G. {Fazio},
  \& H.~A. {MacEwen}, 91433X, \dodoi{10.1117/12.2056689}

\bibitem[{{Planck Collaboration} {et~al.}(2020){Planck Collaboration},
  {Aghanim}, {Akrami}, {Ashdown}, {Aumont}, {Baccigalupi}, {Ballardini},
  {Banday}, {Barreiro}, {Bartolo}, {Basak}, {Battye}, {Benabed}, {Bernard},
  {Bersanelli}, {Bielewicz}, {Bock}, {Bond}, {Borrill}, {Bouchet}, {Boulanger},
  {Bucher}, {Burigana}, {Butler}, {Calabrese}, {Cardoso}, {Carron},
  {Challinor}, {Chiang}, {Chluba}, {Colombo}, {Combet}, {Contreras}, {Crill},
  {Cuttaia}, {de Bernardis}, {de Zotti}, {Delabrouille}, {Delouis}, {Di
  Valentino}, {Diego}, {Dor{\'e}}, {Douspis}, {Ducout}, {Dupac}, {Dusini},
  {Efstathiou}, {Elsner}, {En{\ss}lin}, {Eriksen}, {Fantaye}, {Farhang},
  {Fergusson}, {Fernandez-Cobos}, {Finelli}, {Forastieri}, {Frailis},
  {Fraisse}, {Franceschi}, {Frolov}, {Galeotta}, {Galli}, {Ganga},
  {G{\'e}nova-Santos}, {Gerbino}, {Ghosh}, {Gonz{\'a}lez-Nuevo}, {G{\'o}rski},
  {Gratton}, {Gruppuso}, {Gudmundsson}, {Hamann}, {Handley}, {Hansen},
  {Herranz}, {Hildebrandt}, {Hivon}, {Huang}, {Jaffe}, {Jones}, {Karakci},
  {Keih{\"a}nen}, {Keskitalo}, {Kiiveri}, {Kim}, {Kisner}, {Knox},
  {Krachmalnicoff}, {Kunz}, {Kurki-Suonio}, {Lagache}, {Lamarre}, {Lasenby},
  {Lattanzi}, {Lawrence}, {Le Jeune}, {Lemos}, {Lesgourgues}, {Levrier},
  {Lewis}, {Liguori}, {Lilje}, {Lilley}, {Lindholm}, {L{\'o}pez-Caniego},
  {Lubin}, {Ma}, {Mac{\'\i}as-P{\'e}rez}, {Maggio}, {Maino}, {Mandolesi},
  {Mangilli}, {Marcos-Caballero}, {Maris}, {Martin}, {Martinelli},
  {Mart{\'\i}nez-Gonz{\'a}lez}, {Matarrese}, {Mauri}, {McEwen}, {Meinhold},
  {Melchiorri}, {Mennella}, {Migliaccio}, {Millea}, {Mitra},
  {Miville-Desch{\^e}nes}, {Molinari}, {Montier}, {Morgante}, {Moss}, {Natoli},
  {N{\o}rgaard-Nielsen}, {Pagano}, {Paoletti}, {Partridge}, {Patanchon},
  {Peiris}, {Perrotta}, {Pettorino}, {Piacentini}, {Polastri}, {Polenta},
  {Puget}, {Rachen}, {Reinecke}, {Remazeilles}, {Renzi}, {Rocha}, {Rosset},
  {Roudier}, {Rubi{\~n}o-Mart{\'\i}n}, {Ruiz-Granados}, {Salvati}, {Sandri},
  {Savelainen}, {Scott}, {Shellard}, {Sirignano}, {Sirri}, {Spencer},
  {Sunyaev}, {Suur-Uski}, {Tauber}, {Tavagnacco}, {Tenti}, {Toffolatti},
  {Tomasi}, {Trombetti}, {Valenziano}, {Valiviita}, {Van Tent}, {Vibert},
  {Vielva}, {Villa}, {Vittorio}, {Wandelt}, {Wehus}, {White}, {White},
  {Zacchei}, \& {Zonca}}]{Planck:2020}
{Planck Collaboration}, {Aghanim}, N., {Akrami}, Y., {et~al.} 2020, \aap, 641,
  A6, \dodoi{10.1051/0004-6361/201833910}

\bibitem[{{Popesso} {et~al.}(2023){Popesso}, {Concas}, {Cresci}, {Belli},
  {Rodighiero}, {Inami}, {Dickinson}, {Ilbert}, {Pannella}, \&
  {Elbaz}}]{Popesso:2023}
{Popesso}, P., {Concas}, A., {Cresci}, G., {et~al.} 2023, \mnras, 519, 1526,
  \dodoi{10.1093/mnras/stac3214}

\bibitem[{{Quadri} {et~al.}(2007){Quadri}, {Marchesini}, {van Dokkum},
  {Gawiser}, {Franx}, {Lira}, {Rudnick}, {Urry}, {Maza}, {Kriek}, {Barrientos},
  {Blanc}, {Castander}, {Christlein}, {Coppi}, {Hall}, {Herrera}, {Infante},
  {Taylor}, {Treister}, \& {Willis}}]{Quadri:2007}
{Quadri}, R., {Marchesini}, D., {van Dokkum}, P., {et~al.} 2007, \aj, 134,
  1103, \dodoi{10.1086/520330}

\bibitem[{{Rieke} {et~al.}(2005){Rieke}, {Kelly}, \& {Horner}}]{Rieke:2005}
{Rieke}, M.~J., {Kelly}, D., \& {Horner}, S. 2005, in Society of Photo-Optical
  Instrumentation Engineers (SPIE) Conference Series, Vol. 5904, Cryogenic
  Optical Systems and Instruments XI, ed. J.~B. {Heaney} \& L.~G. {Burriesci},
  1--8, \dodoi{10.1117/12.615554}

\bibitem[{{Rieke} {et~al.}(2023{\natexlab{a}}){Rieke}, {Kelly}, {Misselt},
  {Stansberry}, {Boyer}, {Beatty}, {Egami}, {Florian}, {Greene}, {Hainline},
  {Leisenring}, {Roellig}, {Schlawin}, {Sun}, {Tinnin}, {Williams}, {Willmer},
  {Wilson}, {Clark}, {Rohrbach}, {Brooks}, {Canipe}, {Correnti}, {DiFelice},
  {Gennaro}, {Girard}, {Hartig}, {Hilbert}, {Koekemoer}, {Nikolov}, {Pirzkal},
  {Rest}, {Robberto}, {Sunnquist}, {Telfer}, {Wu}, {Ferry}, {Lewis}, {Baum},
  {Beichman}, {Doyon}, {Dressler}, {Eisenstein}, {Ferrarese}, {Hodapp},
  {Horner}, {Jaffe}, {Johnstone}, {Krist}, {Martin}, {McCarthy}, {Meyer},
  {Rieke}, {Trauger}, \& {Young}}]{Rieke:2022}
{Rieke}, M.~J., {Kelly}, D.~M., {Misselt}, K., {et~al.} 2023{\natexlab{a}},
  \pasp, 135, 028001, \dodoi{10.1088/1538-3873/acac53}

\bibitem[{{Rieke} {et~al.}(2023{\natexlab{b}}){Rieke}, {Robertson},
  {Tacchella}, {Hainline}, {Johnson}, {Hausan}, {Ji}, {Willmer}, {Eisenstein},
  {Pusk{\`a}s}, {Alberts}, {Arribas}, {Baker}, {Baum}, {Bhatawdekar},
  {Bonaventura}, {Boyett}, {Bunker}, {Cameron}, {Carniani}, {Charlot},
  {Chevallard}, {Chen}, {Curti}, {Curtis-Lake}, {Danhaive}, {DeCoursey},
  {Dressler}, {Egami}, {Endsley}, {Helton}, {Hviding}, {Kumari}, {Looser},
  {Lyu}, {Maiolino}, {Maseda}, {Nelson}, {Rieke}, {Rix}, {Sandles}, {Saxena},
  {Sharpe}, {Shivaei}, {Skarbinski}, {Smit}, {Stark}, {Stone}, {Suess}, {Sun},
  {Topping}, {Uebler}, {Villanueva}, {Wallace}, {Williams}, {Willott},
  {Whitler}, {Witstok}, \& {Woodrum}}]{Rieke:2023}
{Rieke}, M.~J., {Robertson}, B.~E., {Tacchella}, S., {et~al.}
  2023{\natexlab{b}}, arXiv e-prints, arXiv:2306.02466,
  \dodoi{10.48550/arXiv.2306.02466}

\bibitem[{{Rigby} {et~al.}(2022){Rigby}, {Perrin}, {McElwain}, {Kimble},
  {Friedman}, {Lallo}, {Doyon}, {Feinberg}, {Ferruit}, {Glasse}, \&
  et~al.}]{Rigby:2022}
{Rigby}, J., {Perrin}, M., {McElwain}, M., {et~al.} 2022, arXiv e-prints,
  arXiv:2207.05632.
\newblock \doarXiv{2207.05632}

\bibitem[{{Robertson} {et~al.}(2023){Robertson}, {Tacchella}, {Johnson},
  {Hainline}, {Whitler}, {Eisenstein}, {Endsley}, {Rieke}, {Stark}, {Alberts},
  {Dressler}, {Egami}, {Hausen}, {Rieke}, {Shivaei}, {Williams}, {Willmer},
  {Arribas}, {Bonaventura}, {Bunker}, {Cameron}, {Carniani}, {Charlot},
  {Chevallard}, {Curti}, {Curtis-Lake}, {D'Eugenio}, {Jakobsen}, {Looser},
  {L{\"u}tzgendorf}, {Maiolino}, {Maseda}, {Rawle}, {Rix}, {Smit}, {{\"U}bler},
  {Willott}, {Witstok}, {Baum}, {Bhatawdekar}, {Boyett}, {Chen}, {de Graaff},
  {Florian}, {Helton}, {Hviding}, {Ji}, {Kumari}, {Lyu}, {Nelson}, {Sandles},
  {Saxena}, {Suess}, {Sun}, {Topping}, \& {Wallace}}]{Robertson:2022}
{Robertson}, B.~E., {Tacchella}, S., {Johnson}, B.~D., {et~al.} 2023, Nature
  Astronomy, 7, 611, \dodoi{10.1038/s41550-023-01921-1}

\bibitem[{{Ruel} {et~al.}(2014){Ruel}, {Bazin}, {Bayliss}, {Brodwin}, {Foley},
  {Stalder}, {Aird}, {Armstrong}, {Ashby}, {Bautz}, {Benson}, {Bleem},
  {Bocquet}, {Carlstrom}, {Chang}, {Chapman}, {Cho}, {Clocchiatti}, {Crawford},
  {Crites}, {de Haan}, {Desai}, {Dobbs}, {Dudley}, {Forman}, {George},
  {Gladders}, {Gonzalez}, {Halverson}, {Harrington}, {High}, {Holder},
  {Holzapfel}, {Hrubes}, {Jones}, {Joy}, {Keisler}, {Knox}, {Lee}, {Leitch},
  {Liu}, {Lueker}, {Luong-Van}, {Mantz}, {Marrone}, {McDonald}, {McMahon},
  {Mehl}, {Meyer}, {Mocanu}, {Mohr}, {Montroy}, {Murray}, {Natoli},
  {Nurgaliev}, {Padin}, {Plagge}, {Pryke}, {Reichardt}, {Rest}, {Ruhl},
  {Saliwanchik}, {Saro}, {Sayre}, {Schaffer}, {Shaw}, {Shirokoff}, {Song},
  {{\v{S}}uhada}, {Spieler}, {Stanford}, {Staniszewski}, {Starsk}, {Story},
  {Stubbs}, {van Engelen}, {Vanderlinde}, {Vieira}, {Vikhlinin}, {Williamson},
  {Zahn}, \& {Zenteno}}]{Ruel:2014}
{Ruel}, J., {Bazin}, G., {Bayliss}, M., {et~al.} 2014, \apj, 792, 45,
  \dodoi{10.1088/0004-637X/792/1/45}

\bibitem[{{Shibuya} {et~al.}(2015){Shibuya}, {Ouchi}, \&
  {Harikane}}]{Shibuya:2015}
{Shibuya}, T., {Ouchi}, M., \& {Harikane}, Y. 2015, \apjs, 219, 15,
  \dodoi{10.1088/0067-0049/219/2/15}

\bibitem[{{Skelton} {et~al.}(2014){Skelton}, {Whitaker}, {Momcheva}, {Brammer},
  {van Dokkum}, {Labb{\'e}}, {Franx}, {van der Wel}, {Bezanson}, {Da Cunha},
  {Fumagalli}, {F{\"o}rster Schreiber}, {Kriek}, {Leja}, {Lundgren}, {Magee},
  {Marchesini}, {Maseda}, {Nelson}, {Oesch}, {Pacifici}, {Patel}, {Price},
  {Rix}, {Tal}, {Wake}, \& {Wuyts}}]{Skelton:2014}
{Skelton}, R.~E., {Whitaker}, K.~E., {Momcheva}, I.~G., {et~al.} 2014, \apjs,
  214, 24, \dodoi{10.1088/0067-0049/214/2/24}

\bibitem[{{Speagle}(2020)}]{Speagle:2020}
{Speagle}, J.~S. 2020, \mnras, 493, 3132, \dodoi{10.1093/mnras/staa278}

\bibitem[{{Sun} {et~al.}(2023{\natexlab{a}}){Sun}, {Egami}, {Pirzkal}, {Rieke},
  {Baum}, {Boyer}, {Boyett}, {Bunker}, {Cameron}, {Curti}, {Eisenstein},
  {Gennaro}, {Greene}, {Jaffe}, {Kelly}, {Koekemoer}, {Kumari}, {Maiolino},
  {Maseda}, {Perna}, {Rest}, {Robertson}, {Schlawin}, {Smit}, {Stansberry},
  {Sunnquist}, {Tacchella}, {Williams}, \& {Willmer}}]{Sun:2022b}
{Sun}, F., {Egami}, E., {Pirzkal}, N., {et~al.} 2023{\natexlab{a}}, \apj, 953,
  53, \dodoi{10.3847/1538-4357/acd53c}

\bibitem[{{Sun} {et~al.}(2023{\natexlab{b}}){Sun}, {Helton}, {Egami},
  {Hainline}, {Rieke}, {Willmer}, {Eisenstein}, {Johnson}, {Rieke},
  {Robertson}, {Tacchella}, {Alberts}, {Baker}, {Bhatawdekar}, {Boyett},
  {Bunker}, {Charlot}, {Chen}, {Chevallard}, {Curtis-Lake}, {Danhaive},
  {DeCoursey}, {Ji}, {Lyu}, {Maiolino}, {Rujopakarn}, {Sandles}, {Shivaei},
  {Ubler}, {Willott}, \& {Witstok}}]{Sun:2023}
{Sun}, F., {Helton}, J.~M., {Egami}, E., {et~al.} 2023{\natexlab{b}}, arXiv
  e-prints, arXiv:2309.04529, \dodoi{10.48550/arXiv.2309.04529}

\bibitem[{{Tacchella} {et~al.}(2022){Tacchella}, {Finkelstein}, {Bagley},
  {Dickinson}, {Ferguson}, {Giavalisco}, {Graziani}, {Grogin}, {Hathi},
  {Hutchison}, {Jung}, {Koekemoer}, {Larson}, {Papovich}, {Pirzkal},
  {Rojas-Ruiz}, {Song}, {Schneider}, {Somerville}, {Wilkins}, \&
  {Yung}}]{Tacchella:2022b}
{Tacchella}, S., {Finkelstein}, S.~L., {Bagley}, M., {et~al.} 2022, \apj, 927,
  170, \dodoi{10.3847/1538-4357/ac4cad}

\bibitem[{{The Pandas Development Team}(2022)}]{Pandas:2022}
{The Pandas Development Team}. 2022, {pandas-dev/pandas: Pandas}, v1.5.0,
  Zenodo,  Zenodo, \dodoi{10.5281/zenodo.7093122}

\bibitem[{{Urrutia} {et~al.}(2019){Urrutia}, {Wisotzki}, {Kerutt}, {Schmidt},
  {Herenz}, {Klar}, {Saust}, {Werhahn}, {Diener}, {Caruana}, {Krajnovi{\'c}},
  {Bacon}, {Boogaard}, {Brinchmann}, {Enke}, {Maseda}, {Nanayakkara},
  {Richard}, {Steinmetz}, \& {Weilbacher}}]{Urrutia:2019}
{Urrutia}, T., {Wisotzki}, L., {Kerutt}, J., {et~al.} 2019, \aap, 624, A141,
  \dodoi{10.1051/0004-6361/201834656}

\bibitem[{{van der Walt} {et~al.}(2011){van der Walt}, {Colbert}, \&
  {Varoquaux}}]{NumPy:2011}
{van der Walt}, S., {Colbert}, S.~C., \& {Varoquaux}, G. 2011, Computing in
  Science and Engineering, 13, 22, \dodoi{10.1109/MCSE.2011.37}

\bibitem[{{Vazdekis} {et~al.}(2015){Vazdekis}, {Coelho}, {Cassisi},
  {Ricciardelli}, {Falc{\'o}n-Barroso}, {S{\'a}nchez-Bl{\'a}zquez}, {La
  Barbera}, {Beasley}, \& {Pietrinferni}}]{Vazdekis:2015}
{Vazdekis}, A., {Coelho}, P., {Cassisi}, S., {et~al.} 2015, \mnras, 449, 1177,
  \dodoi{10.1093/mnras/stv151}

\bibitem[{{Virtanen} {et~al.}(2020){Virtanen}, {Gommers}, {Oliphant},
  {Haberland}, {Reddy}, {Cournapeau}, {Burovski}, {Peterson}, {Weckesser},
  {Bright}, {van der Walt}, {Brett}, {Wilson}, {Millman}, {Mayorov}, {Nelson},
  {Jones}, {Kern}, {Larson}, {Carey}, {Polat}, {Feng}, {Moore}, {VanderPlas},
  {Laxalde}, {Perktold}, {Cimrman}, {Henriksen}, {Quintero}, {Harris},
  {Archibald}, {Ribeiro}, {Pedregosa}, {van Mulbregt}, \& {SciPy 1. 0
  Contributors}}]{SciPy:2020}
{Virtanen}, P., {Gommers}, R., {Oliphant}, T.~E., {et~al.} 2020, Nature
  Methods, 17, 261, \dodoi{10.1038/s41592-019-0686-2}

\bibitem[{{Walter} {et~al.}(2012){Walter}, {Decarli}, {Carilli}, {Bertoldi},
  {Cox}, {da Cunha}, {Daddi}, {Dickinson}, {Downes}, {Elbaz}, {Ellis}, {Hodge},
  {Neri}, {Riechers}, {Weiss}, {Bell}, {Dannerbauer}, {Krips}, {Krumholz},
  {Lentati}, {Maiolino}, {Menten}, {Rix}, {Robertson}, {Spinrad}, {Stark}, \&
  {Stern}}]{Walter:2012}
{Walter}, F., {Decarli}, R., {Carilli}, C., {et~al.} 2012, \nat, 486, 233,
  \dodoi{10.1038/nature11073}

\bibitem[{{Wang} {et~al.}(2023){Wang}, {Yang}, {Hennawi}, {Fan}, {Sun},
  {Champagne}, {Costa}, {Habouzit}, {Endsley}, {Li}, {Lin}, {Meyer},
  {Schindler}, {Wu}, {Ba{\~n}ados}, {Barth}, {Bhowmick}, {Bieri}, {Blecha},
  {Bosman}, {Cai}, {Colina}, {Connor}, {Davies}, {Decarli}, {De Rosa}, {Drake},
  {Egami}, {Eilers}, {Evans}, {Farina}, {Haiman}, {Jiang}, {Jin}, {Jun},
  {Kakiichi}, {Khusanova}, {Kulkarni}, {Li}, {Liu}, {Loiacono}, {Lupi},
  {Mazzucchelli}, {Onoue}, {Pudoka}, {Rojas-Ruiz}, {Shen}, {Strauss}, {Tee},
  {Trakhtenbrot}, {Trebitsch}, {Venemans}, {Volonteri}, {Walter}, {Xie}, {Yue},
  {Zhang}, {Zhang}, \& {Zou}}]{Wang:2023}
{Wang}, F., {Yang}, J., {Hennawi}, J.~F., {et~al.} 2023, \apjl, 951, L4,
  \dodoi{10.3847/2041-8213/accd6f}

\bibitem[{{Wang} {et~al.}(2021){Wang}, {Hill}, {Chapman}, {Wei{\ss}}, {Scott},
  {Apostolovski}, {Aravena}, {Archipley}, {B{\'e}thermin}, {Canning}, {De
  Breuck}, {Dong}, {Everett}, {Gonzalez}, {Greve}, {Hayward}, {Hezaveh},
  {Jarugula}, {Marrone}, {Phadke}, {Reuter}, {Rotermund}, {Spilker}, \&
  {Vieira}}]{Wang:2021}
{Wang}, G. C.~P., {Hill}, R., {Chapman}, S.~C., {et~al.} 2021, \mnras, 508,
  3754, \dodoi{10.1093/mnras/stab2800}

\bibitem[{{Wang} {et~al.}(2013){Wang}, {Farrah}, {Oliver}, {Amblard},
  {B{\'e}thermin}, {Bock}, {Conley}, {Cooray}, {Halpern}, {Heinis}, {Ibar},
  {Ilbert}, {Ivison}, {Marsden}, {Roseboom}, {Rowan-Robinson}, {Schulz},
  {Smith}, {Viero}, \& {Zemcov}}]{Wang:2013}
{Wang}, L., {Farrah}, D., {Oliver}, S.~J., {et~al.} 2013, \mnras, 431, 648,
  \dodoi{10.1093/mnras/stt190}

\bibitem[{{Waskom}(2021)}]{Waskom:2021}
{Waskom}, M. 2021, The Journal of Open Source Software, 6, 3021,
  \dodoi{10.21105/joss.03021}

\bibitem[{{Webb} {et~al.}(2020){Webb}, {Balogh}, {Leja}, {van der Burg},
  {Rudnick}, {Muzzin}, {Boak}, {Cerulo}, {Gilbank}, {Lidman}, {Old},
  {Pintos-Castro}, {McGee}, {Shipley}, {Biviano}, {Chan}, {Cooper}, {De Lucia},
  {Demarco}, {Forrest}, {Jablonka}, {Kukstas}, {McCarthy}, {McNab}, {Nantais},
  {Noble}, {Poggianti}, {Reeves}, {Vulcani}, {Wilson}, {Yee}, \&
  {Zaritsky}}]{Webb:2020}
{Webb}, K., {Balogh}, M.~L., {Leja}, J., {et~al.} 2020, \mnras, 498, 5317,
  \dodoi{10.1093/mnras/staa2752}

\bibitem[{{Whitaker} {et~al.}(2011){Whitaker}, {Labb{\'e}}, {van Dokkum},
  {Brammer}, {Kriek}, {Marchesini}, {Quadri}, {Franx}, {Muzzin}, {Williams},
  {Bezanson}, {Illingworth}, {Lee}, {Lundgren}, {Nelson}, {Rudnick}, {Tal}, \&
  {Wake}}]{Whitaker:2011}
{Whitaker}, K.~E., {Labb{\'e}}, I., {van Dokkum}, P.~G., {et~al.} 2011, \apj,
  735, 86, \dodoi{10.1088/0004-637X/735/2/86}

\bibitem[{{Whitaker} {et~al.}(2019){Whitaker}, {Ashas}, {Illingworth}, {Magee},
  {Leja}, {Oesch}, {van Dokkum}, {Mowla}, {Bouwens}, {Franx}, {Holden},
  {Labb{\'e}}, {Rafelski}, {Teplitz}, \& {Gonzalez}}]{Whitaker:2019}
{Whitaker}, K.~E., {Ashas}, M., {Illingworth}, G., {et~al.} 2019, \apjs, 244,
  16, \dodoi{10.3847/1538-4365/ab3853}

\bibitem[{{White} \& {Rees}(1978)}]{White:1978}
{White}, S.~D.~M., \& {Rees}, M.~J. 1978, \mnras, 183, 341,
  \dodoi{10.1093/mnras/183.3.341}

\bibitem[{{Williams} {et~al.}(2023){Williams}, {Tacchella}, {Maseda},
  {Robertson}, {Johnson}, {Willott}, {Eisenstein}, {Willmer}, {Ji}, {Hainline},
  {Helton}, {Alberts}, {Baum}, {Bhatawdekar}, {Boyett}, {Bunker}, {Carniani},
  {Charlot}, {Chevallard}, {Curtis-Lake}, {de Graaf}, {Egami}, {Franx},
  {Kumari}, {Maiolino}, {Nelson}, {Rieke}, {Sandles}, {Shivaei}, {Simmonds},
  {Smit}, {Suess}, {Sun}, {Ubler}, \& {Witstok}}]{Williams:2023}
{Williams}, C.~C., {Tacchella}, S., {Maseda}, M.~V., {et~al.} 2023, arXiv
  e-prints, arXiv:2301.09780, \dodoi{10.48550/arXiv.2301.09780}

\bibitem[{{Zabludoff} {et~al.}(1996){Zabludoff}, {Zaritsky}, {Lin}, {Tucker},
  {Hashimoto}, {Shectman}, {Oemler}, \& {Kirshner}}]{Zabludoff:1996}
{Zabludoff}, A.~I., {Zaritsky}, D., {Lin}, H., {et~al.} 1996, \apj, 466, 104,
  \dodoi{10.1086/177495}

\end{thebibliography}
\appendix

\section{Cutout Images and Grism Spectra for Final Spectroscopic Sample}
\label{AppendixOne}

\begin{figure}
\figurenum{A1}
\label{figset:A1}
\plotone{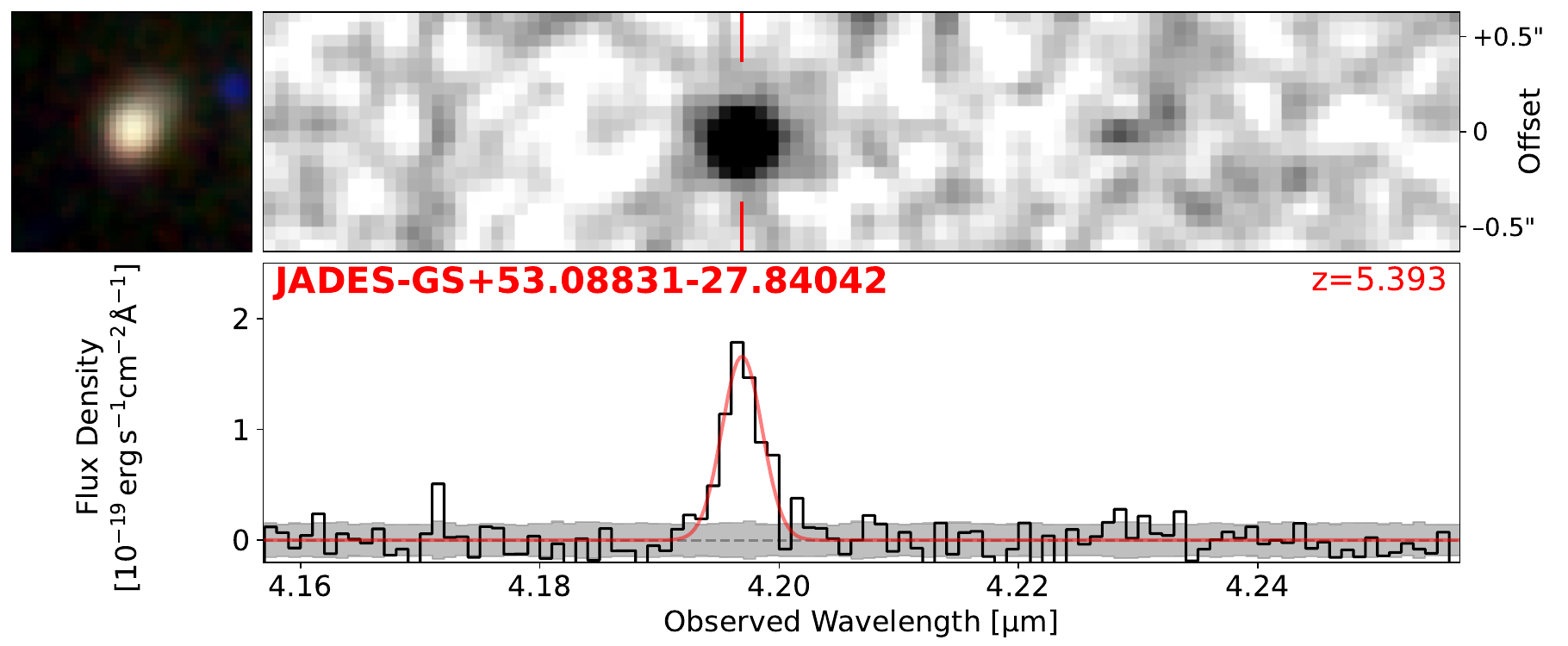}
\caption{The NIRCam cutout images alongside the continuum subtracted 2d and 1d grism spectra of JADES-GS+53.08831-27.84042 at $z=5.393$. The upper-left panel shows the $1.2^{\prime\prime} \times 1.2^{\prime\prime}$ F444W--F277W--F150W RGB thumbnail. The upper-right panel shows the extracted 2d spectrum around the H$\alpha$ emission line detection, indicated by the solid red line. The lower-right panel instead shows the extracted 1d spectrum around the H$\alpha$ emission line detection alongside the best-fit Gaussian profile given by the solid red line. The JADES ID and confirmed spectroscopic redshift are given in the lower-right panel for each galaxy. The complete figure set is available in the online journal.}
\label{fig:appendix1}
\end{figure}

\begin{figure}
\figurenum{A2}
\label{figset:A2}
\plotone{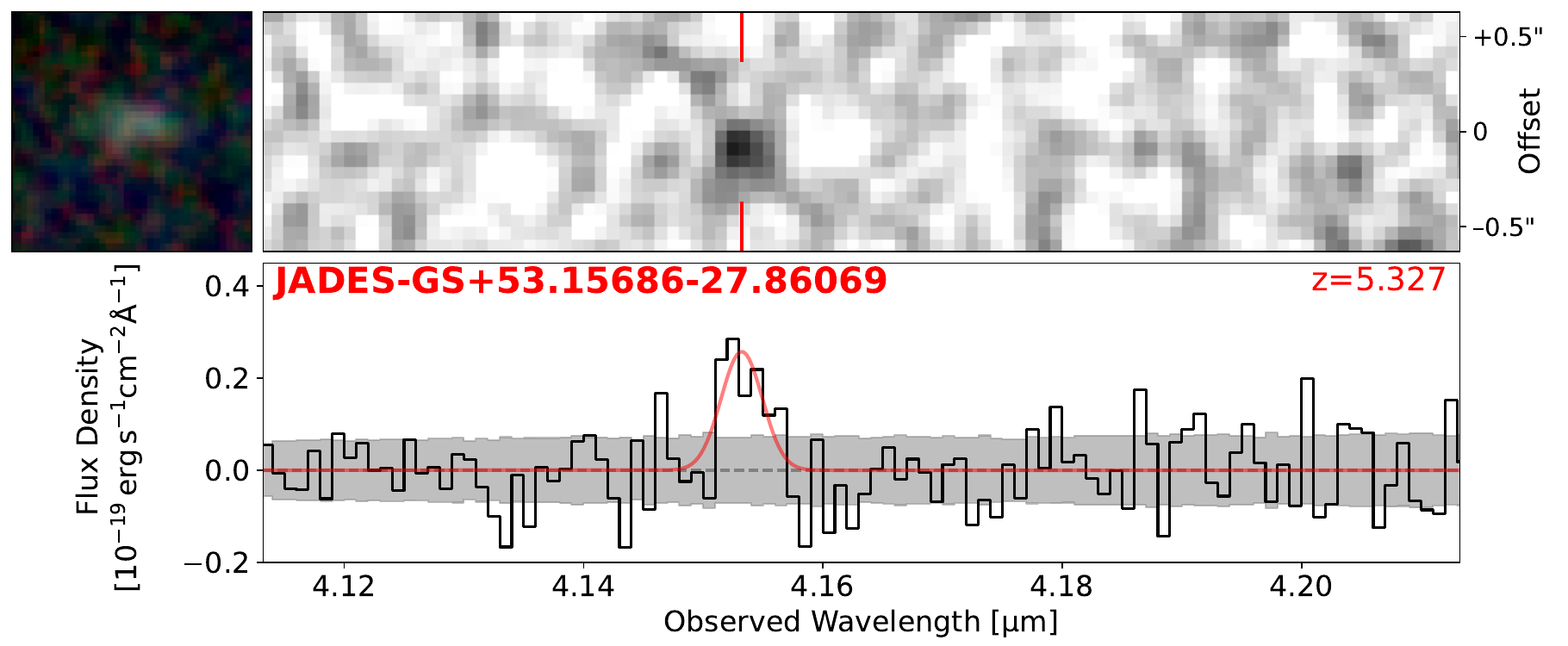}
\caption{The NIRCam cutout images alongside the continuum subtracted 2d and 1d grism spectra of JADES-GS+53.15686-27.86069 at $z=5.327$, just like in Figure~\ref{fig:appendix1}. The complete figure set is available in the online journal.}
\label{fig:appendix2}
\end{figure}

Figure Sets~\ref{figset:A1} and \ref{figset:A2} show the cutout images (see Section~\ref{SectionTwoOne}) alongside the continuum subtracted 2d and 1d extracted spectra (see Section~\ref{SectionTwoThree}) for the $81$ objects that are part of the final spectroscopic sample described in Section~\ref{SectionTwoThree}. The $39$ confirmed members of the overdensity are given by Figure~\ref{fig:appendix1} while the $42$ confirmed members of the field are given by Figure~\ref{fig:appendix2}. For each galaxy, the upper-left panel shows the $1.2^{\prime\prime} \times 1.2^{\prime\prime}$ F444W-F277W-F150W RGB thumbnail. The upper-right panel shows the extracted 2d spectrum around the $\mathrm{H} \alpha$ emission line detection, indicated by the solid red line. The lower-right panel instead shows the extracted 1d spectrum around the $\mathrm{H} \alpha$ emission line detection alongside the best-fit Gaussian profile given by the solid red line. The JADES ID and confirmed spectroscopic redshift are given in the lower-right panel for each galaxy. 

\figsetstart
\figsetnum{A1}
\figsettitle{NIRCam images and grism spectra for the $N = 39$ confirmed members of the overdensity.}

\figsetgrpstart
\figsetgrpnum{A1.1}
\figsetgrptitle{Ha}
\figsetplot{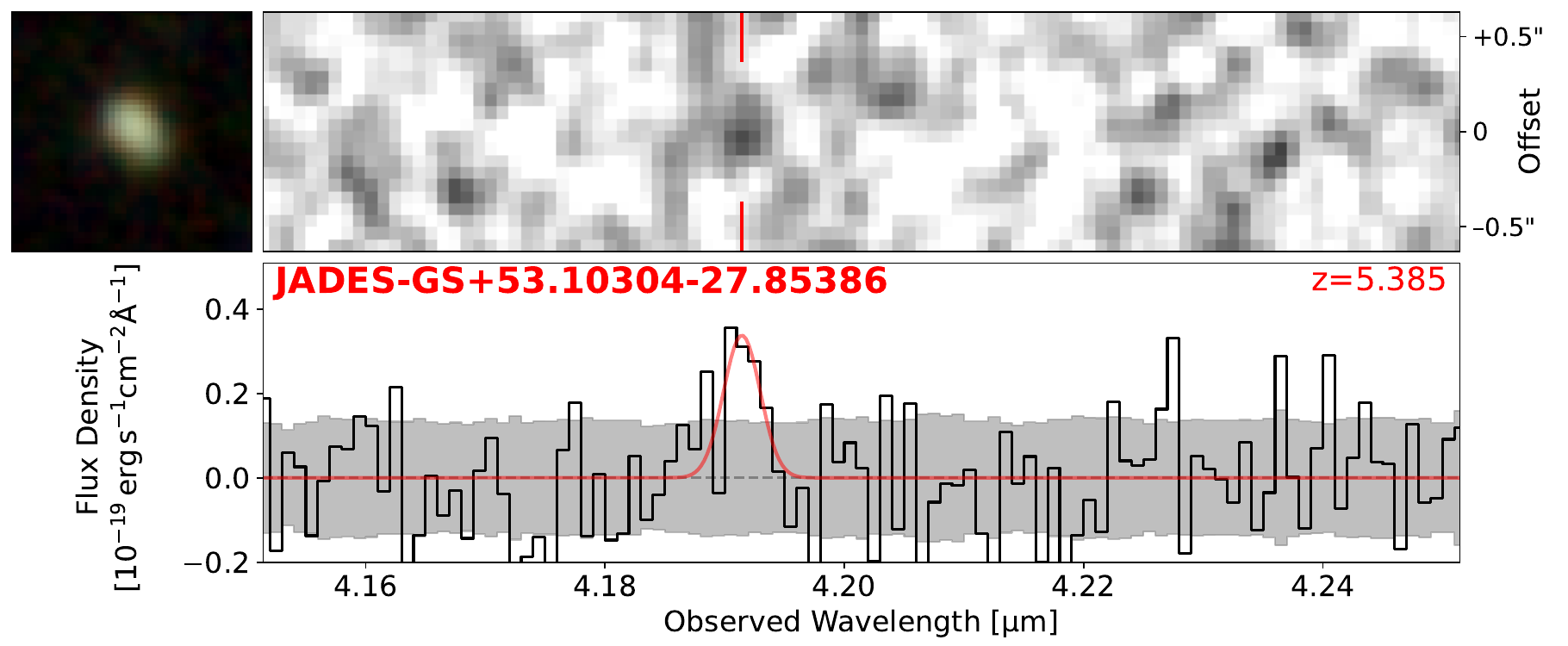}
\figsetgrpnote{NIRCam cutout images alongside the continuum-subtracted 2d and 1d grism spectra of JADES-GS+53.10304-27.85386 at $z = 5.385$, with $\mathrm{H} \alpha$ detected at $3.4\sigma$.}
\figsetgrpend

\figsetgrpstart
\figsetgrpnum{A1.2}
\figsetgrptitle{Ha}
\figsetplot{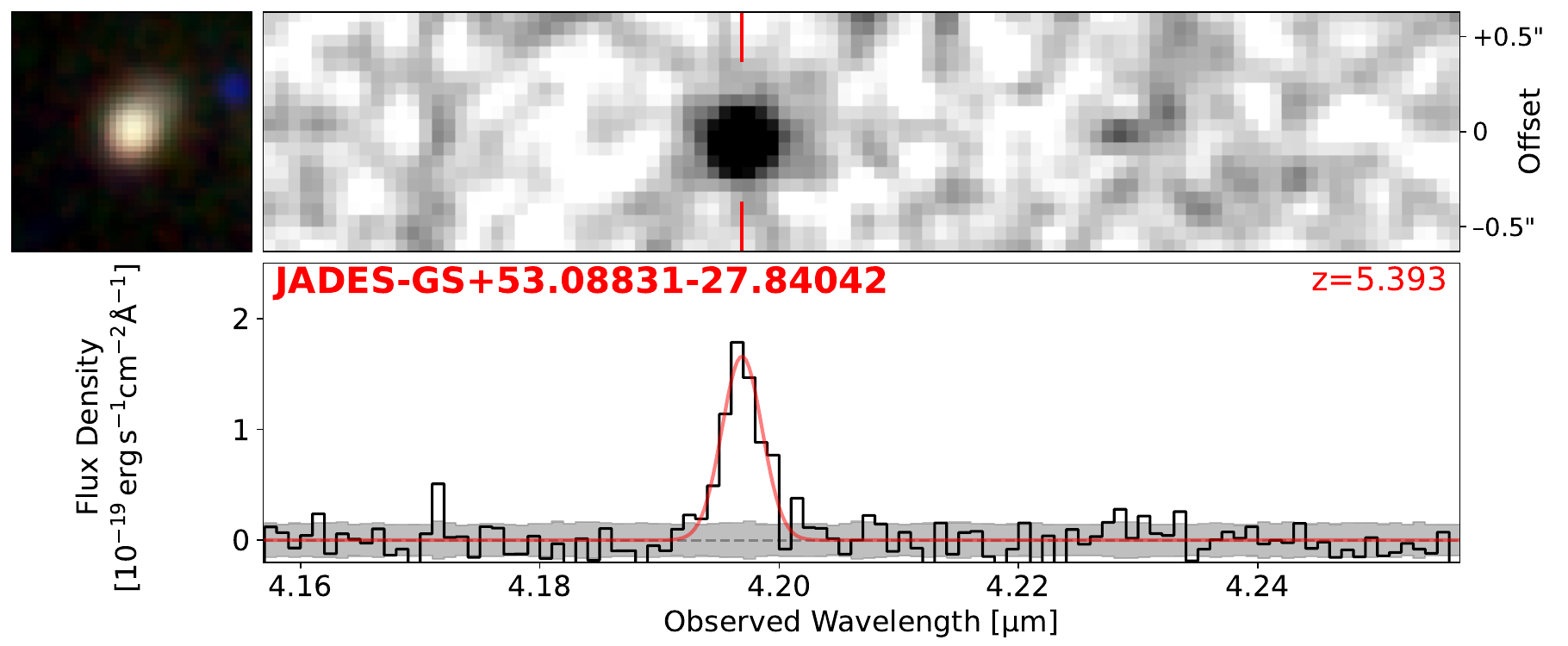}
\figsetgrpnote{NIRCam cutout images alongside the continuum-subtracted 2d and 1d grism spectra of JADES-GS+53.08831-27.84042 at $z = 5.393$, with $\mathrm{H} \alpha$ detected at $15.6\sigma$.}
\figsetgrpend

\figsetgrpstart
\figsetgrpnum{A1.3}
\figsetgrptitle{Ha}
\figsetplot{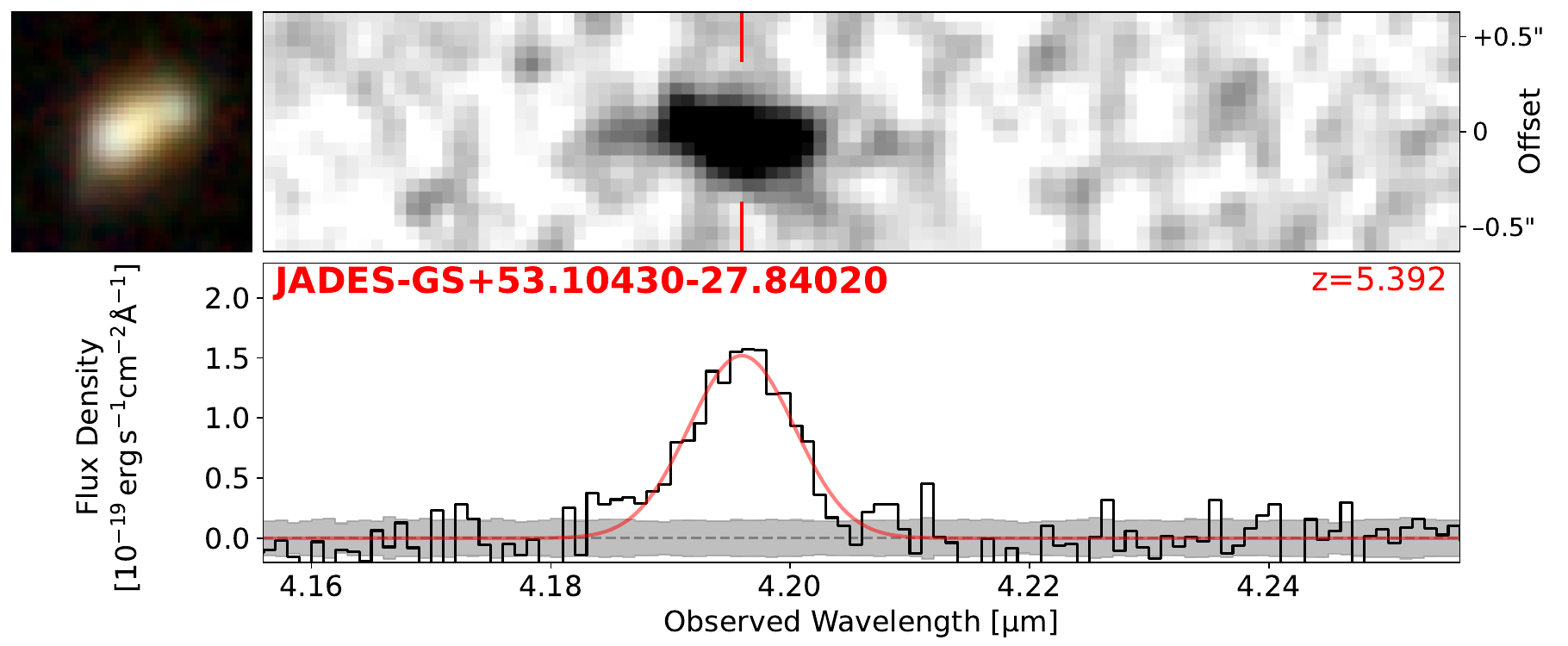}
\figsetgrpnote{NIRCam cutout images alongside the continuum-subtracted 2d and 1d grism spectra of JADES-GS+53.10430-27.84020 at $z = 5.392$, with $\mathrm{H} \alpha$ detected at $23.6\sigma$.}
\figsetgrpend

\figsetgrpstart
\figsetgrpnum{A1.4}
\figsetgrptitle{Ha}
\figsetplot{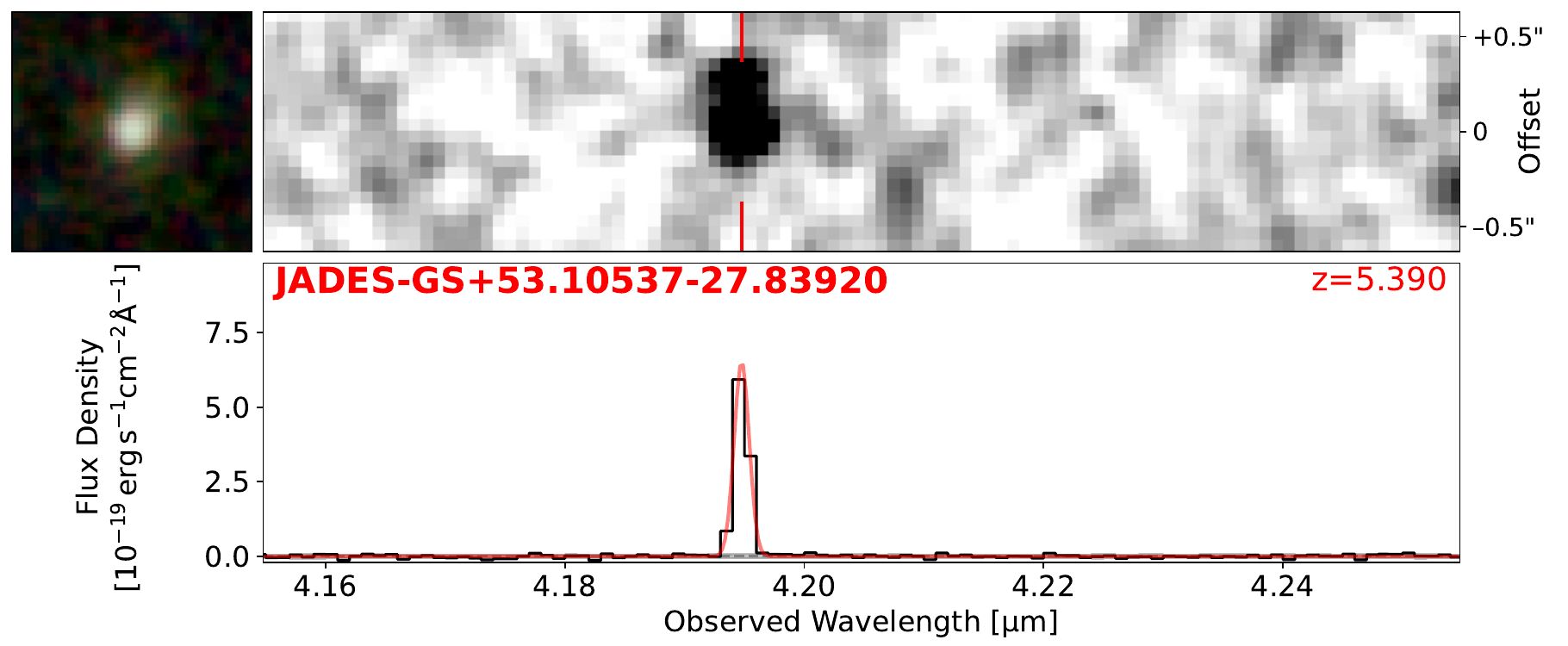}
\figsetgrpnote{NIRCam cutout images alongside the continuum-subtracted 2d and 1d grism spectra of JADES-GS+53.10537-27.83920 at $z = 5.390$, with $\mathrm{H} \alpha$ detected at $78.1\sigma$.}
\figsetgrpend

\figsetgrpstart
\figsetgrpnum{A1.5}
\figsetgrptitle{Ha}
\figsetplot{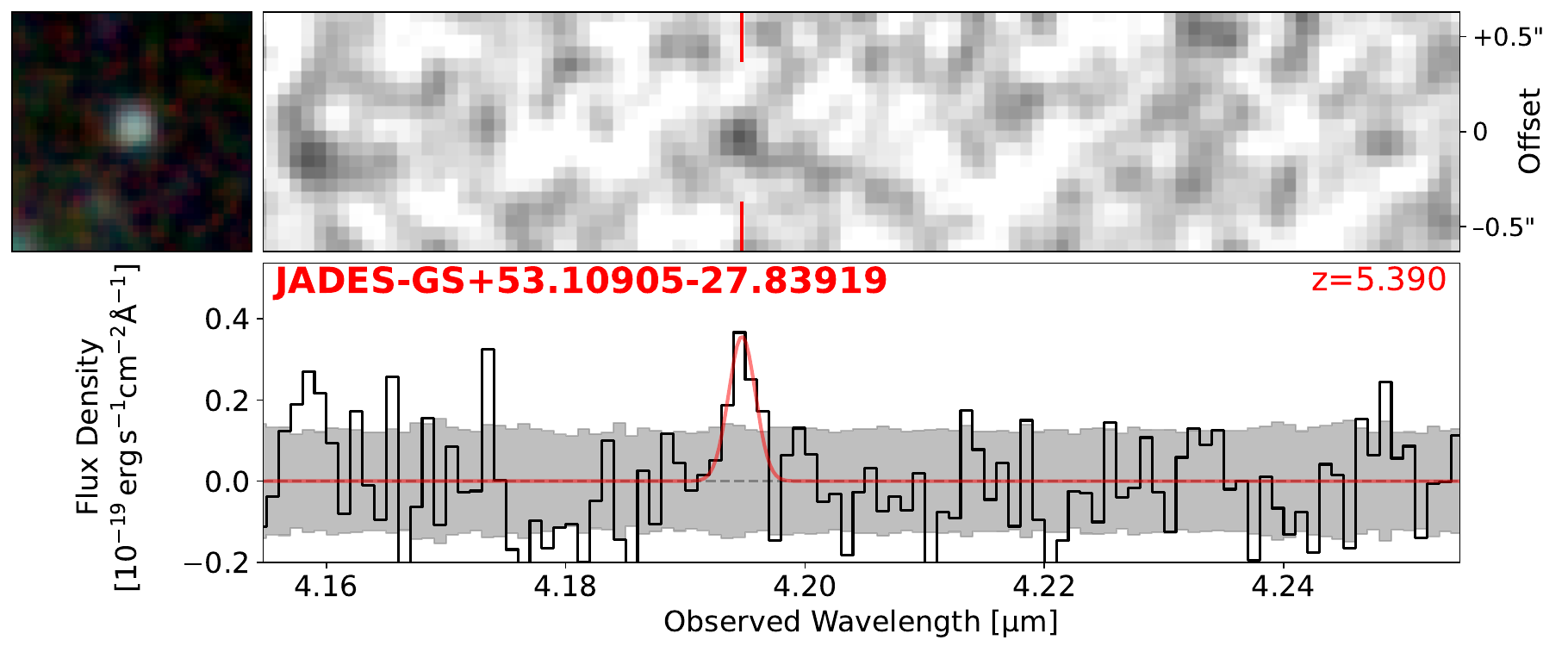}
\figsetgrpnote{NIRCam cutout images alongside the continuum-subtracted 2d and 1d grism spectra of JADES-GS+53.10905-27.83919 at $z = 5.390$, with $\mathrm{H} \alpha$ detected at $3.1\sigma$.}
\figsetgrpend

\figsetgrpstart
\figsetgrpnum{A1.6}
\figsetgrptitle{Ha}
\figsetplot{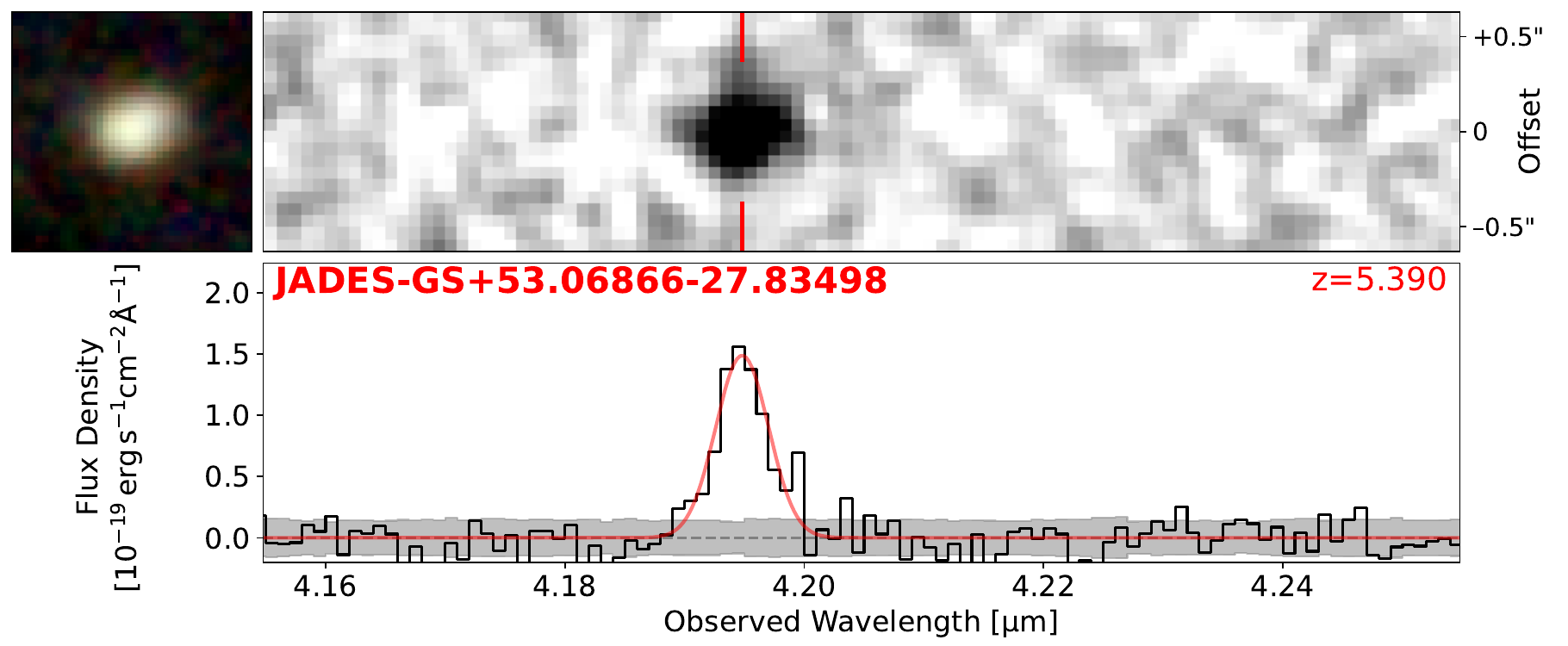}
\figsetgrpnote{NIRCam cutout images alongside the continuum-subtracted 2d and 1d grism spectra of JADES-GS+53.06866-27.83498 at $z = 5.390$, with $\mathrm{H} \alpha$ detected at $16.4\sigma$.}
\figsetgrpend

\figsetgrpstart
\figsetgrpnum{A1.7}
\figsetgrptitle{Ha}
\figsetplot{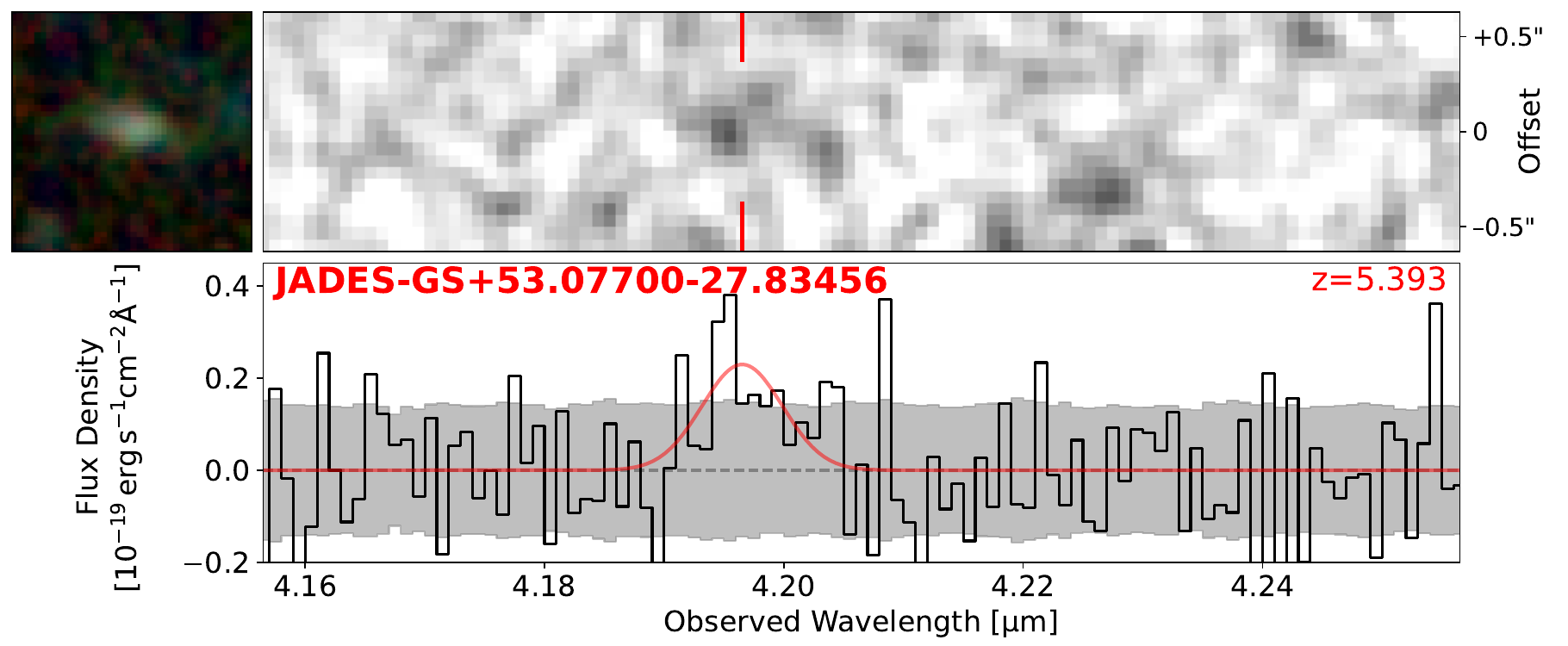}
\figsetgrpnote{NIRCam cutout images alongside the continuum-subtracted 2d and 1d grism spectra of JADES-GS+53.07700-27.83456 at $z = 5.393$, with $\mathrm{H} \alpha$ detected at $3.2\sigma$.}
\figsetgrpend

\figsetgrpstart
\figsetgrpnum{A1.8}
\figsetgrptitle{Ha}
\figsetplot{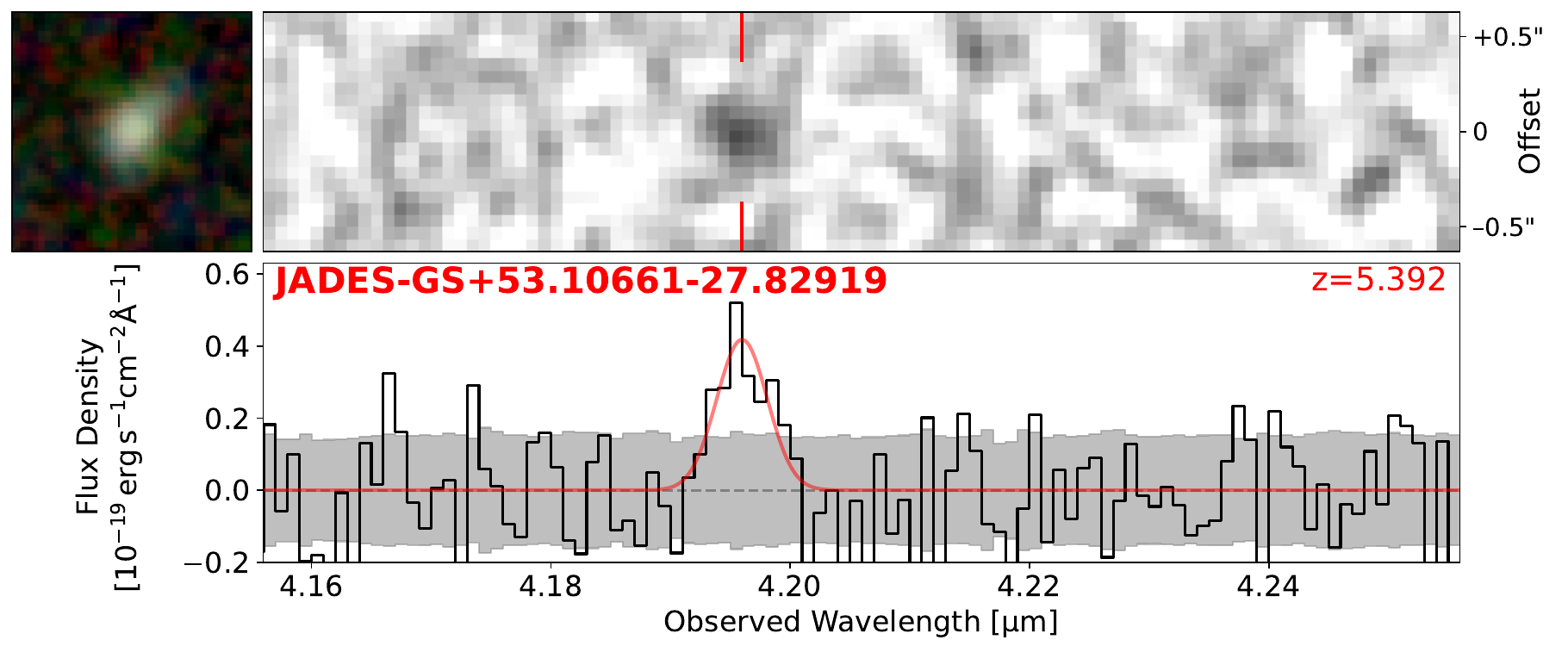}
\figsetgrpnote{NIRCam cutout images alongside the continuum-subtracted 2d and 1d grism spectra of JADES-GS+53.10661-27.82919 at $z = 5.392$, with $\mathrm{H} \alpha$ detected at $4.4\sigma$.}
\figsetgrpend

\figsetgrpstart
\figsetgrpnum{A1.9}
\figsetgrptitle{Ha}
\figsetplot{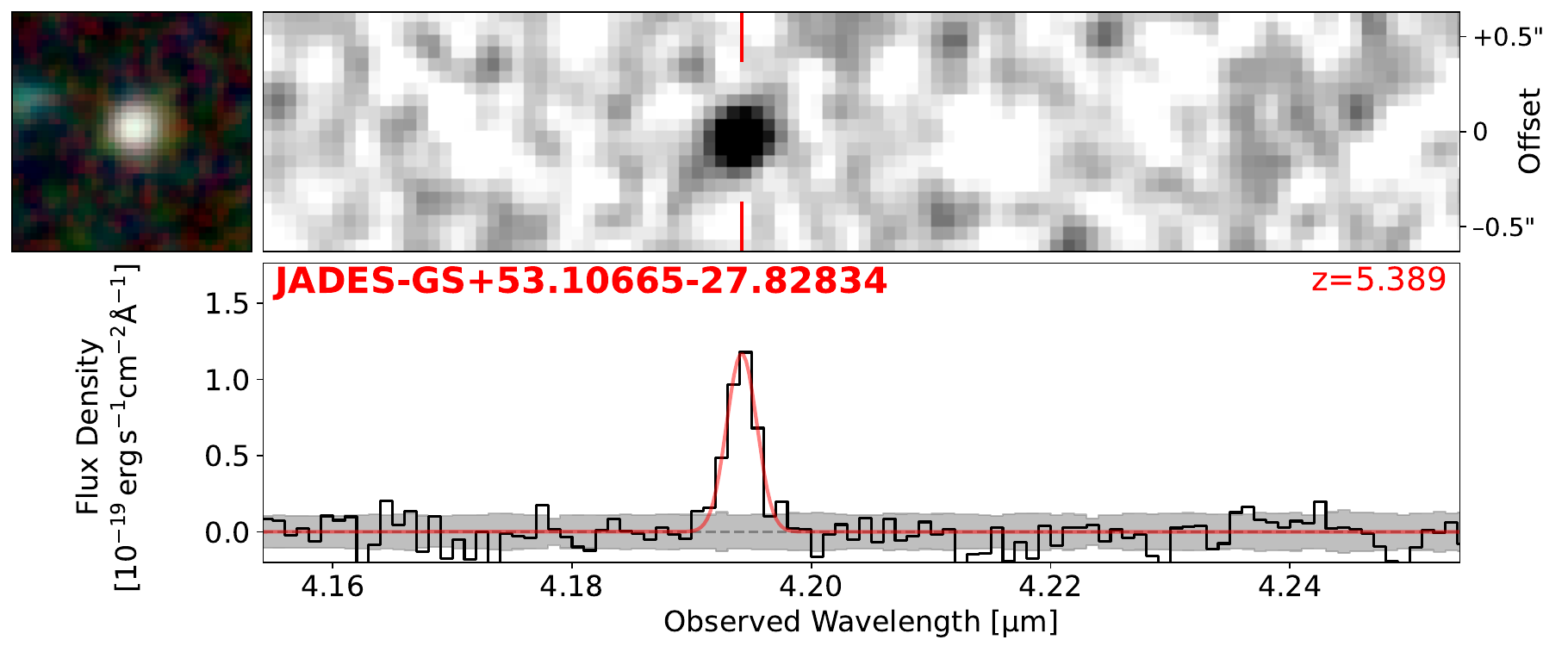}
\figsetgrpnote{NIRCam cutout images alongside the continuum-subtracted 2d and 1d grism spectra of JADES-GS+53.10665-27.82834 at $z = 5.389$, with $\mathrm{H} \alpha$ detected at $12.1\sigma$.}
\figsetgrpend

\figsetgrpstart
\figsetgrpnum{A1.10}
\figsetgrptitle{Ha}
\figsetplot{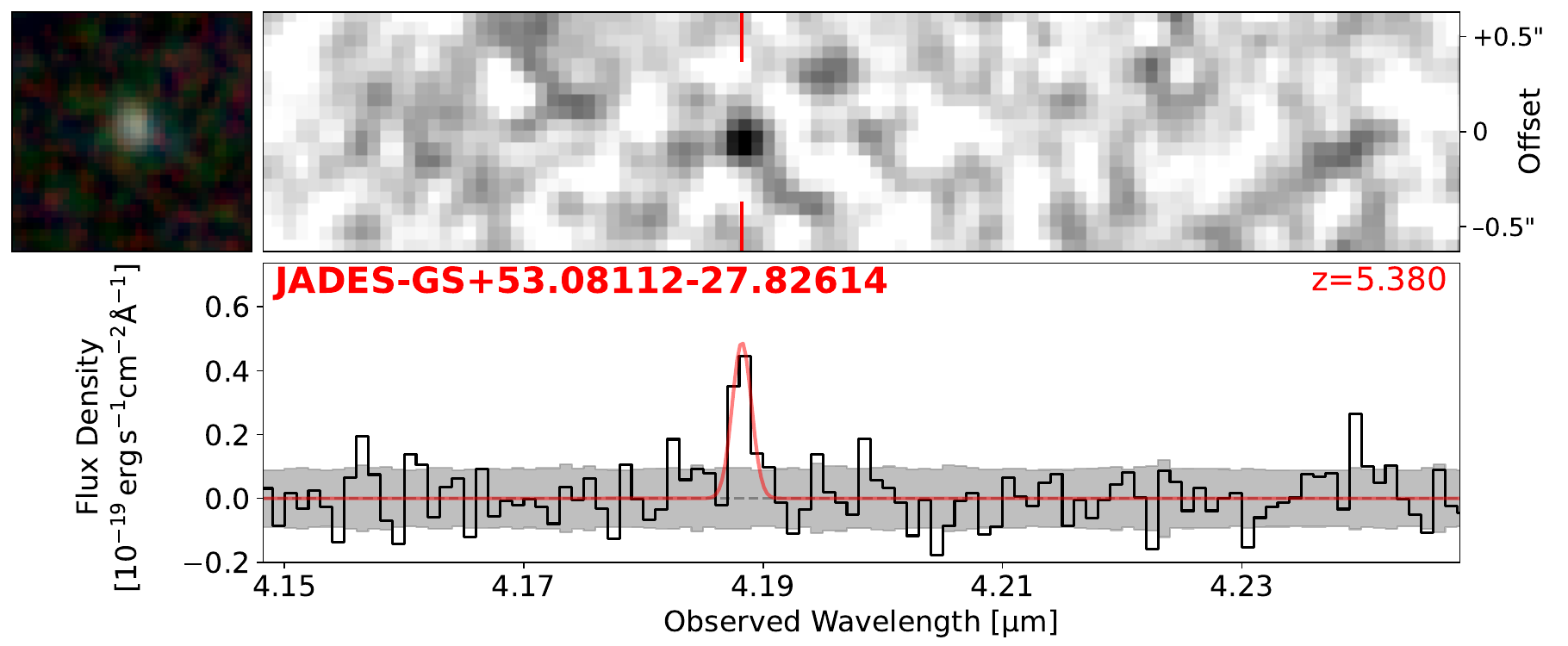}
\figsetgrpnote{NIRCam cutout images alongside the continuum-subtracted 2d and 1d grism spectra of JADES-GS+53.08112-27.82614 at $z = 5.380$, with $\mathrm{H} \alpha$ detected at $5.1\sigma$.}
\figsetgrpend

\figsetgrpstart
\figsetgrpnum{A1.11}
\figsetgrptitle{Ha}
\figsetplot{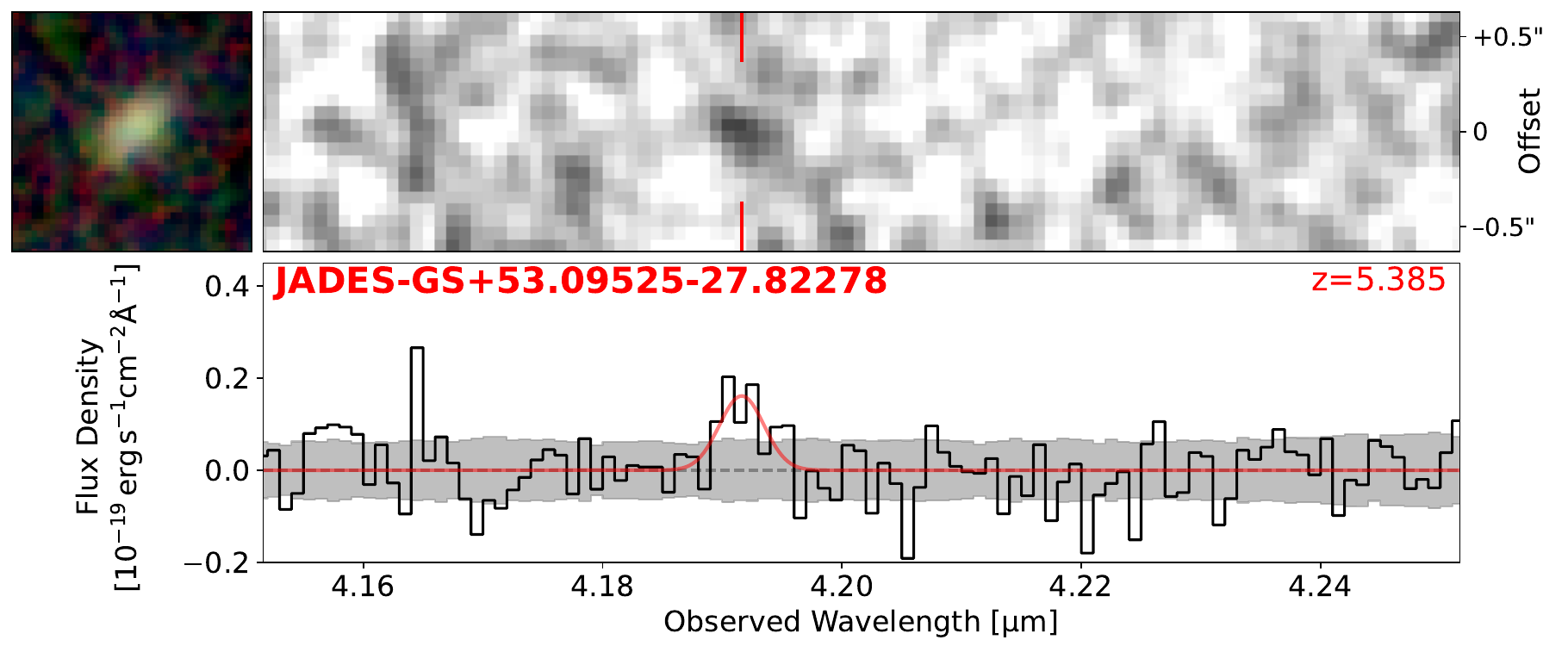}
\figsetgrpnote{NIRCam cutout images alongside the continuum-subtracted 2d and 1d grism spectra of JADES-GS+53.09525-27.82278 at $z = 5.385$, with $\mathrm{H} \alpha$ detected at $3.8\sigma$.}
\figsetgrpend

\figsetgrpstart
\figsetgrpnum{A1.12}
\figsetgrptitle{Ha}
\figsetplot{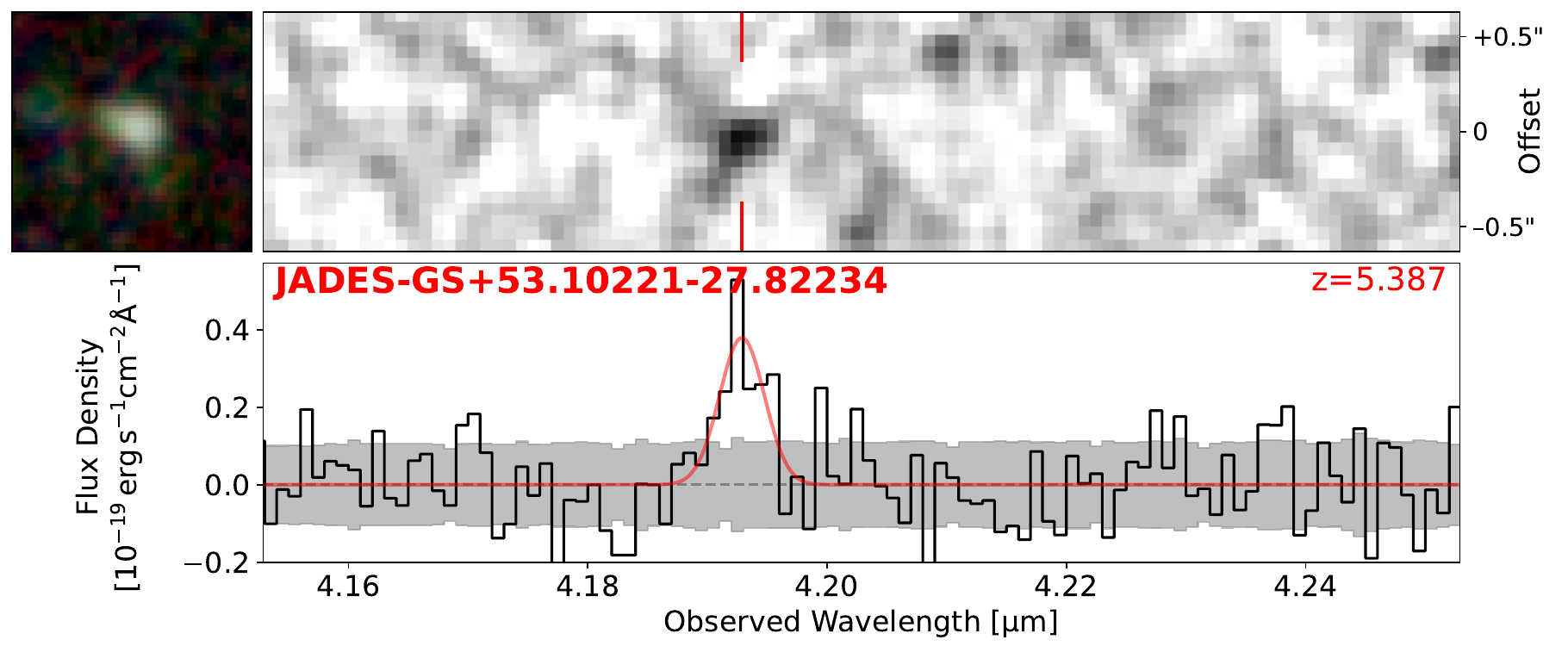}
\figsetgrpnote{NIRCam cutout images alongside the continuum-subtracted 2d and 1d grism spectra of JADES-GS+53.10221-27.82234 at $z = 5.387$, with $\mathrm{H} \alpha$ detected at $5.1\sigma$.}
\figsetgrpend

\figsetgrpstart
\figsetgrpnum{A1.13}
\figsetgrptitle{Ha}
\figsetplot{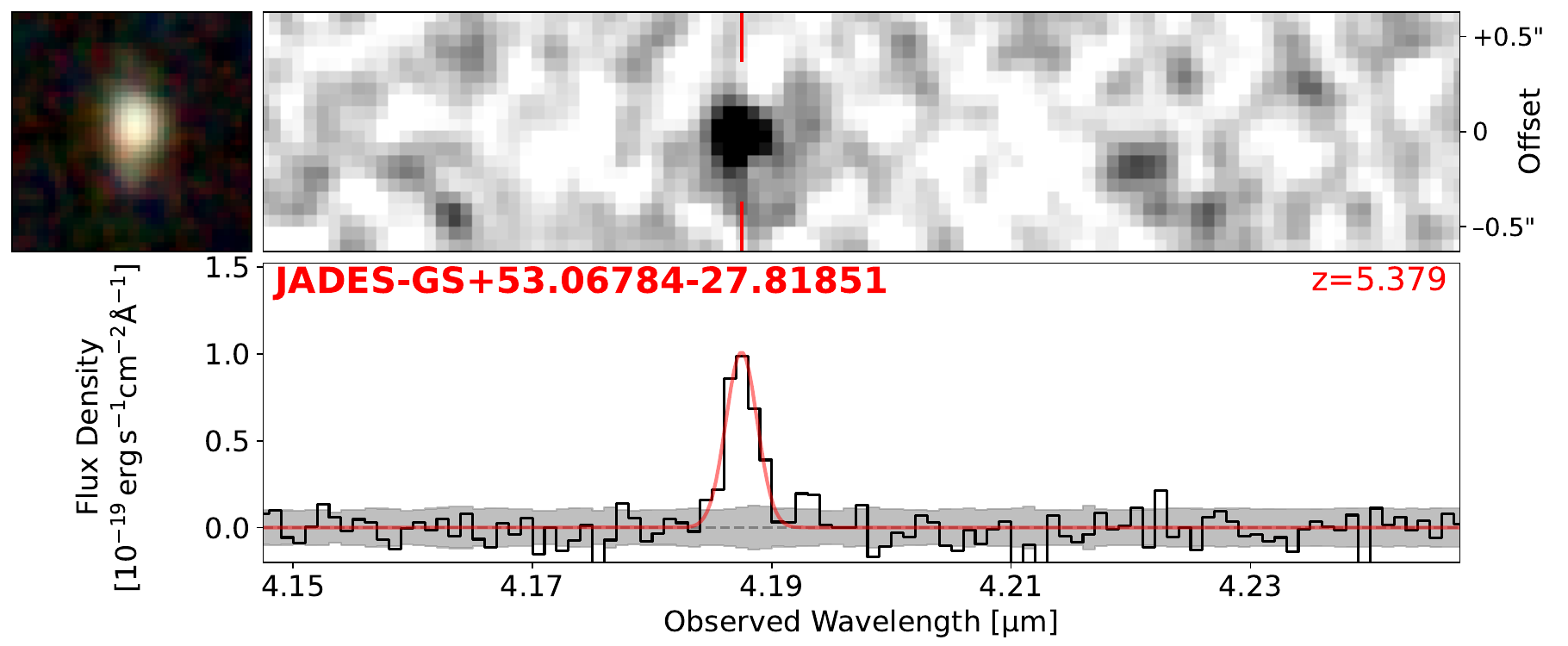}
\figsetgrpnote{NIRCam cutout images alongside the continuum-subtracted 2d and 1d grism spectra of JADES-GS+53.06784-27.81851 at $z = 5.379$, with $\mathrm{H} \alpha$ detected at $11.0\sigma$.}
\figsetgrpend

\figsetgrpstart
\figsetgrpnum{A1.14}
\figsetgrptitle{Ha}
\figsetplot{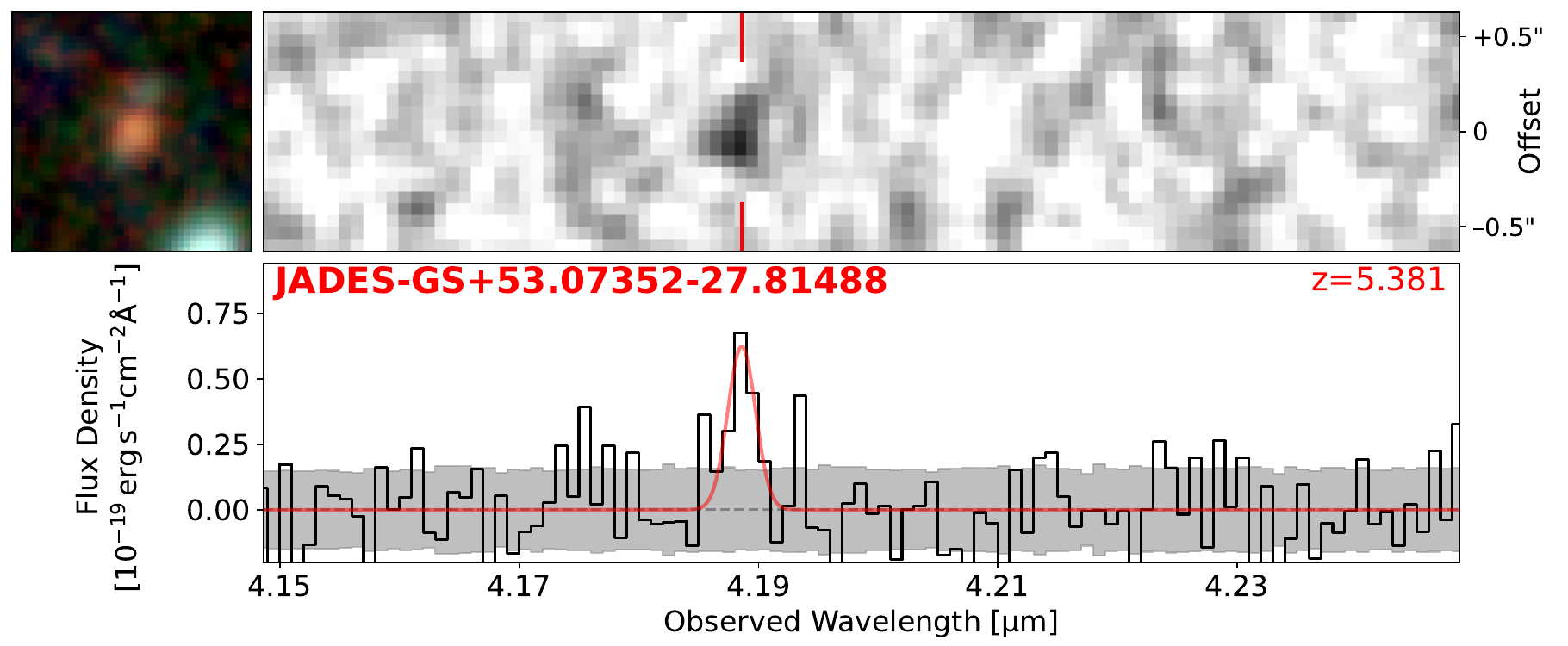}
\figsetgrpnote{NIRCam cutout images alongside the continuum-subtracted 2d and 1d grism spectra of JADES-GS+53.07352-27.81488 at $z = 5.381$, with $\mathrm{H} \alpha$ detected at $4.7\sigma$.}
\figsetgrpend

\figsetgrpstart
\figsetgrpnum{A1.15}
\figsetgrptitle{Ha}
\figsetplot{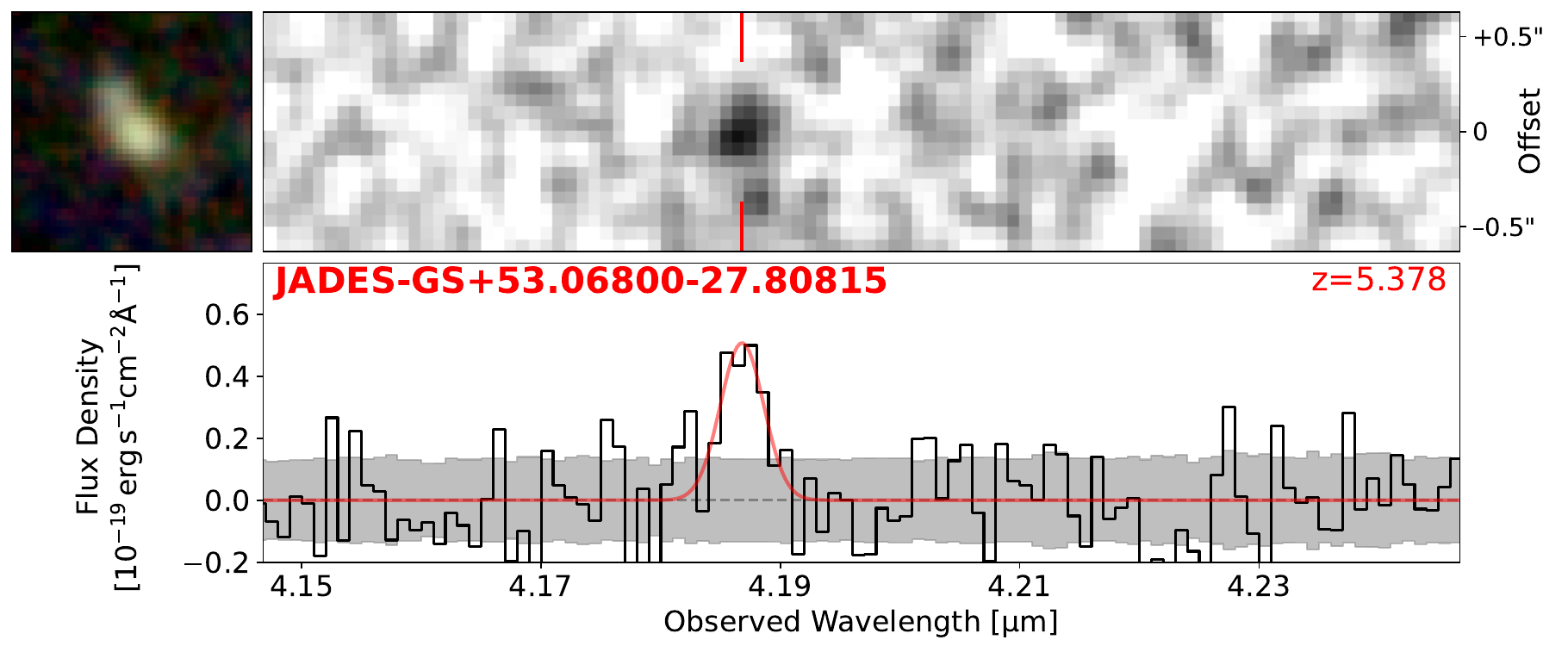}
\figsetgrpnote{NIRCam cutout images alongside the continuum-subtracted 2d and 1d grism spectra of JADES-GS+53.06800-27.80815 at $z = 5.378$, with $\mathrm{H} \alpha$ detected at $5.5\sigma$.}
\figsetgrpend

\figsetgrpstart
\figsetgrpnum{A1.16}
\figsetgrptitle{Ha}
\figsetplot{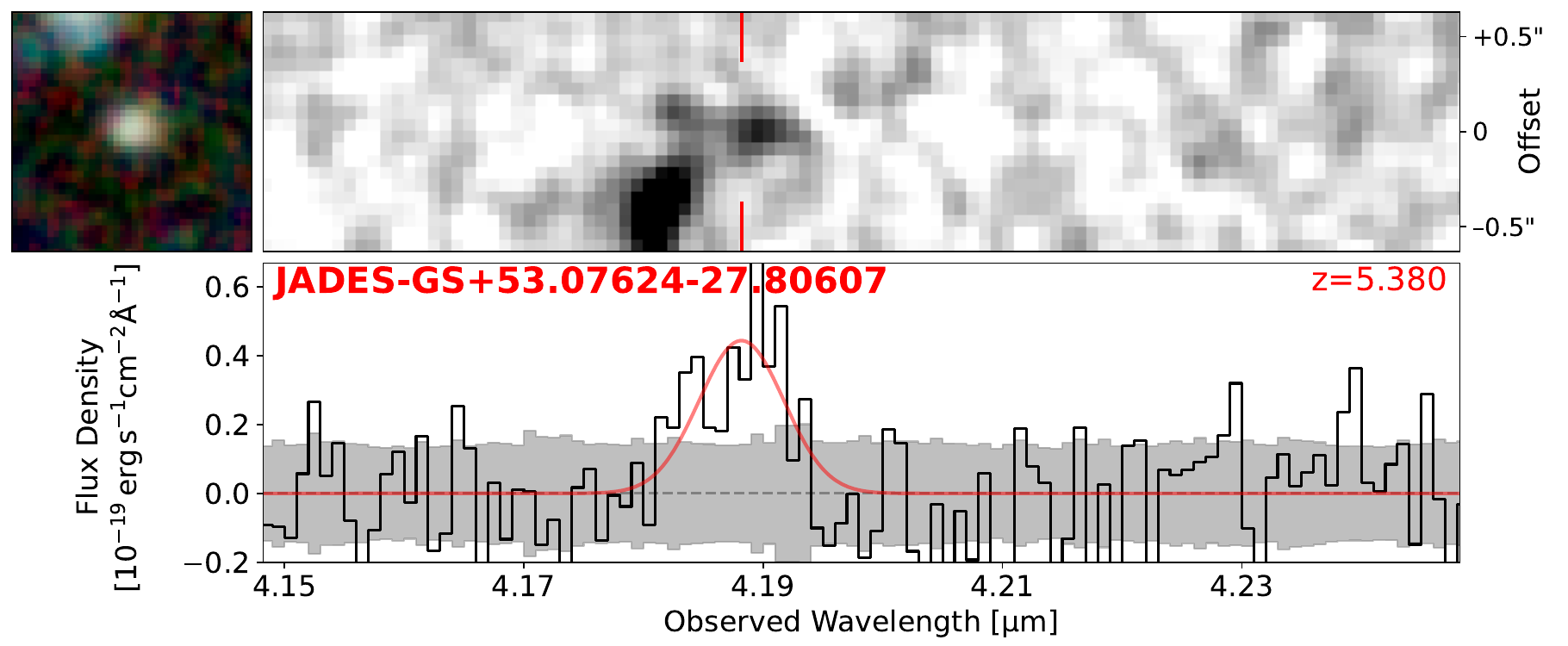}
\figsetgrpnote{NIRCam cutout images alongside the continuum-subtracted 2d and 1d grism spectra of JADES-GS+53.07624-27.80607 at $z = 5.380$, with $\mathrm{H} \alpha$ detected at $5.7\sigma$.}
\figsetgrpend

\figsetgrpstart
\figsetgrpnum{A1.17}
\figsetgrptitle{Ha}
\figsetplot{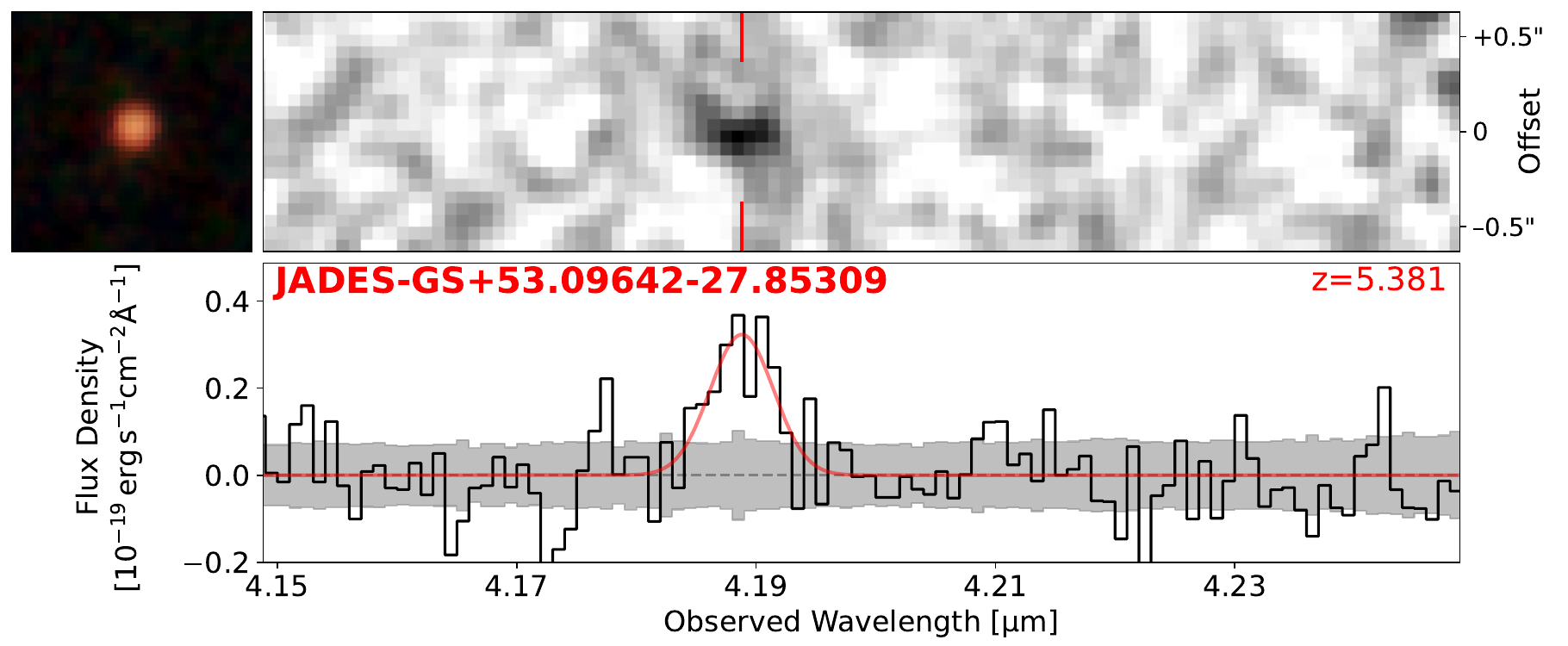}
\figsetgrpnote{NIRCam cutout images alongside the continuum-subtracted 2d and 1d grism spectra of JADES-GS+53.09642-27.85309 at $z = 5.381$, with $\mathrm{H} \alpha$ detected at $7.5\sigma$.}
\figsetgrpend

\figsetgrpstart
\figsetgrpnum{A1.18}
\figsetgrptitle{Ha}
\figsetplot{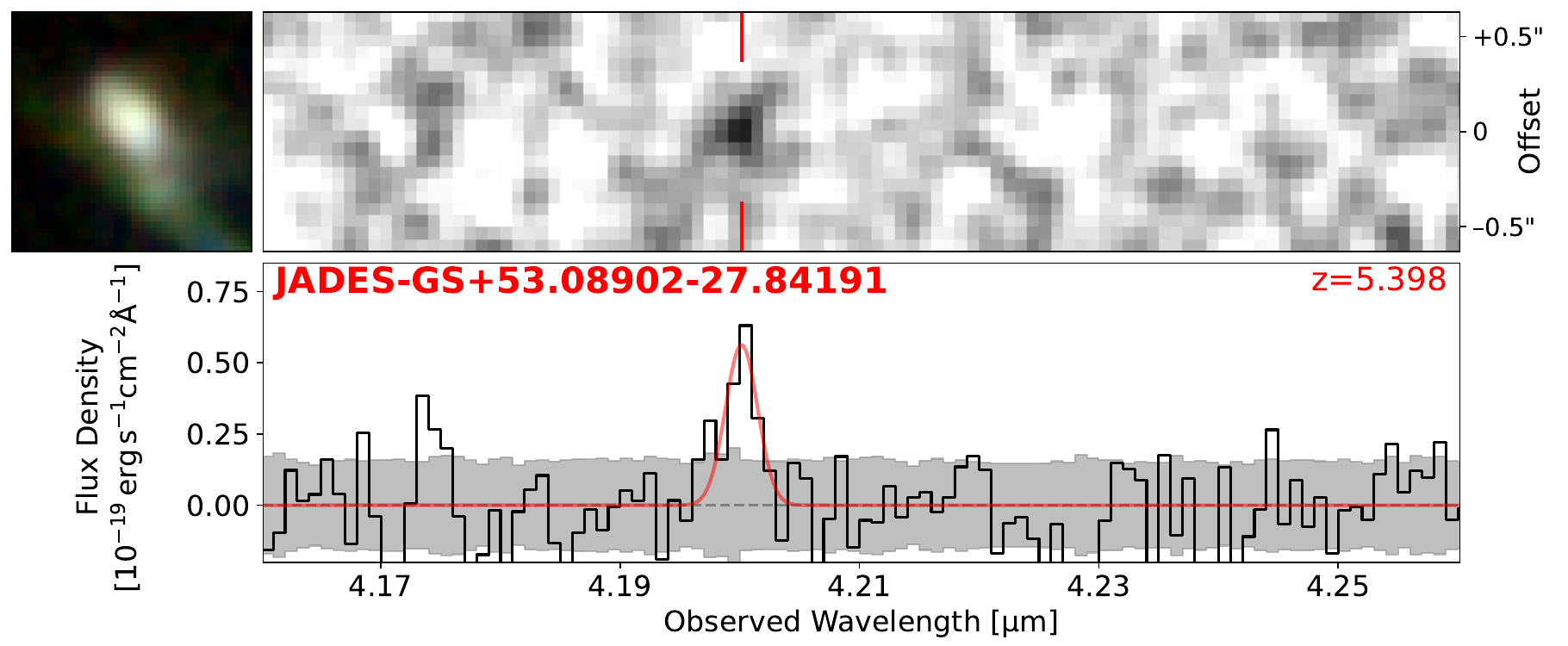}
\figsetgrpnote{NIRCam cutout images alongside the continuum-subtracted 2d and 1d grism spectra of JADES-GS+53.08902-27.84191 at $z = 5.398$, with $\mathrm{H} \alpha$ detected at $4.2\sigma$.}
\figsetgrpend

\figsetgrpstart
\figsetgrpnum{A1.19}
\figsetgrptitle{Ha}
\figsetplot{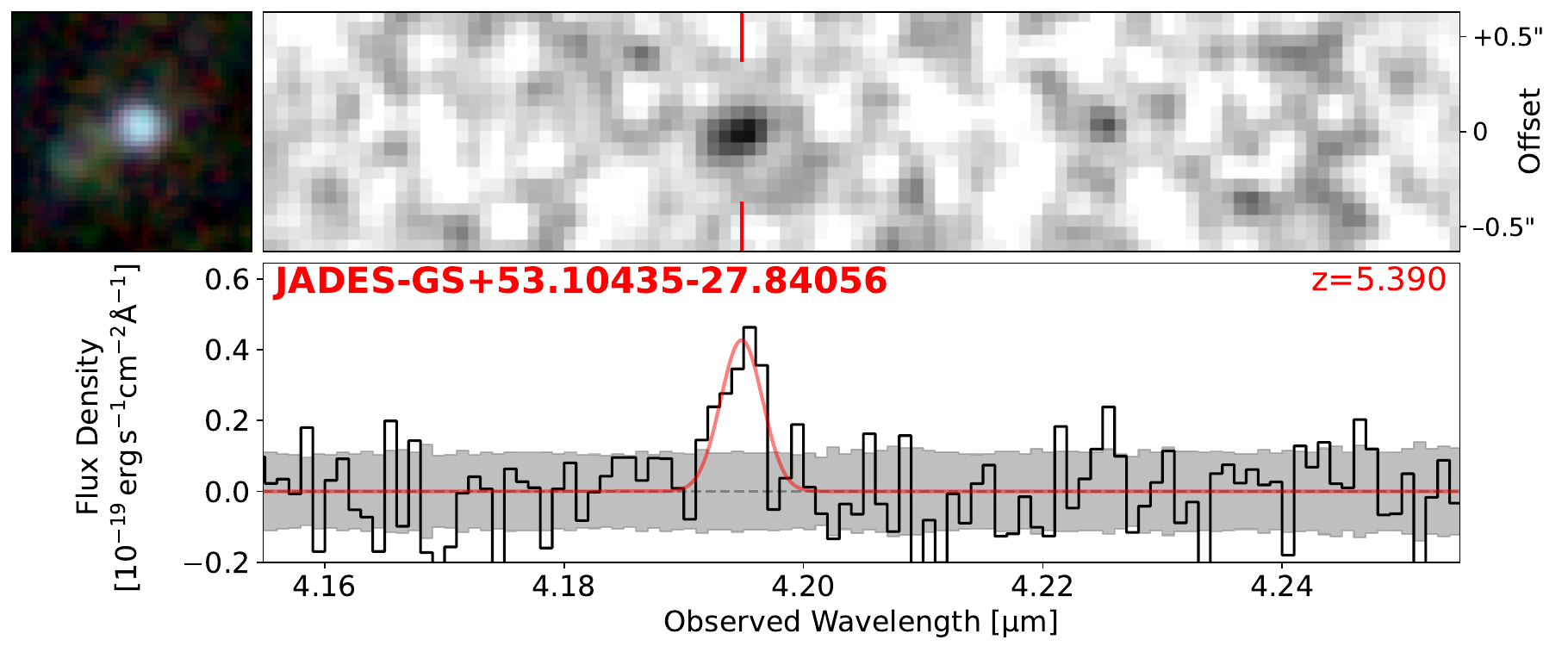}
\figsetgrpnote{NIRCam cutout images alongside the continuum-subtracted 2d and 1d grism spectra of JADES-GS+53.10435-27.84056 at $z = 5.390$, with $\mathrm{H} \alpha$ detected at $5.6\sigma$.}
\figsetgrpend

\figsetgrpstart
\figsetgrpnum{A1.20}
\figsetgrptitle{Ha}
\figsetplot{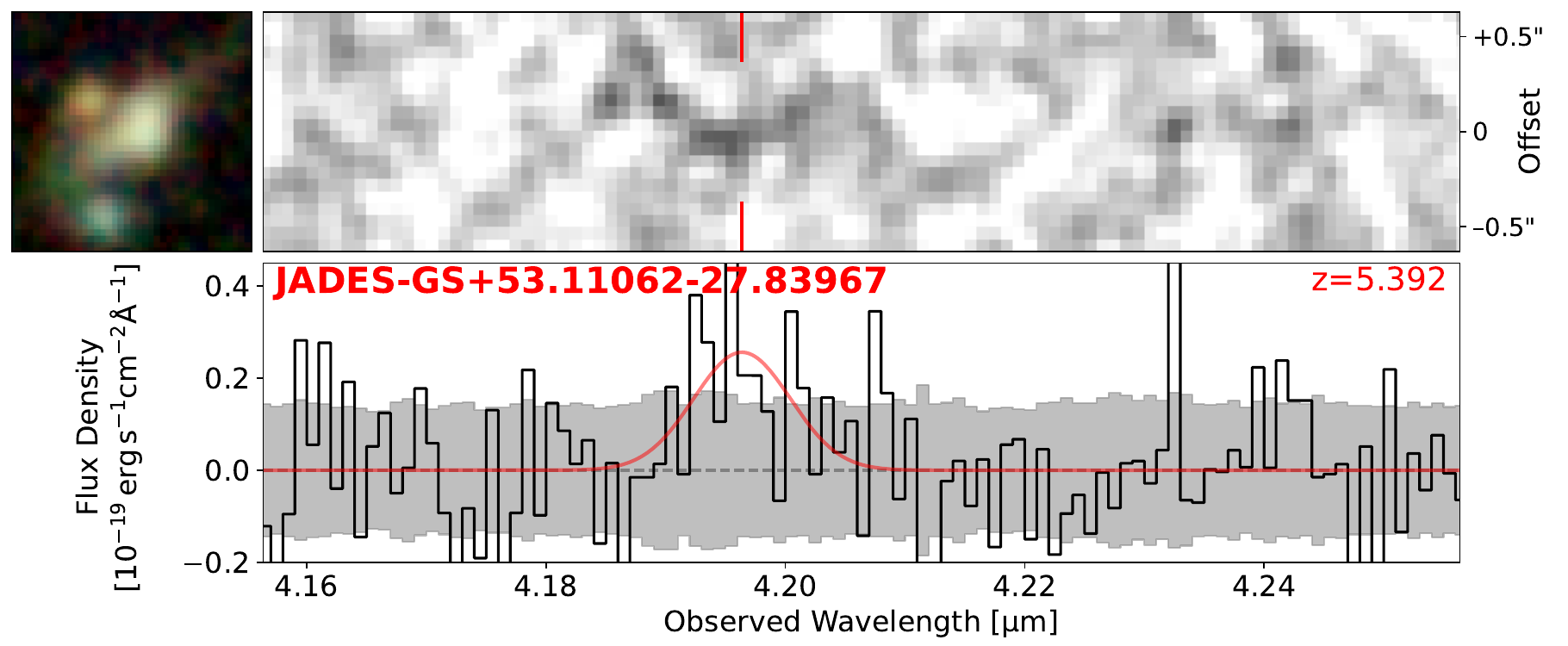}
\figsetgrpnote{NIRCam cutout images alongside the continuum-subtracted 2d and 1d grism spectra of JADES-GS+53.11062-27.83967 at $z = 5.392$, with $\mathrm{H} \alpha$ detected at $3.6\sigma$.}
\figsetgrpend

\figsetgrpstart
\figsetgrpnum{A1.21}
\figsetgrptitle{Ha}
\figsetplot{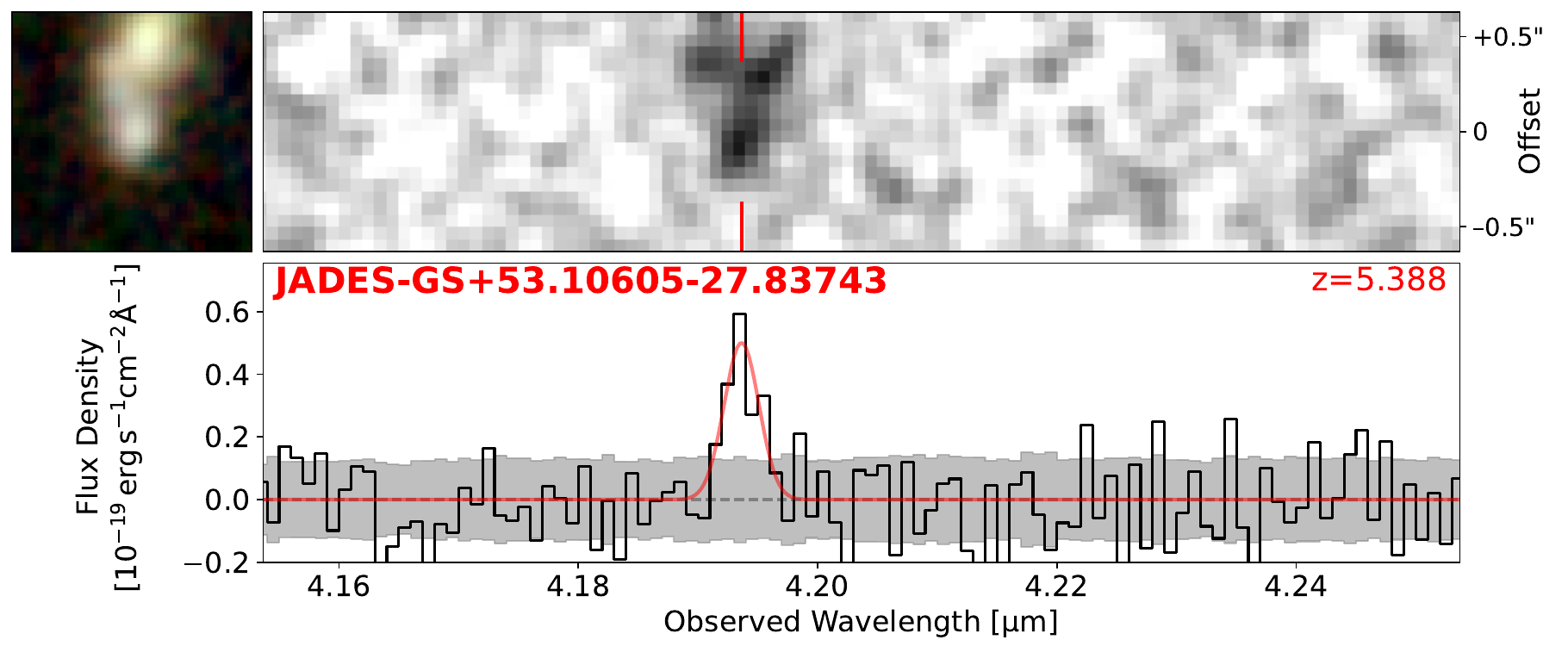}
\figsetgrpnote{NIRCam cutout images alongside the continuum-subtracted 2d and 1d grism spectra of JADES-GS+53.10605-27.83743 at $z = 5.388$, with $\mathrm{H} \alpha$ detected at $5.1\sigma$.}
\figsetgrpend

\figsetgrpstart
\figsetgrpnum{A1.22}
\figsetgrptitle{Ha}
\figsetplot{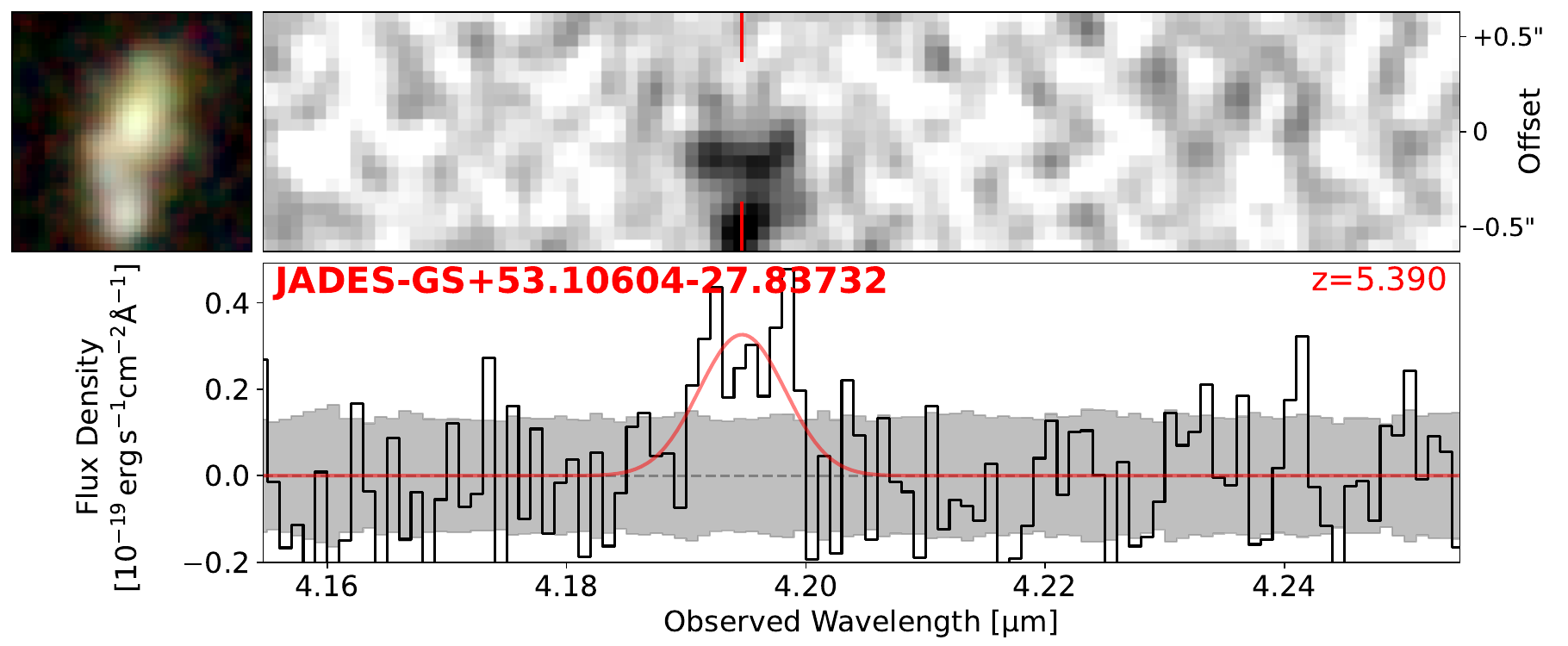}
\figsetgrpnote{NIRCam cutout images alongside the continuum-subtracted 2d and 1d grism spectra of JADES-GS+53.10604-27.83732 at $z = 5.390$, with $\mathrm{H} \alpha$ detected at $4.9\sigma$.}
\figsetgrpend

\figsetgrpstart
\figsetgrpnum{A1.23}
\figsetgrptitle{Ha}
\figsetplot{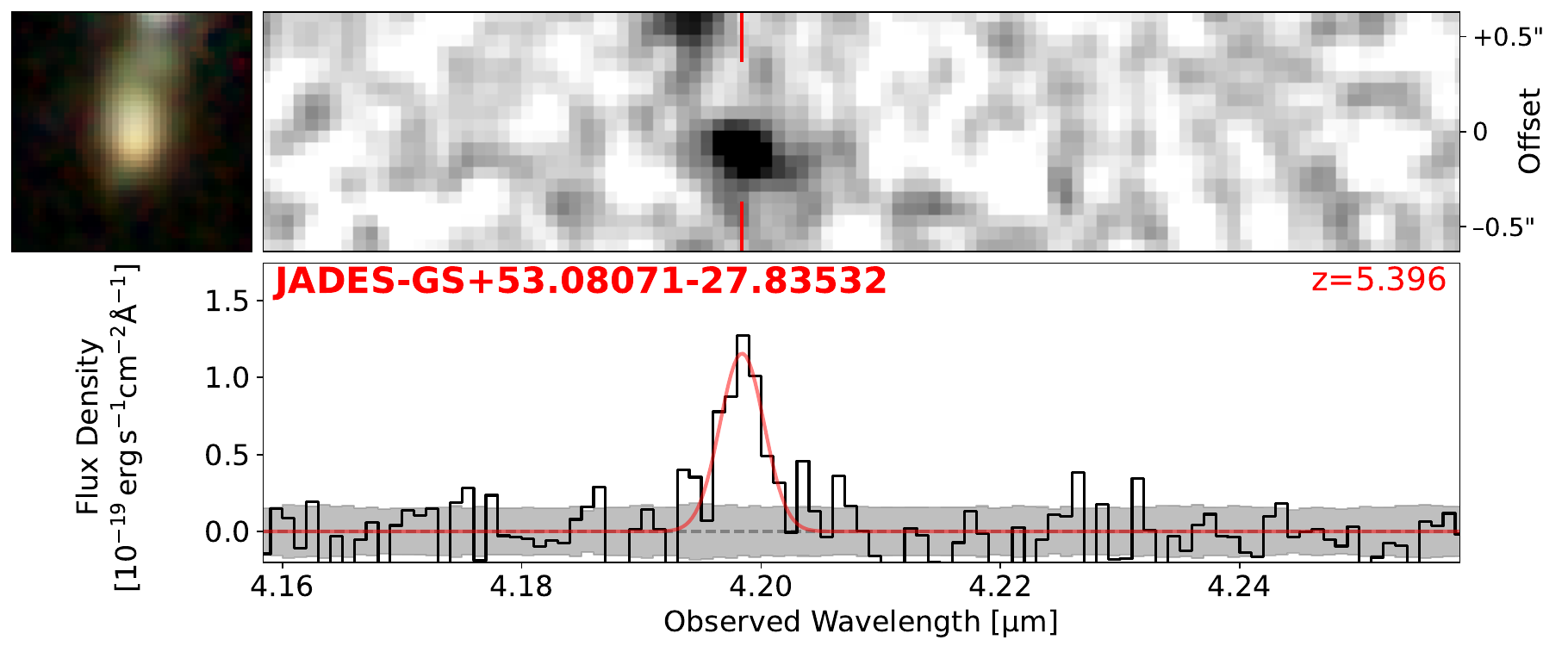}
\figsetgrpnote{NIRCam cutout images alongside the continuum-subtracted 2d and 1d grism spectra of JADES-GS+53.08071-27.83532 at $z = 5.396$, with $\mathrm{H} \alpha$ detected at $10.2\sigma$.}
\figsetgrpend

\figsetgrpstart
\figsetgrpnum{A1.24}
\figsetgrptitle{Ha}
\figsetplot{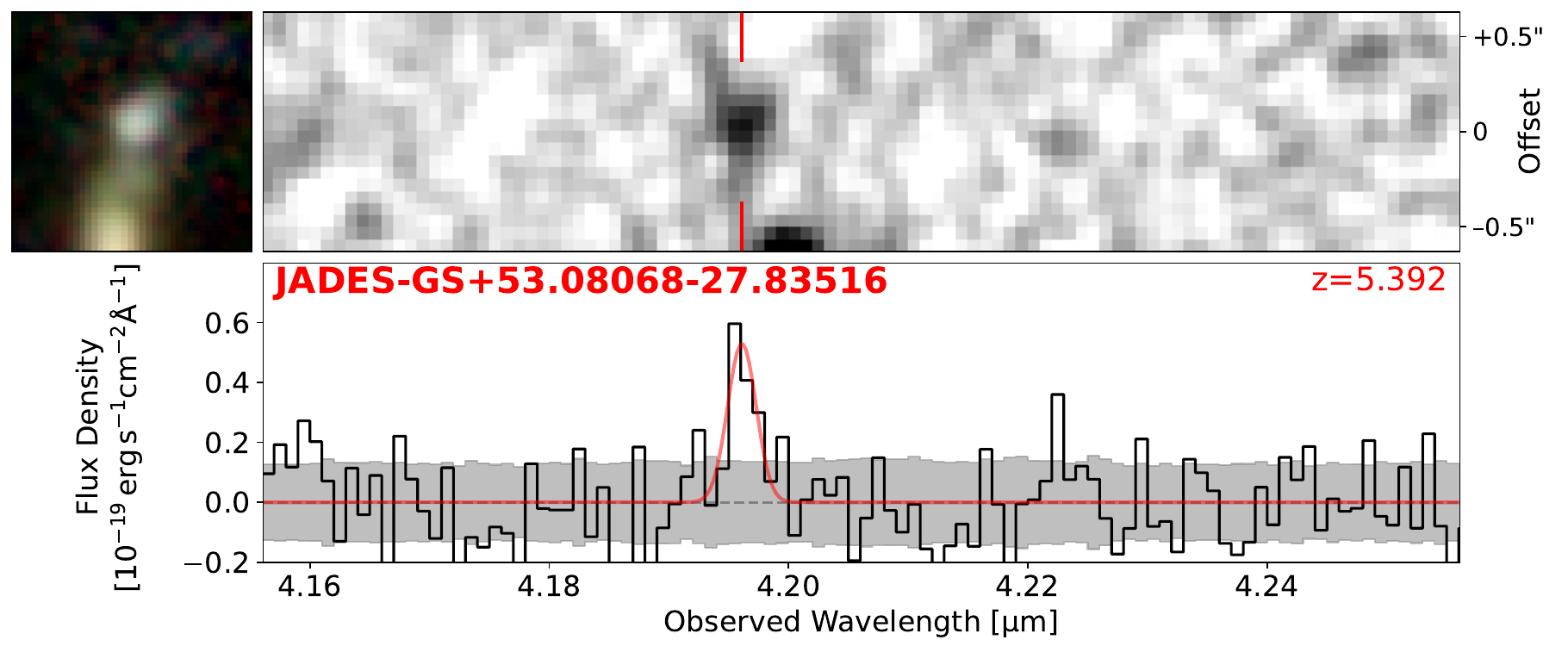}
\figsetgrpnote{NIRCam cutout images alongside the continuum-subtracted 2d and 1d grism spectra of JADES-GS+53.08068-27.83516 at $z = 5.392$, with $\mathrm{H} \alpha$ detected at $4.5\sigma$.}
\figsetgrpend

\figsetgrpstart
\figsetgrpnum{A1.25}
\figsetgrptitle{Ha}
\figsetplot{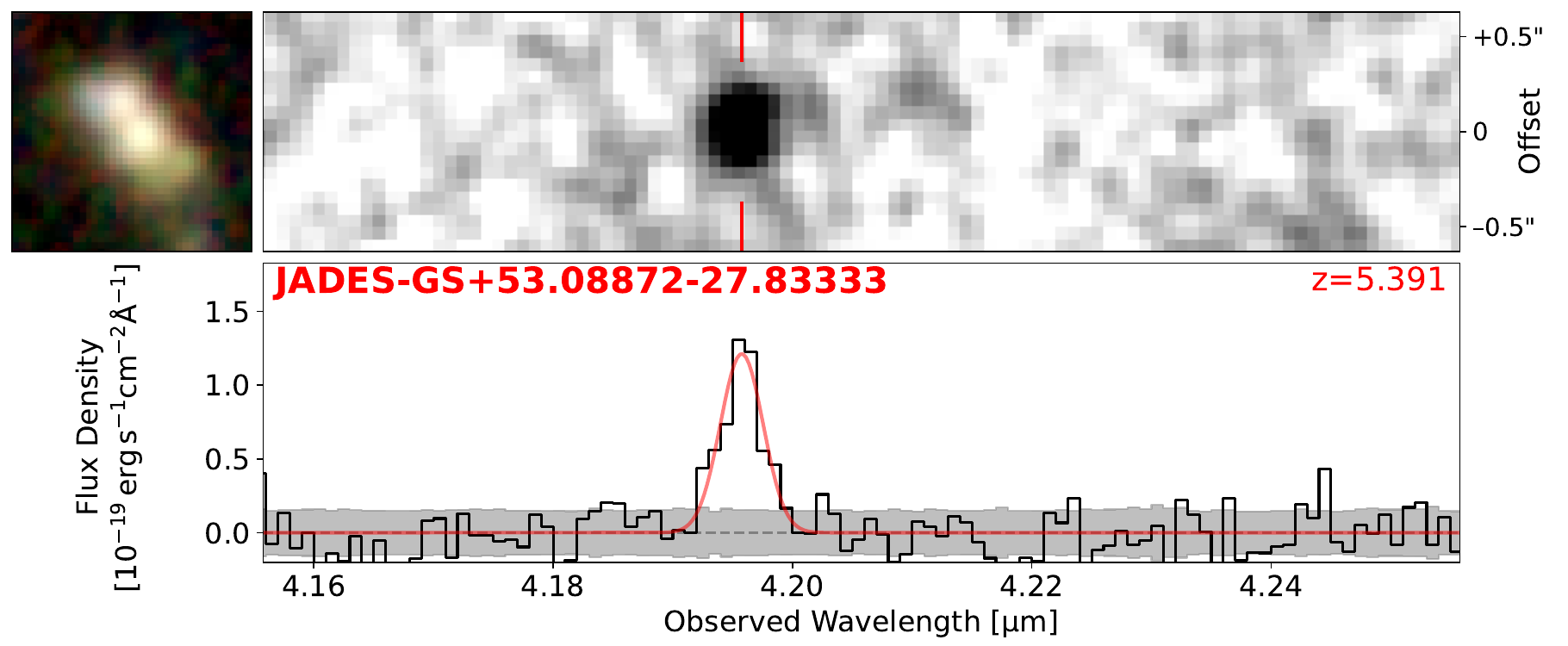}
\figsetgrpnote{NIRCam cutout images alongside the continuum-subtracted 2d and 1d grism spectra of JADES-GS+53.08872-27.83333 at $z = 5.391$, with $\mathrm{H} \alpha$ detected at $11.4\sigma$.}
\figsetgrpend

\figsetgrpstart
\figsetgrpnum{A1.26}
\figsetgrptitle{Ha}
\figsetplot{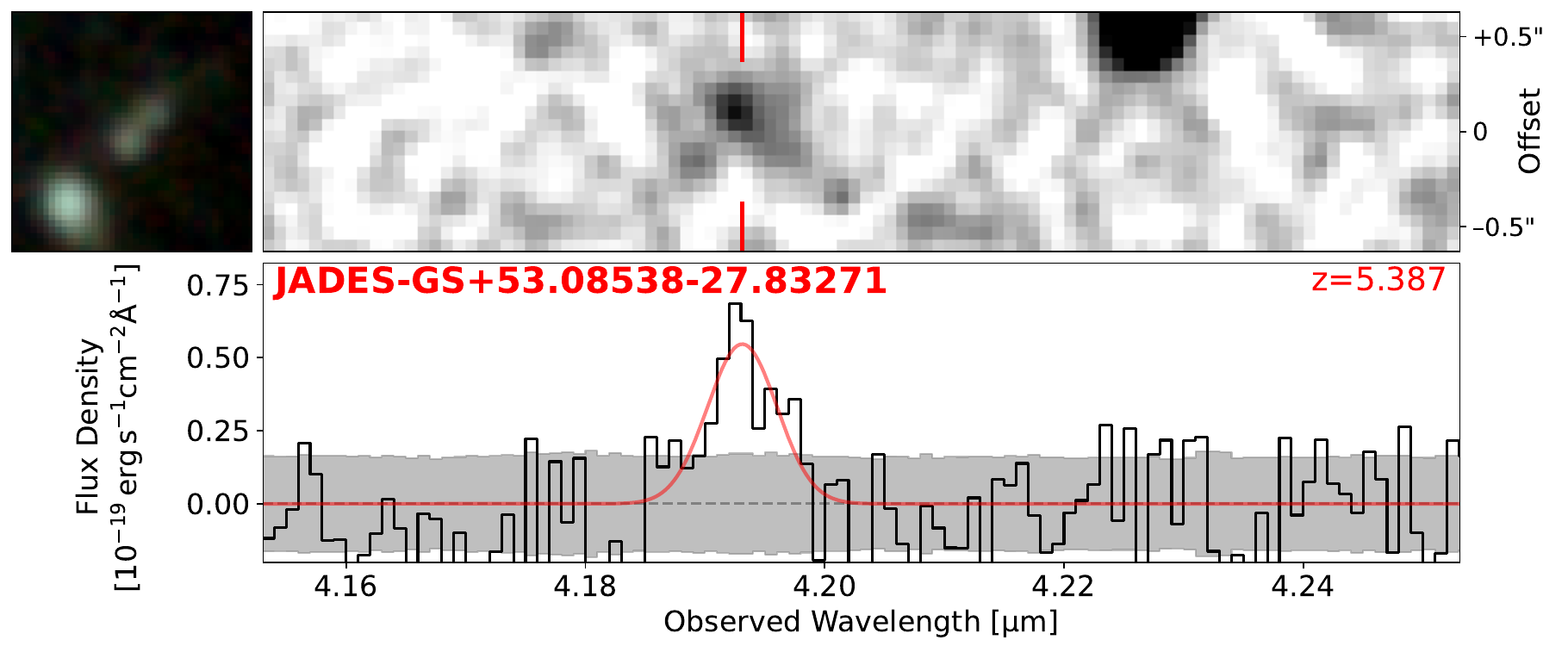}
\figsetgrpnote{NIRCam cutout images alongside the continuum-subtracted 2d and 1d grism spectra of JADES-GS+53.08538-27.83271 at $z = 5.387$, with $\mathrm{H} \alpha$ detected at $6.1\sigma$.}
\figsetgrpend

\figsetgrpstart
\figsetgrpnum{A1.27}
\figsetgrptitle{Ha}
\figsetplot{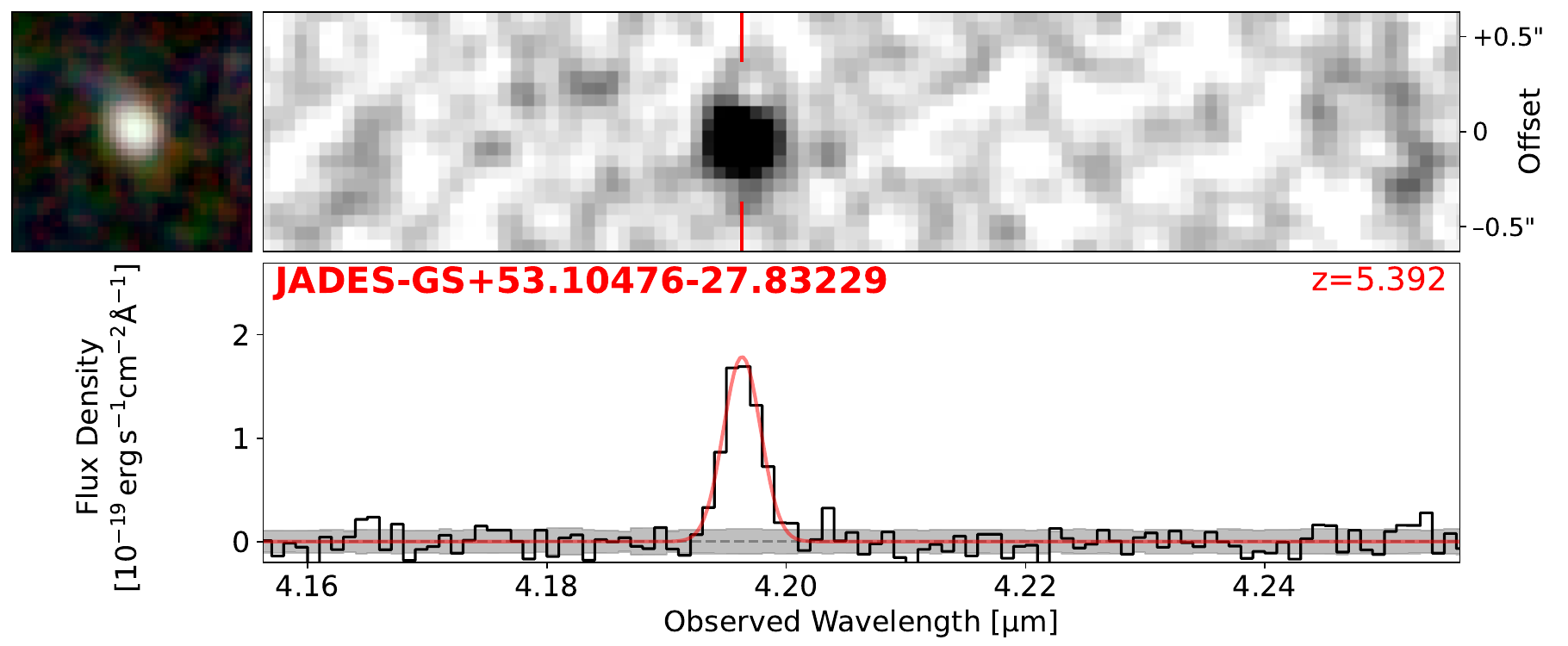}
\figsetgrpnote{NIRCam cutout images alongside the continuum-subtracted 2d and 1d grism spectra of JADES-GS+53.10476-27.83229 at $z = 5.392$, with $\mathrm{H} \alpha$ detected at $20.8\sigma$.}
\figsetgrpend

\figsetgrpstart
\figsetgrpnum{A1.28}
\figsetgrptitle{Ha}
\figsetplot{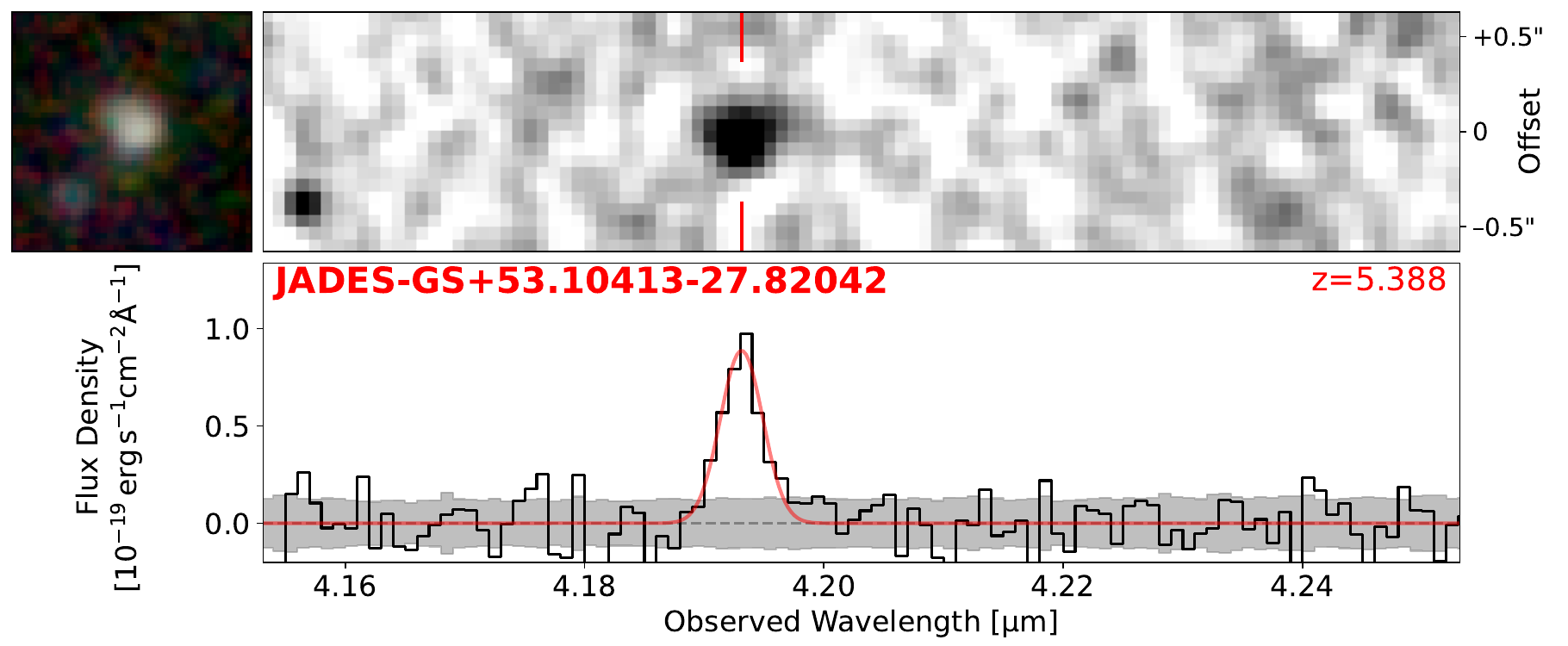}
\figsetgrpnote{NIRCam cutout images alongside the continuum-subtracted 2d and 1d grism spectra of JADES-GS+53.10413-27.82042 at $z = 5.388$, with $\mathrm{H} \alpha$ detected at $10.1\sigma$.}
\figsetgrpend

\figsetgrpstart
\figsetgrpnum{A1.29}
\figsetgrptitle{Ha}
\figsetplot{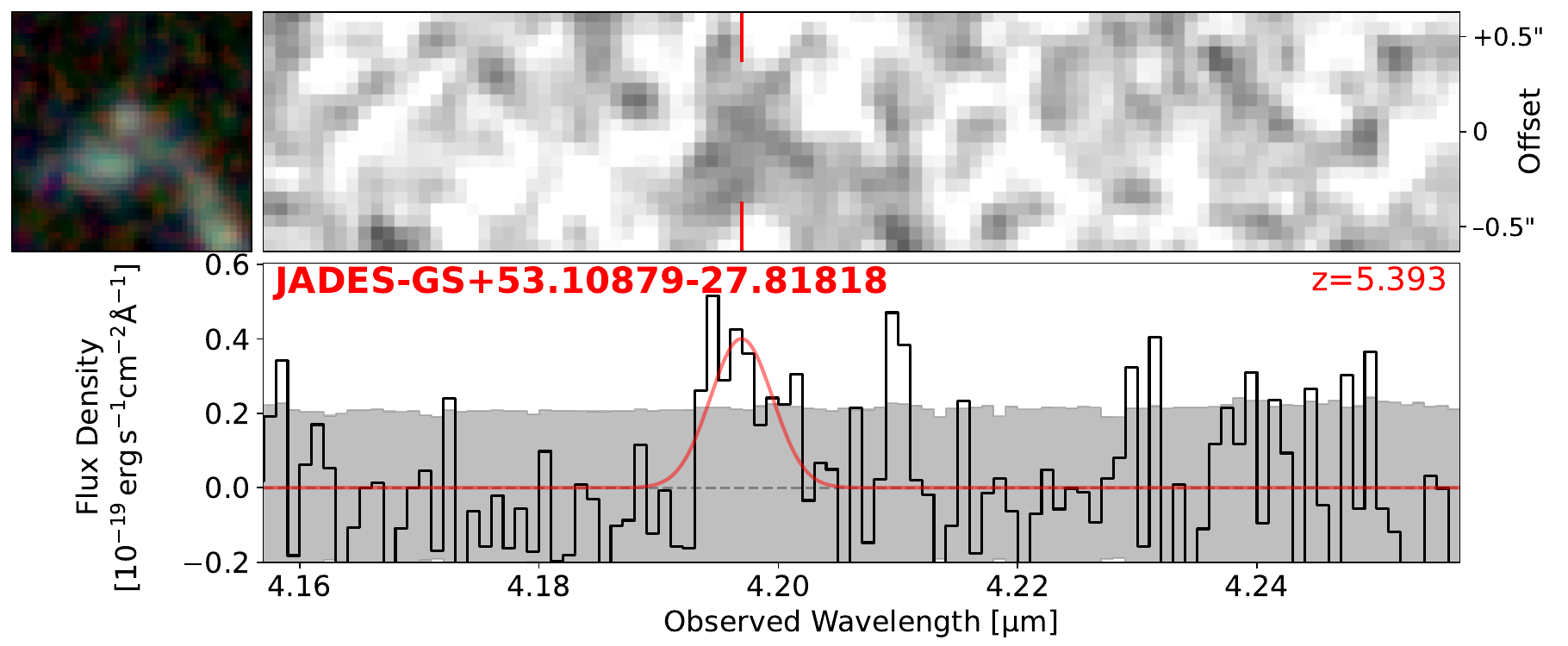}
\figsetgrpnote{NIRCam cutout images alongside the continuum-subtracted 2d and 1d grism spectra of JADES-GS+53.10879-27.81818 at $z = 5.393$, with $\mathrm{H} \alpha$ detected at $3.2\sigma$.}
\figsetgrpend

\figsetgrpstart
\figsetgrpnum{A1.30}
\figsetgrptitle{Ha}
\figsetplot{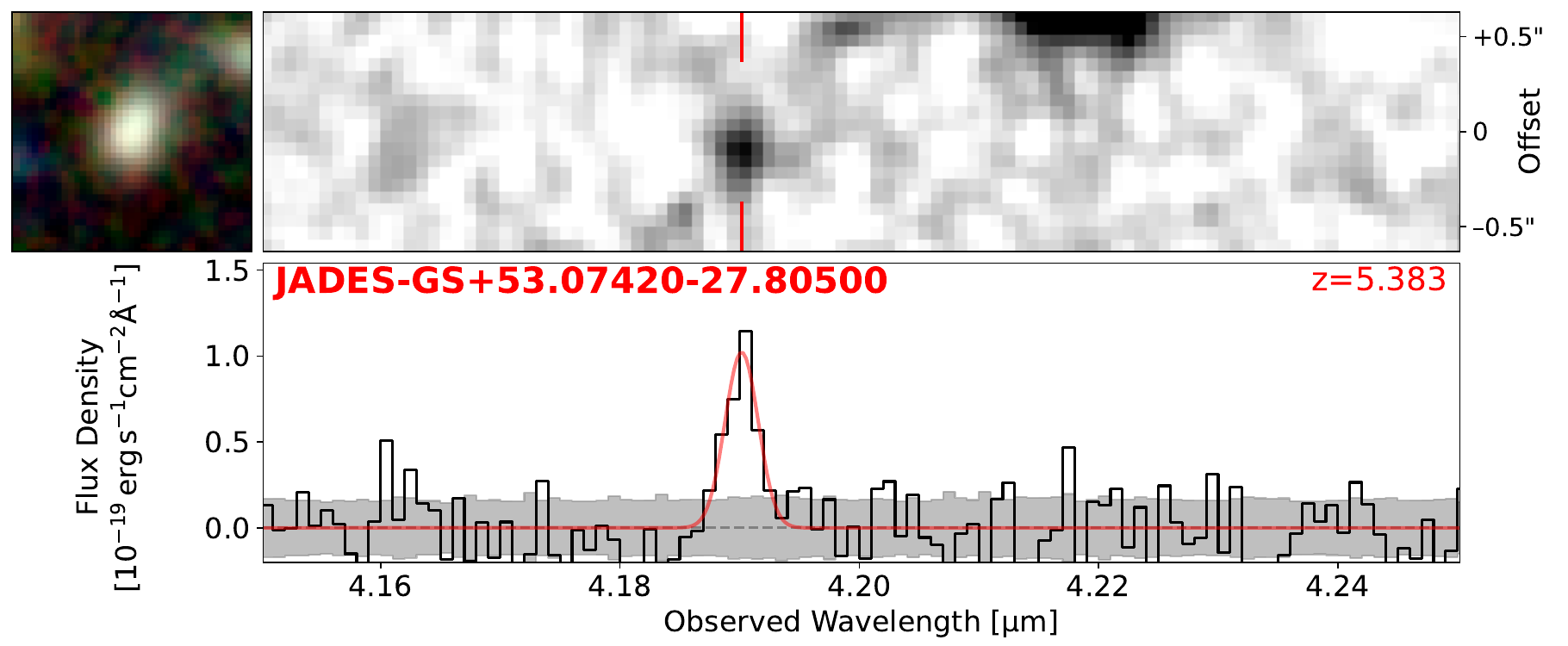}
\figsetgrpnote{NIRCam cutout images alongside the continuum-subtracted 2d and 1d grism spectra of JADES-GS+53.07420-27.80500 at $z = 5.383$, with $\mathrm{H} \alpha$ detected at $7.6\sigma$.}
\figsetgrpend

\figsetgrpstart
\figsetgrpnum{A1.31}
\figsetgrptitle{Ha}
\figsetplot{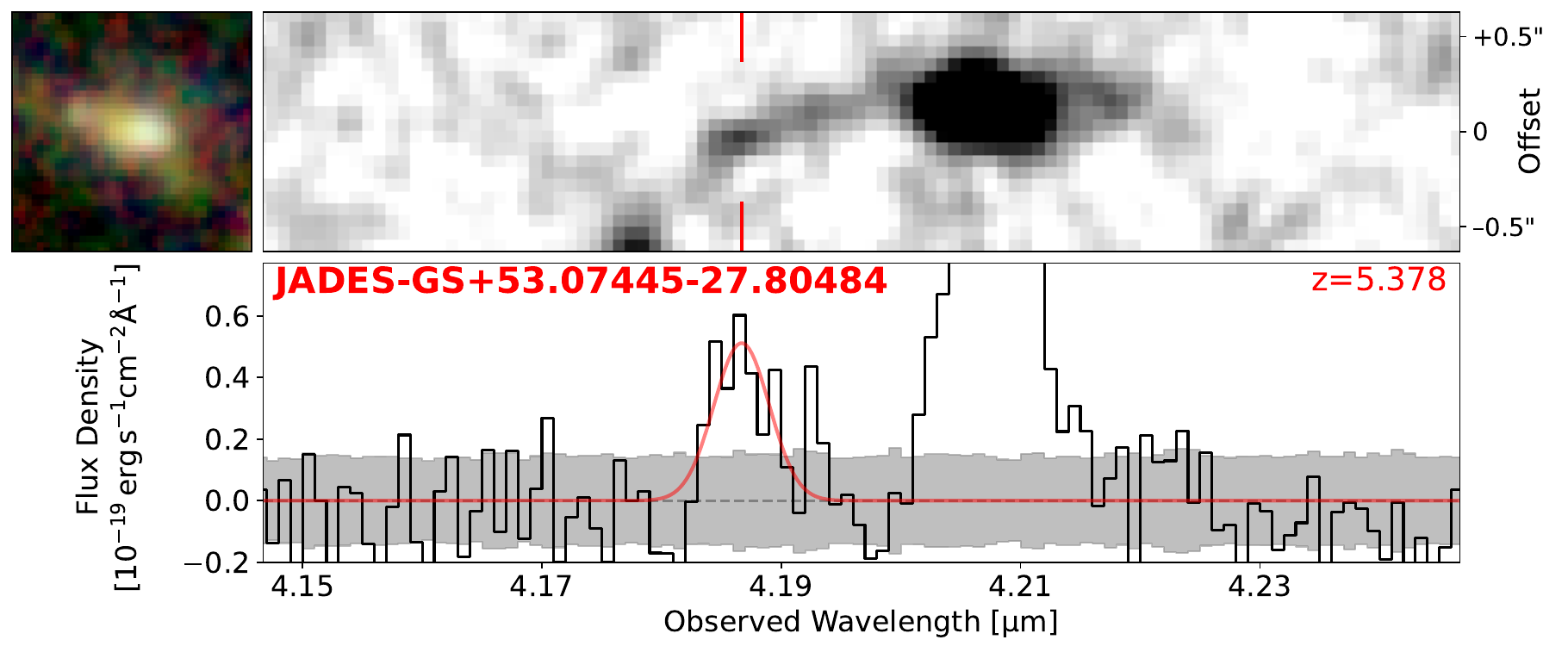}
\figsetgrpnote{NIRCam cutout images alongside the continuum-subtracted 2d and 1d grism spectra of JADES-GS+53.07445-27.80484 at $z = 5.378$, with $\mathrm{H} \alpha$ detected at $5.7\sigma$.}
\figsetgrpend

\figsetgrpstart
\figsetgrpnum{A1.32}
\figsetgrptitle{Ha}
\figsetplot{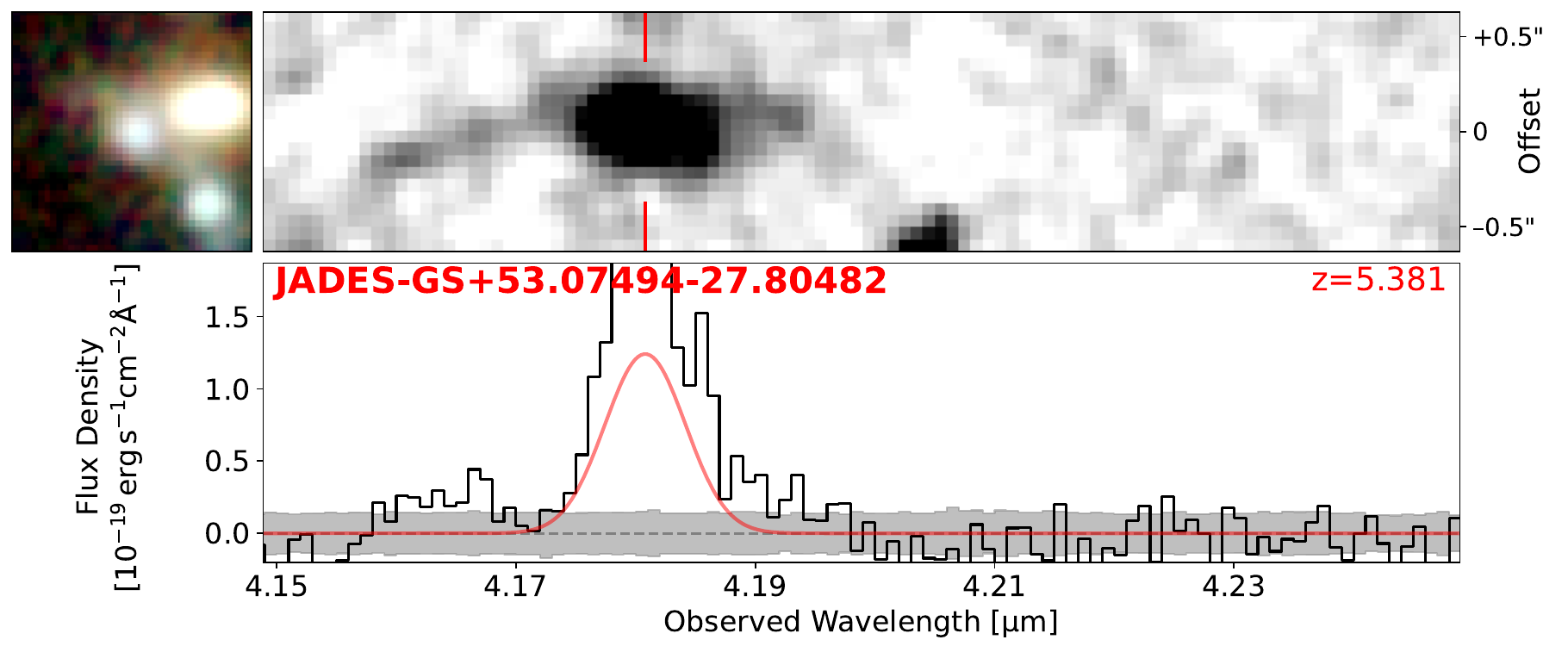}
\figsetgrpnote{NIRCam cutout images alongside the continuum-subtracted 2d and 1d grism spectra of JADES-GS+53.07494-27.80482 at $z = 5.381$, with $\mathrm{H} \alpha$ detected at $34.8\sigma$.}
\figsetgrpend

\figsetgrpstart
\figsetgrpnum{A1.33}
\figsetgrptitle{Ha}
\figsetplot{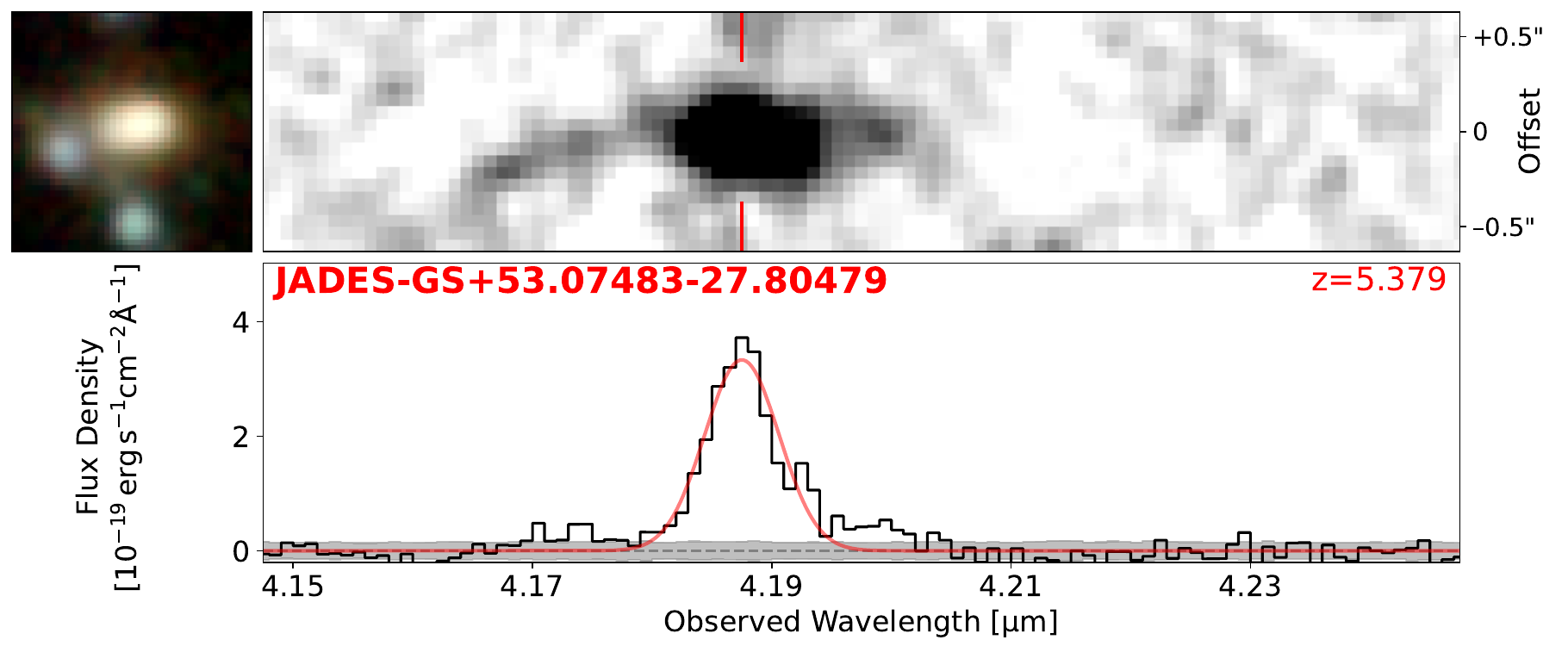}
\figsetgrpnote{NIRCam cutout images alongside the continuum-subtracted 2d and 1d grism spectra of JADES-GS+53.07483-27.80479 at $z = 5.379$, with $\mathrm{H} \alpha$ detected at $41.7\sigma$.}
\figsetgrpend

\figsetgrpstart
\figsetgrpnum{A1.34}
\figsetgrptitle{Ha}
\figsetplot{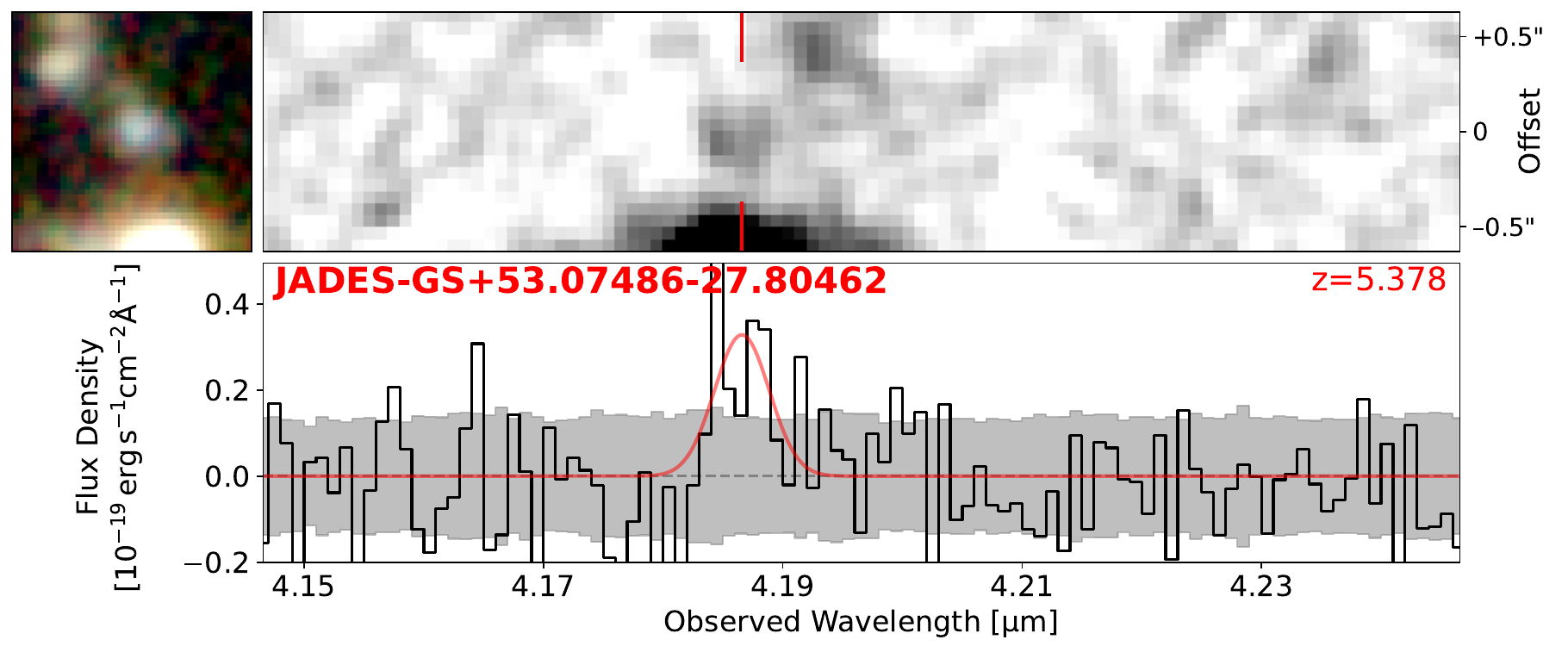}
\figsetgrpnote{NIRCam cutout images alongside the continuum-subtracted 2d and 1d grism spectra of JADES-GS+53.07486-27.80462 at $z = 5.378$, with $\mathrm{H} \alpha$ detected at $3.8\sigma$.}
\figsetgrpend

\figsetgrpstart
\figsetgrpnum{A1.35}
\figsetgrptitle{Ha}
\figsetplot{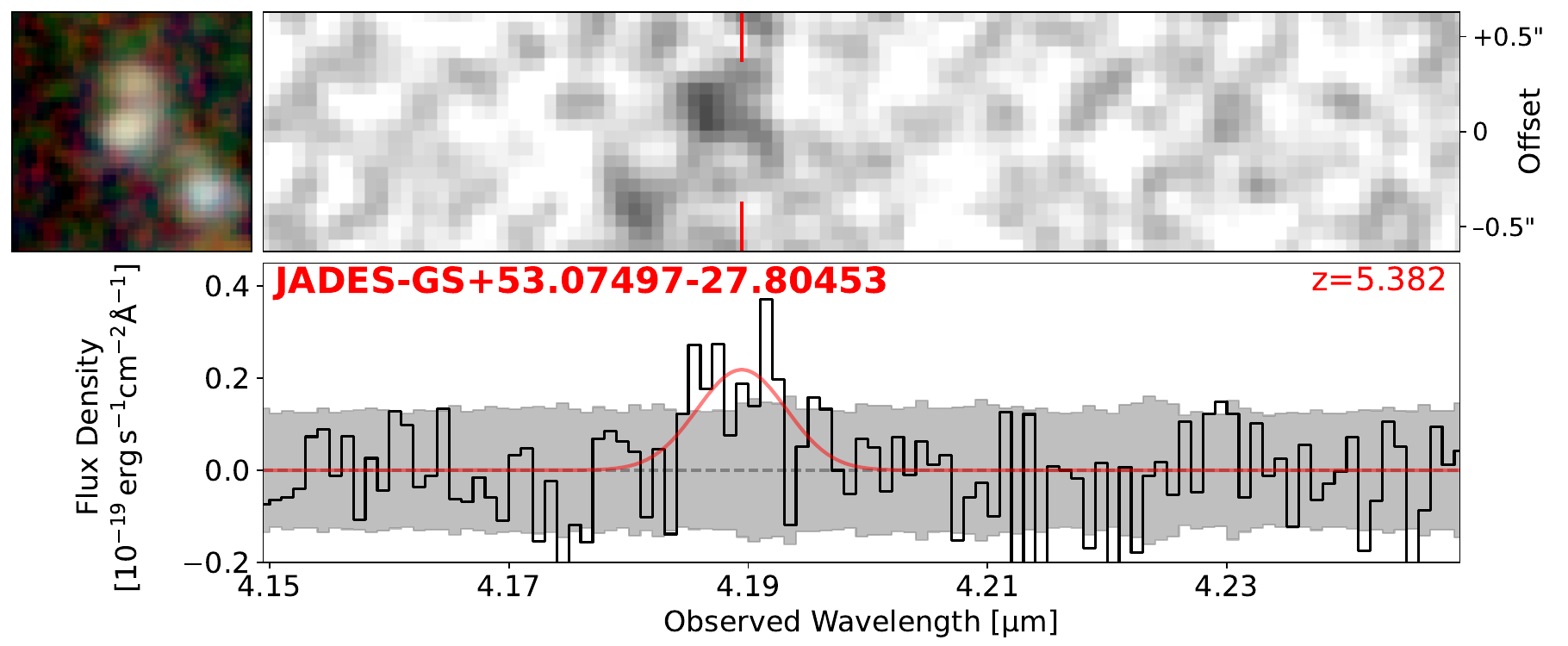}
\figsetgrpnote{NIRCam cutout images alongside the continuum-subtracted 2d and 1d grism spectra of JADES-GS+53.07497-27.80453 at $z = 5.382$, with $\mathrm{H} \alpha$ detected at $3.4\sigma$.}
\figsetgrpend

\figsetgrpstart
\figsetgrpnum{A1.36}
\figsetgrptitle{Ha}
\figsetplot{figures/aastex_specfigs/spec_ID201006.pdf}
\figsetgrpnote{NIRCam cutout images alongside the continuum-subtracted 2d and 1d grism spectra of JADES-GS+53.07496-27.80447 at $z = 5.378$, with $\mathrm{H} \alpha$ detected at $4.2\sigma$.}
\figsetgrpend

\figsetgrpstart
\figsetgrpnum{A1.37}
\figsetgrptitle{Ha}
\figsetplot{figures/aastex_specfigs/spec_ID201158.pdf}
\figsetgrpnote{NIRCam cutout images alongside the continuum-subtracted 2d and 1d grism spectra of JADES-GS+53.07500-27.80421 at $z = 5.378$, with $\mathrm{H} \alpha$ detected at $5.1\sigma$.}
\figsetgrpend

\figsetgrpstart
\figsetgrpnum{A1.38}
\figsetgrptitle{Ha}
\figsetplot{figures/aastex_specfigs/spec_ID201252.pdf}
\figsetgrpnote{NIRCam cutout images alongside the continuum-subtracted 2d and 1d grism spectra of JADES-GS+53.07408-27.80401 at $z = 5.374$, with $\mathrm{H} \alpha$ detected at $19.0\sigma$.}
\figsetgrpend

\figsetgrpstart
\figsetgrpnum{A1.39}
\figsetgrptitle{Ha}
\figsetplot{figures/aastex_specfigs/spec_ID203138.pdf}
\figsetgrpnote{NIRCam cutout images alongside the continuum-subtracted 2d and 1d grism spectra of JADES-GS+53.07876-27.79750 at $z = 5.384$, with $\mathrm{H} \alpha$ detected at $3.1\sigma$.}
\figsetgrpend

\figsetend

\figsetstart
\figsetnum{A2}
\figsettitle{NIRCam images and grism spectra for the $N = 42$ confirmed members of the field.}

\figsetgrpstart
\figsetgrpnum{A2.1}
\figsetgrptitle{Ha}
\figsetplot{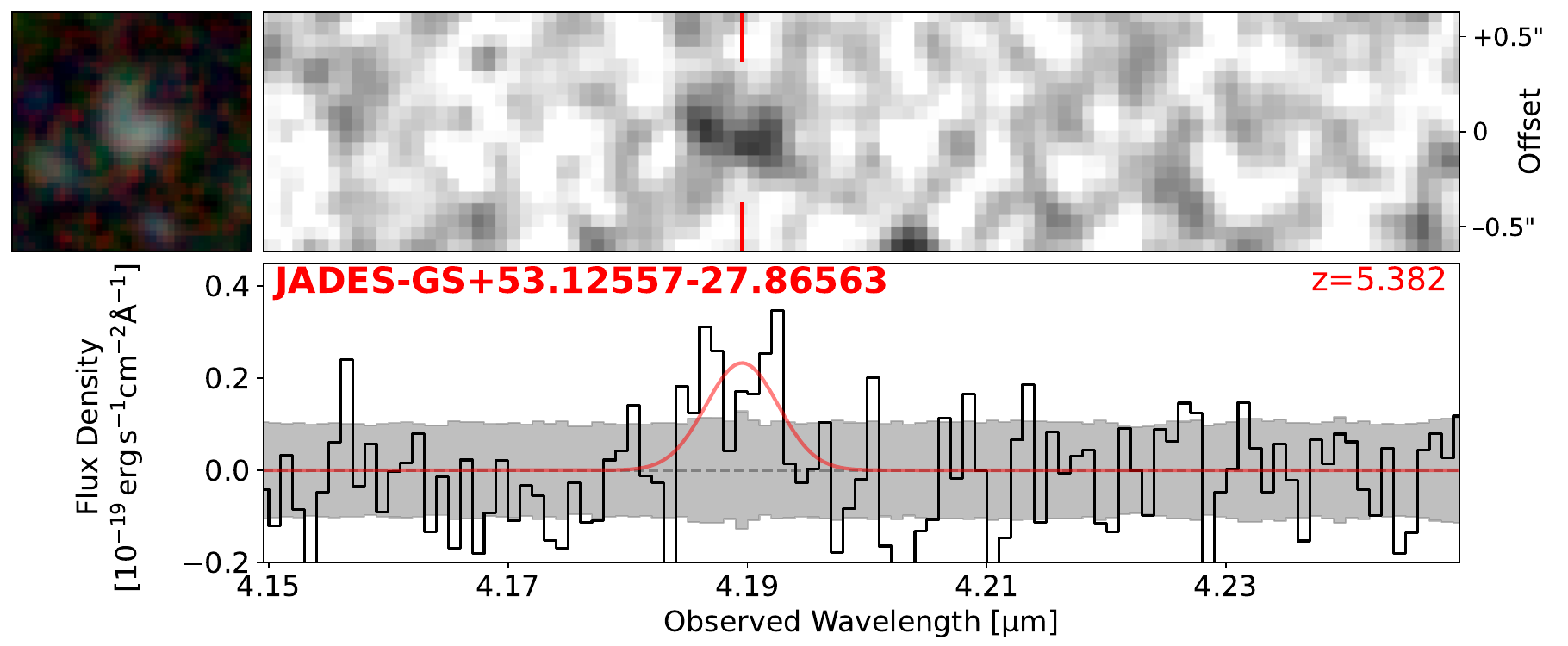}
\figsetgrpnote{NIRCam cutout images alongside the continuum-subtracted 2d and 1d grism spectra of JADES-GS+53.12557-27.86563 at $z = 5.382$, with $\mathrm{H} \alpha$ detected at $4.1\sigma$.}
\figsetgrpend

\figsetgrpstart
\figsetgrpnum{A2.2}
\figsetgrptitle{Ha}
\figsetplot{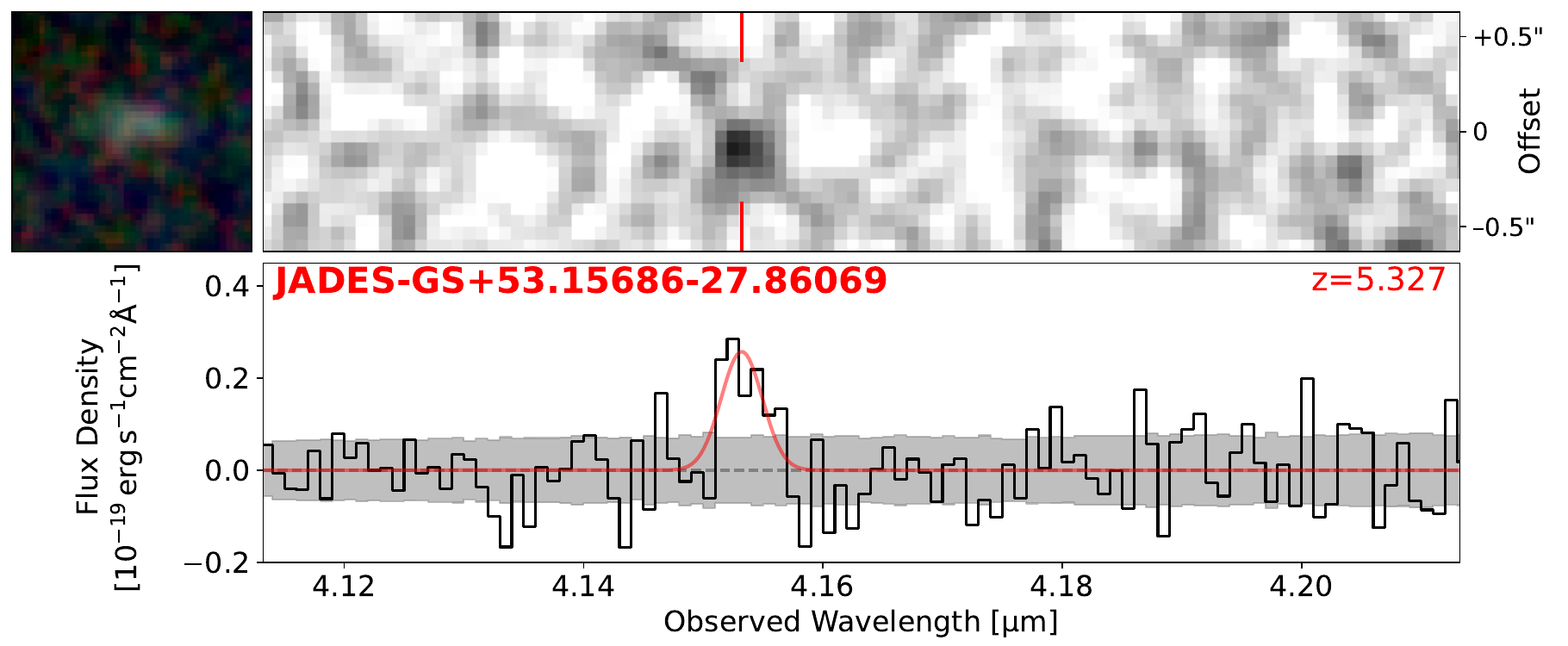}
\figsetgrpnote{NIRCam cutout images alongside the continuum-subtracted 2d and 1d grism spectra of JADES-GS+53.15686-27.86069 at $z = 5.327$, with $\mathrm{H} \alpha$ detected at $4.9\sigma$.}
\figsetgrpend

\figsetgrpstart
\figsetgrpnum{A2.3}
\figsetgrptitle{Ha}
\figsetplot{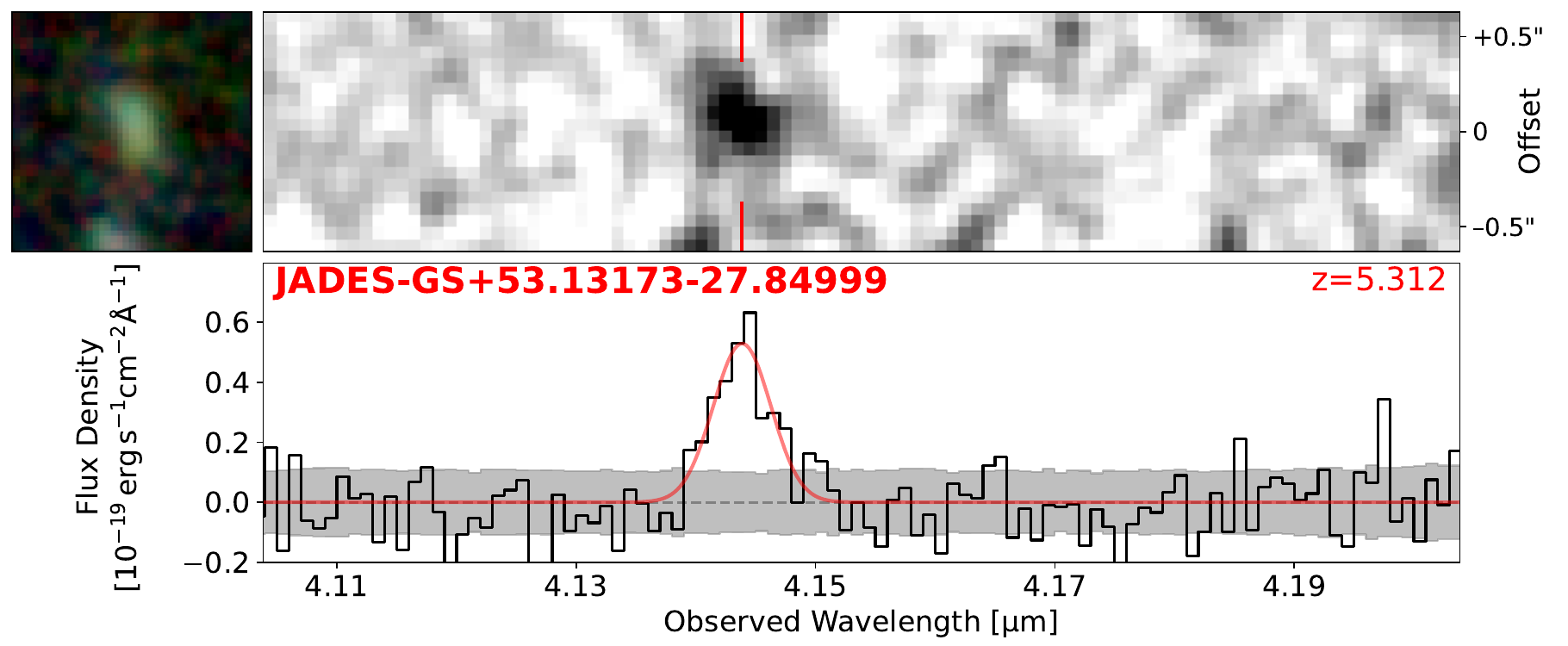}
\figsetgrpnote{NIRCam cutout images alongside the continuum-subtracted 2d and 1d grism spectra of JADES-GS+53.13173-27.84999 at $z = 5.312$, with $\mathrm{H} \alpha$ detected at $8.6\sigma$.}
\figsetgrpend

\figsetgrpstart
\figsetgrpnum{A2.4}
\figsetgrptitle{Ha}
\figsetplot{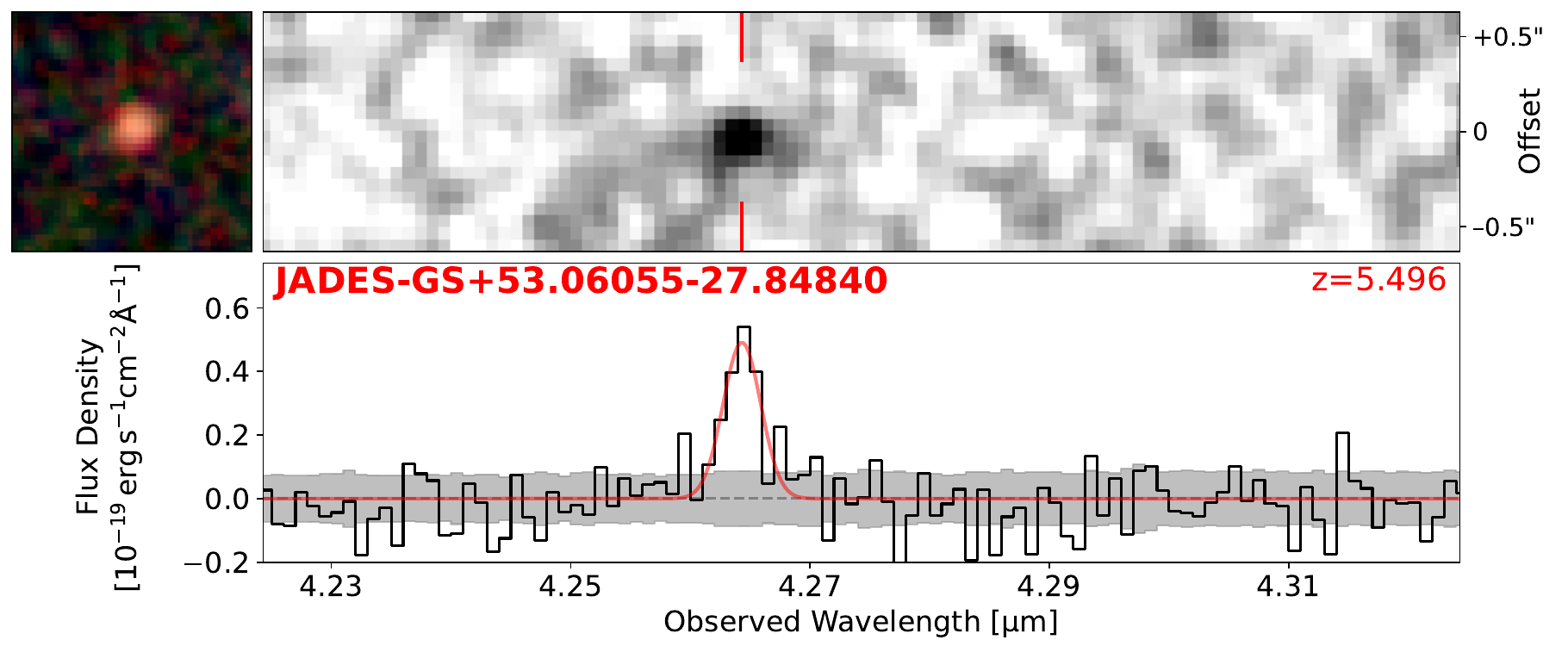}
\figsetgrpnote{NIRCam cutout images alongside the continuum-subtracted 2d and 1d grism spectra of JADES-GS+53.06055-27.84840 at $z = 5.496$, with $\mathrm{H} \alpha$ detected at $8.1\sigma$.}
\figsetgrpend

\figsetgrpstart
\figsetgrpnum{A2.5}
\figsetgrptitle{Ha}
\figsetplot{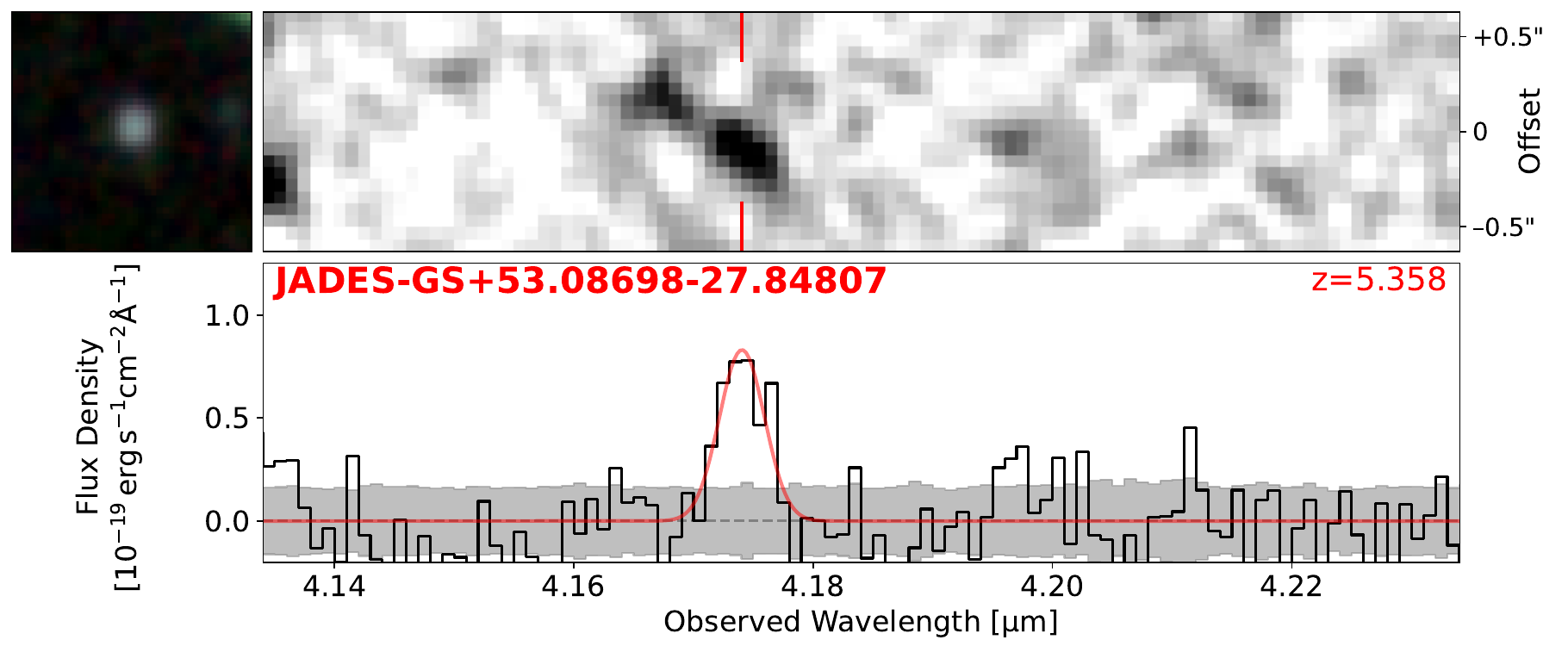}
\figsetgrpnote{NIRCam cutout images alongside the continuum-subtracted 2d and 1d grism spectra of JADES-GS+53.08698-27.84807 at $z = 5.358$, with $\mathrm{H} \alpha$ detected at $7.7\sigma$.}
\figsetgrpend

\figsetgrpstart
\figsetgrpnum{A2.6}
\figsetgrptitle{Ha}
\figsetplot{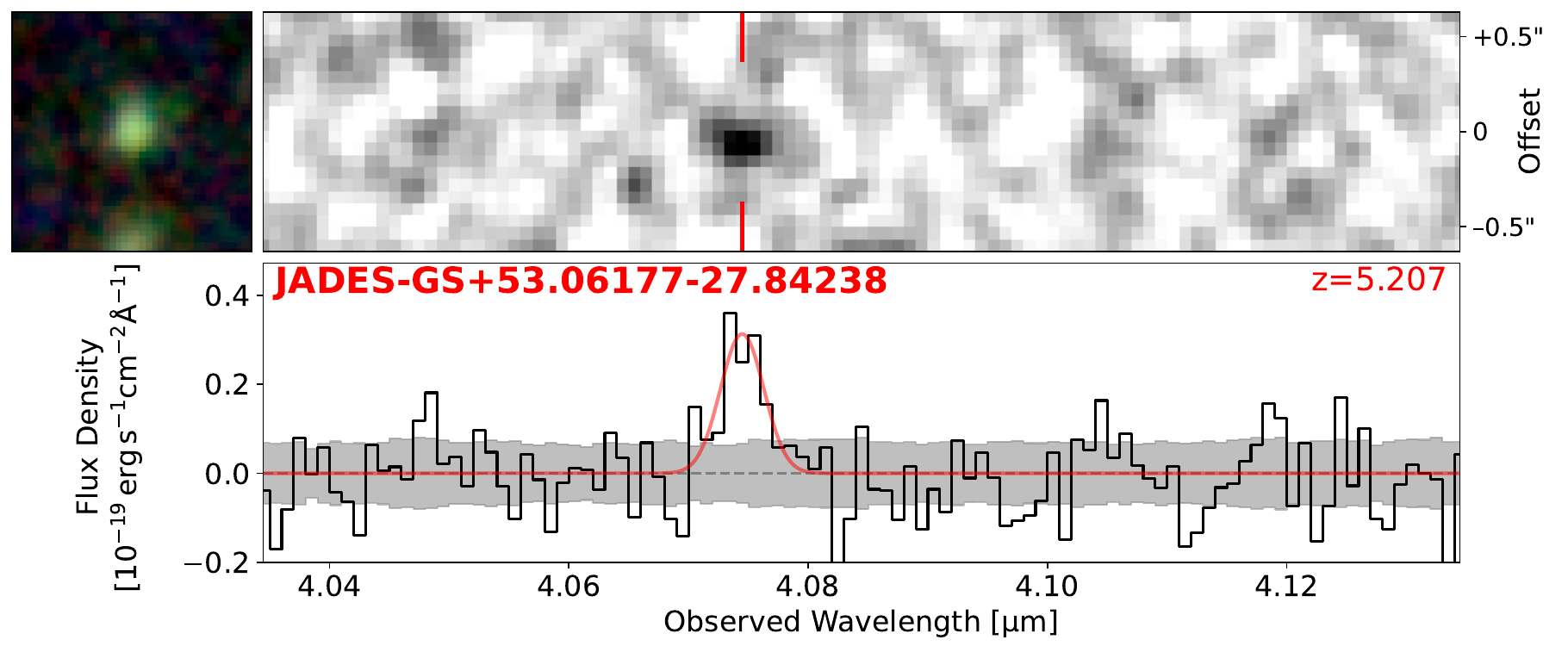}
\figsetgrpnote{NIRCam cutout images alongside the continuum-subtracted 2d and 1d grism spectra of JADES-GS+53.06177-27.84238 at $z = 5.207$, with $\mathrm{H} \alpha$ detected at $6.6\sigma$.}
\figsetgrpend

\figsetgrpstart
\figsetgrpnum{A2.7}
\figsetgrptitle{Ha}
\figsetplot{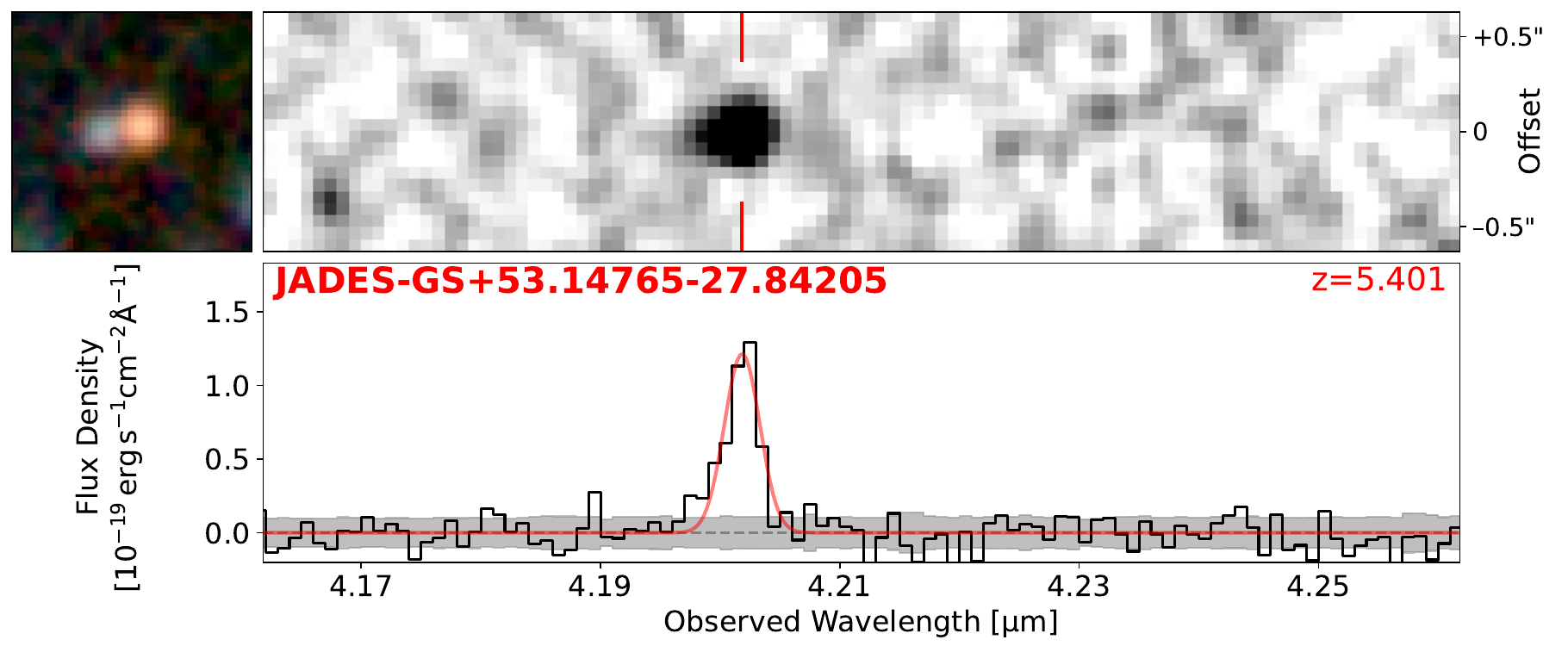}
\figsetgrpnote{NIRCam cutout images alongside the continuum-subtracted 2d and 1d grism spectra of JADES-GS+53.14765-27.84205 at $z = 5.401$, with $\mathrm{H} \alpha$ detected at $4.2\sigma$.}
\figsetgrpend

\figsetgrpstart
\figsetgrpnum{A2.8}
\figsetgrptitle{Ha}
\figsetplot{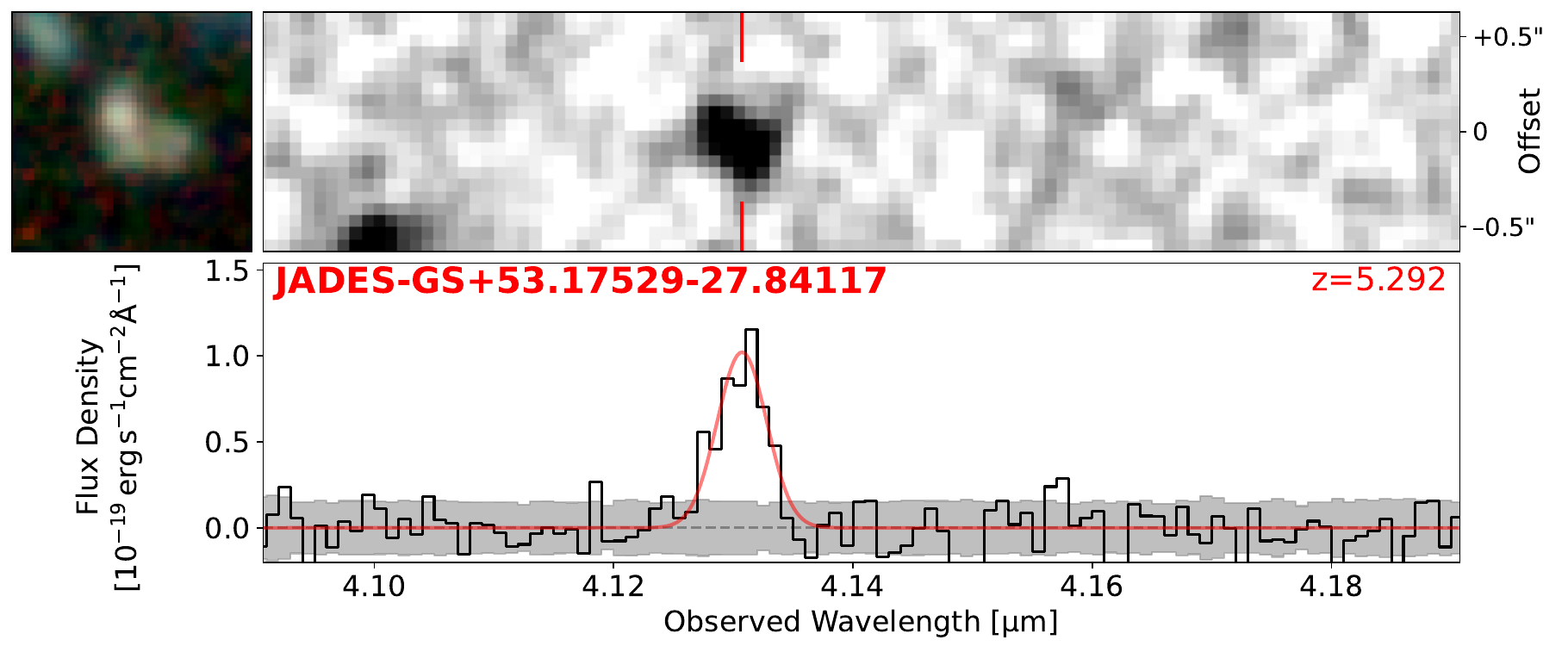}
\figsetgrpnote{NIRCam cutout images alongside the continuum-subtracted 2d and 1d grism spectra of JADES-GS+53.17529-27.84117 at $z = 5.292$, with $\mathrm{H} \alpha$ detected at $10.5\sigma$.}
\figsetgrpend

\figsetgrpstart
\figsetgrpnum{A2.9}
\figsetgrptitle{Ha}
\figsetplot{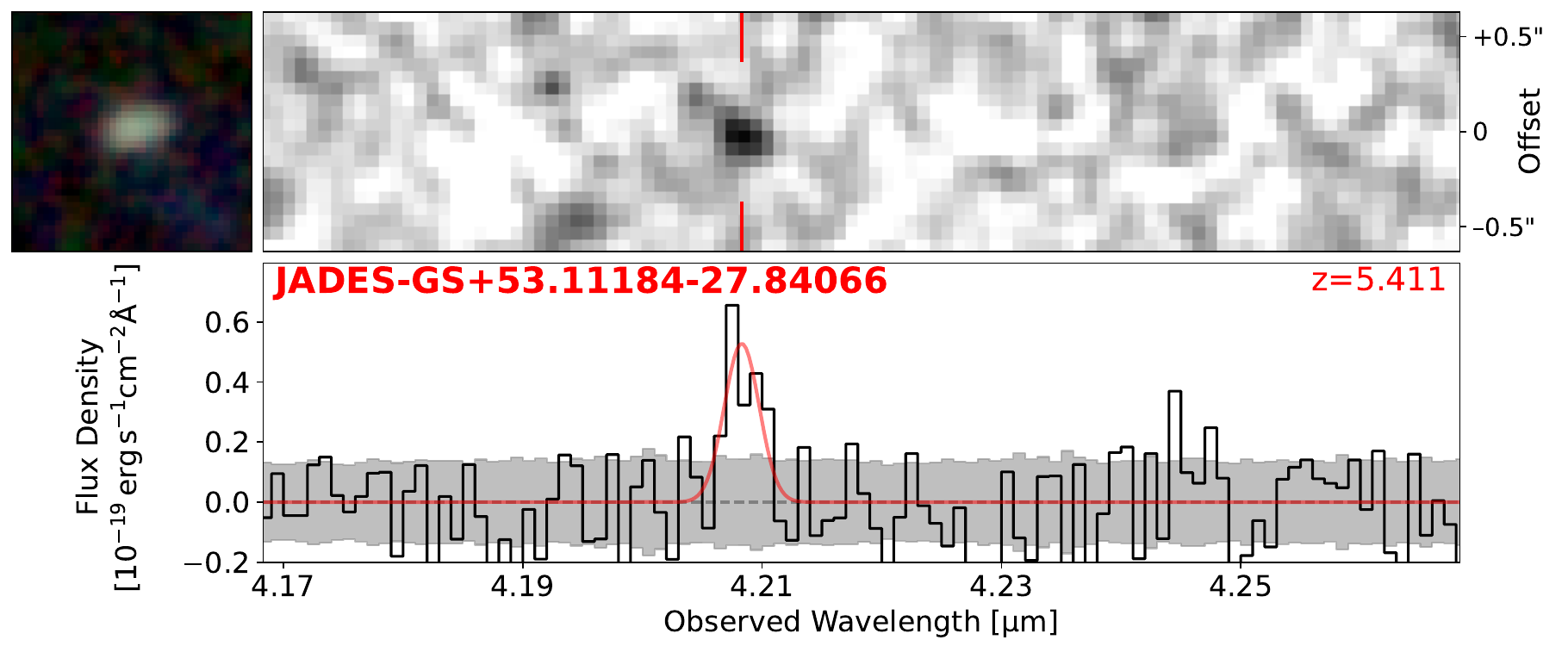}
\figsetgrpnote{NIRCam cutout images alongside the continuum-subtracted 2d and 1d grism spectra of JADES-GS+53.11184-27.84066 at $z = 5.411$, with $\mathrm{H} \alpha$ detected at $4.7\sigma$.}
\figsetgrpend

\figsetgrpstart
\figsetgrpnum{A2.10}
\figsetgrptitle{Ha}
\figsetplot{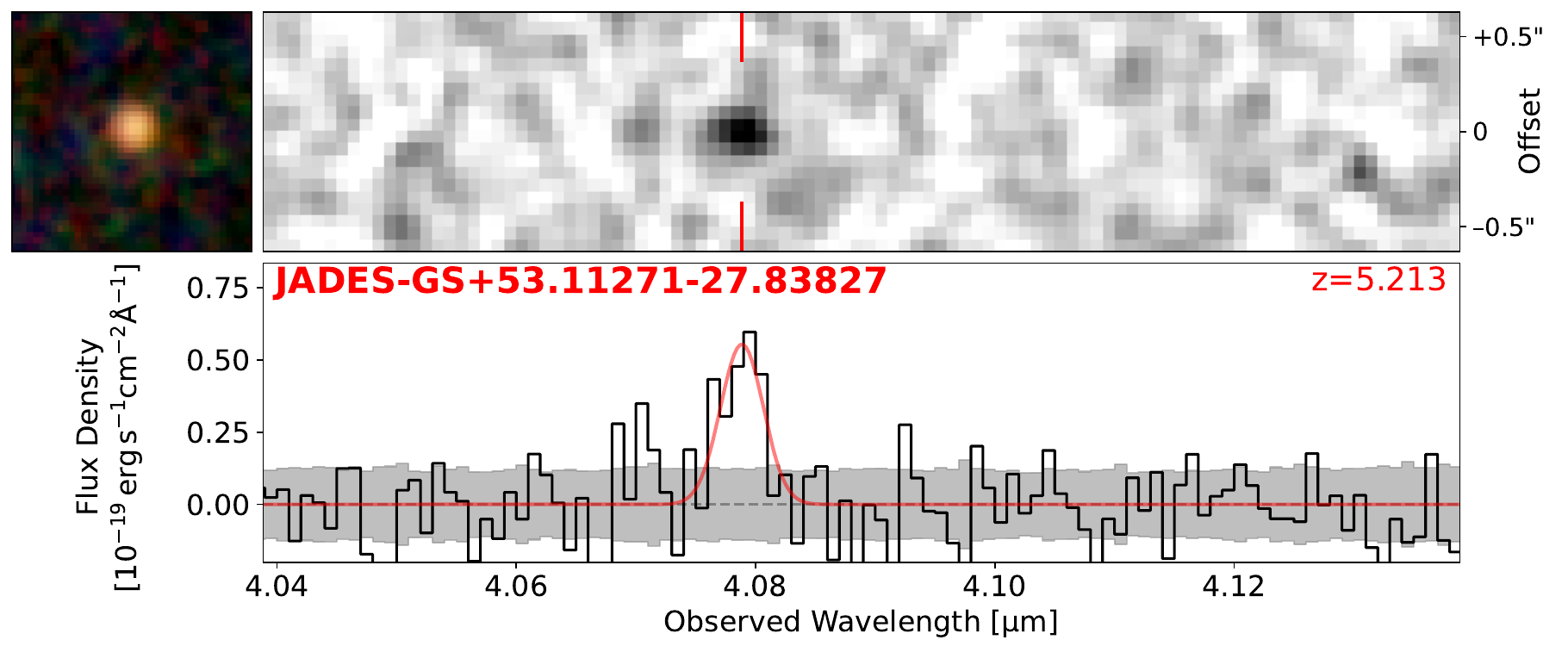}
\figsetgrpnote{NIRCam cutout images alongside the continuum-subtracted 2d and 1d grism spectra of JADES-GS+53.11271-27.83827 at $z = 5.213$, with $\mathrm{H} \alpha$ detected at $6.4\sigma$.}
\figsetgrpend

\figsetgrpstart
\figsetgrpnum{A2.11}
\figsetgrptitle{Ha}
\figsetplot{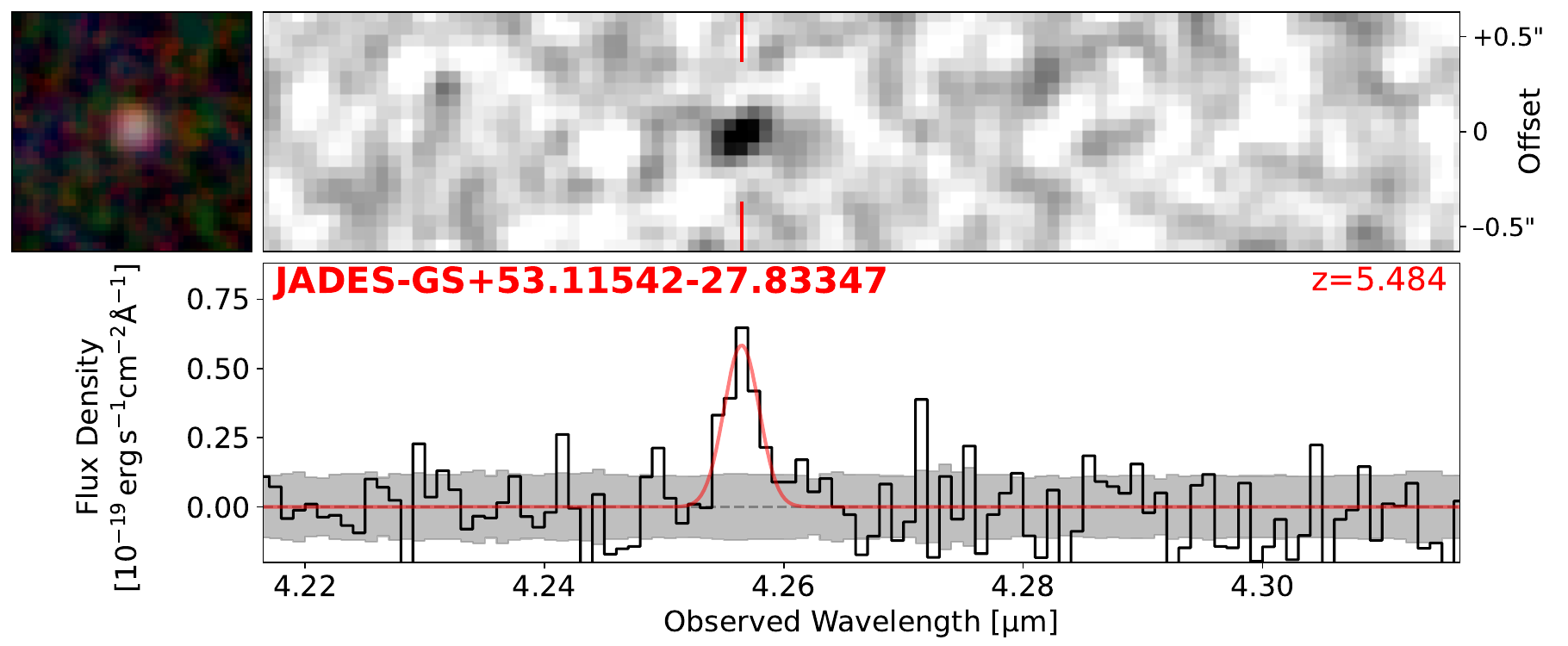}
\figsetgrpnote{NIRCam cutout images alongside the continuum-subtracted 2d and 1d grism spectra of JADES-GS+53.11542-27.83347 at $z = 5.484$, with $\mathrm{H} \alpha$ detected at $6.6\sigma$.}
\figsetgrpend

\figsetgrpstart
\figsetgrpnum{A2.12}
\figsetgrptitle{Ha}
\figsetplot{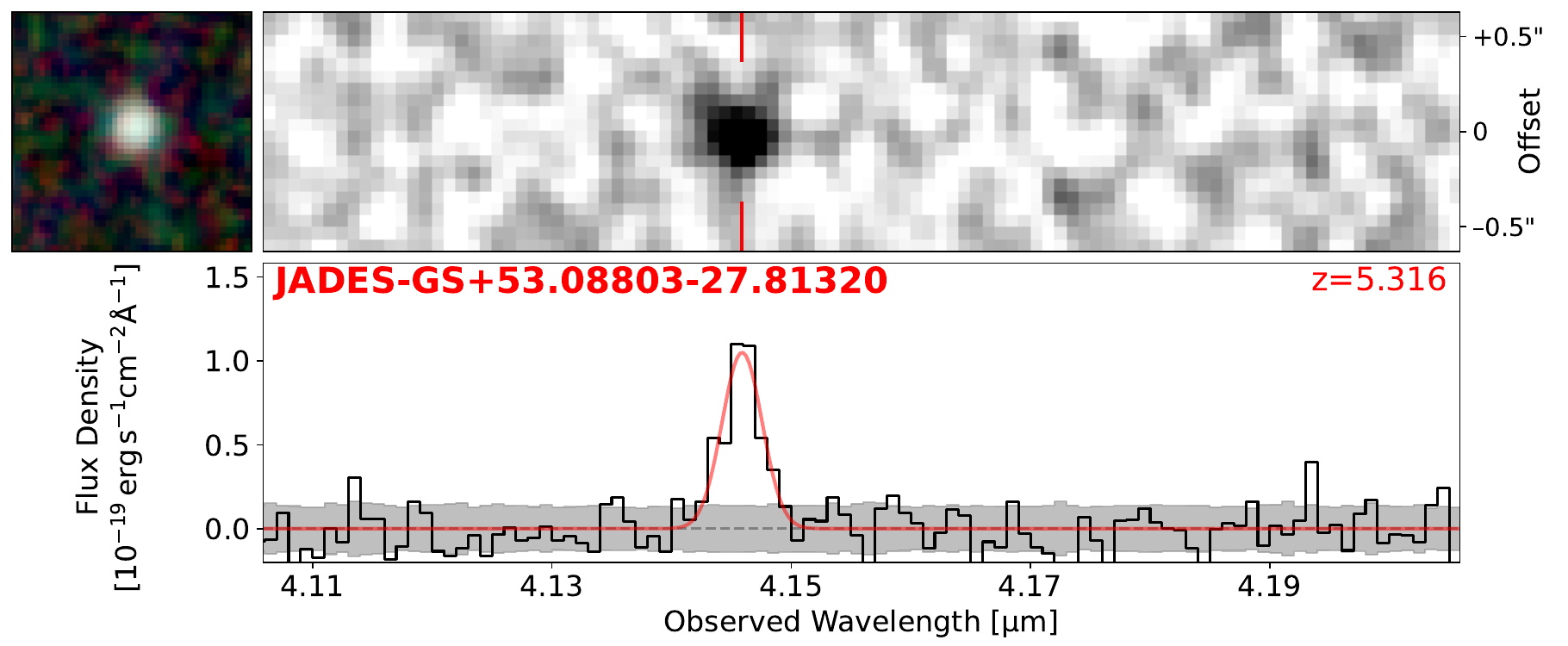}
\figsetgrpnote{NIRCam cutout images alongside the continuum-subtracted 2d and 1d grism spectra of JADES-GS+53.08803-27.81320 at $z = 5.316$, with $\mathrm{H} \alpha$ detected at $10.6\sigma$.}
\figsetgrpend

\figsetgrpstart
\figsetgrpnum{A2.13}
\figsetgrptitle{Ha}
\figsetplot{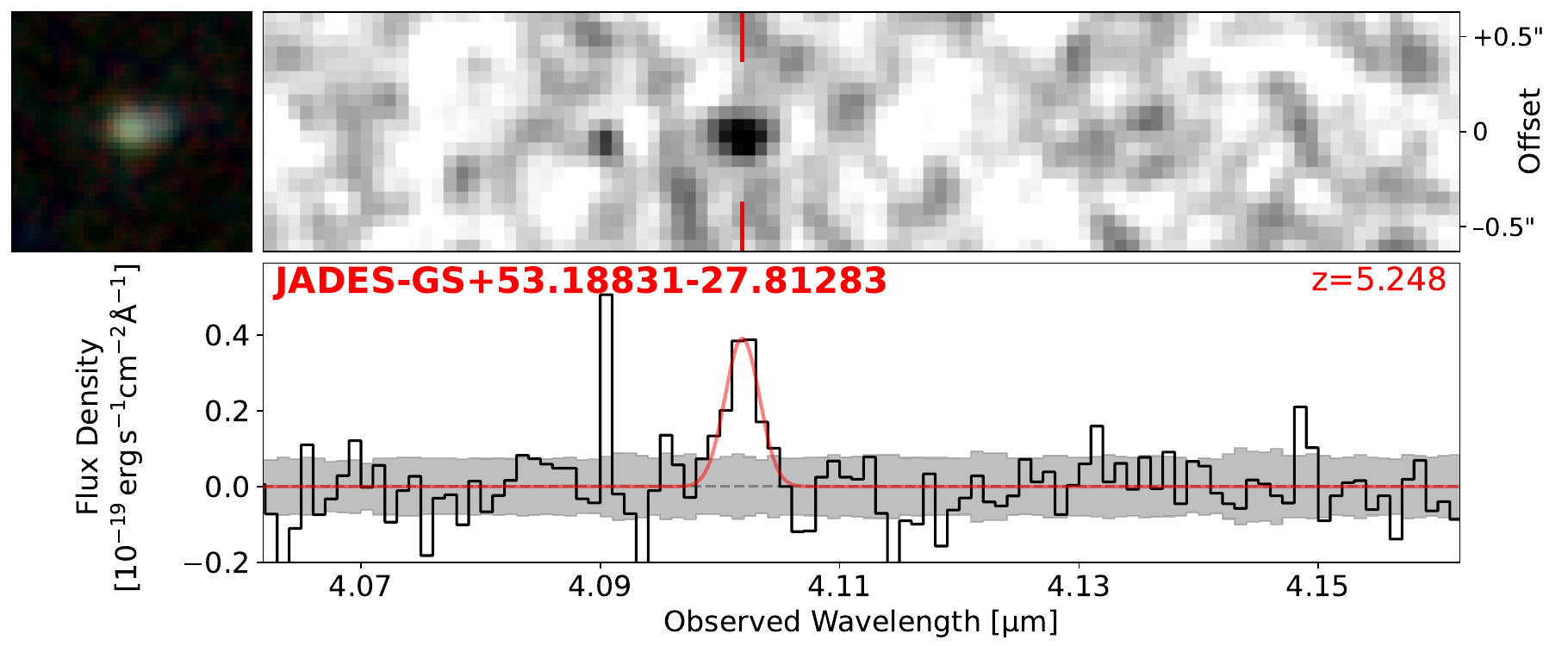}
\figsetgrpnote{NIRCam cutout images alongside the continuum-subtracted 2d and 1d grism spectra of JADES-GS+53.18831-27.81283 at $z = 5.248$, with $\mathrm{H} \alpha$ detected at $6.7\sigma$.}
\figsetgrpend

\figsetgrpstart
\figsetgrpnum{A2.14}
\figsetgrptitle{Ha}
\figsetplot{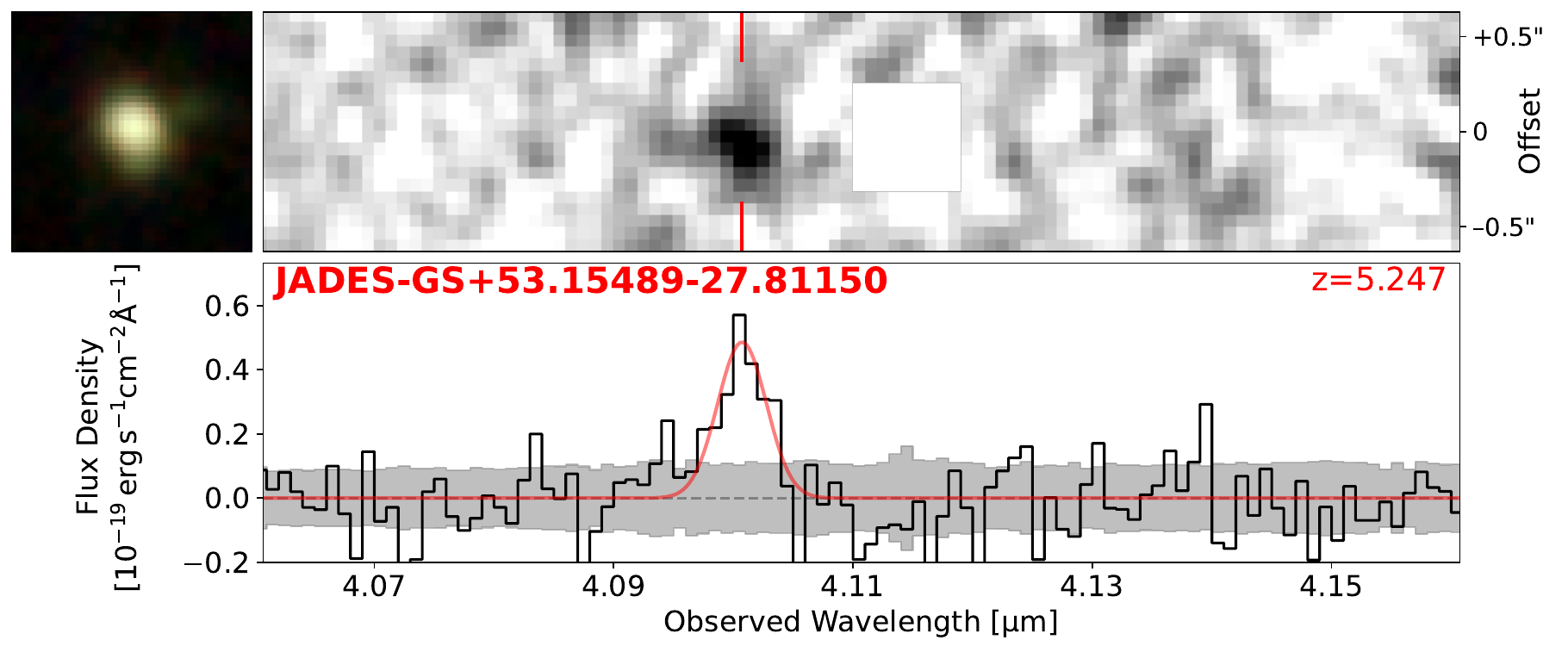}
\figsetgrpnote{NIRCam cutout images alongside the continuum-subtracted 2d and 1d grism spectra of JADES-GS+53.15489-27.81150 at $z = 5.247$, with $\mathrm{H} \alpha$ detected at $7.0\sigma$.}
\figsetgrpend

\figsetgrpstart
\figsetgrpnum{A2.15}
\figsetgrptitle{Ha}
\figsetplot{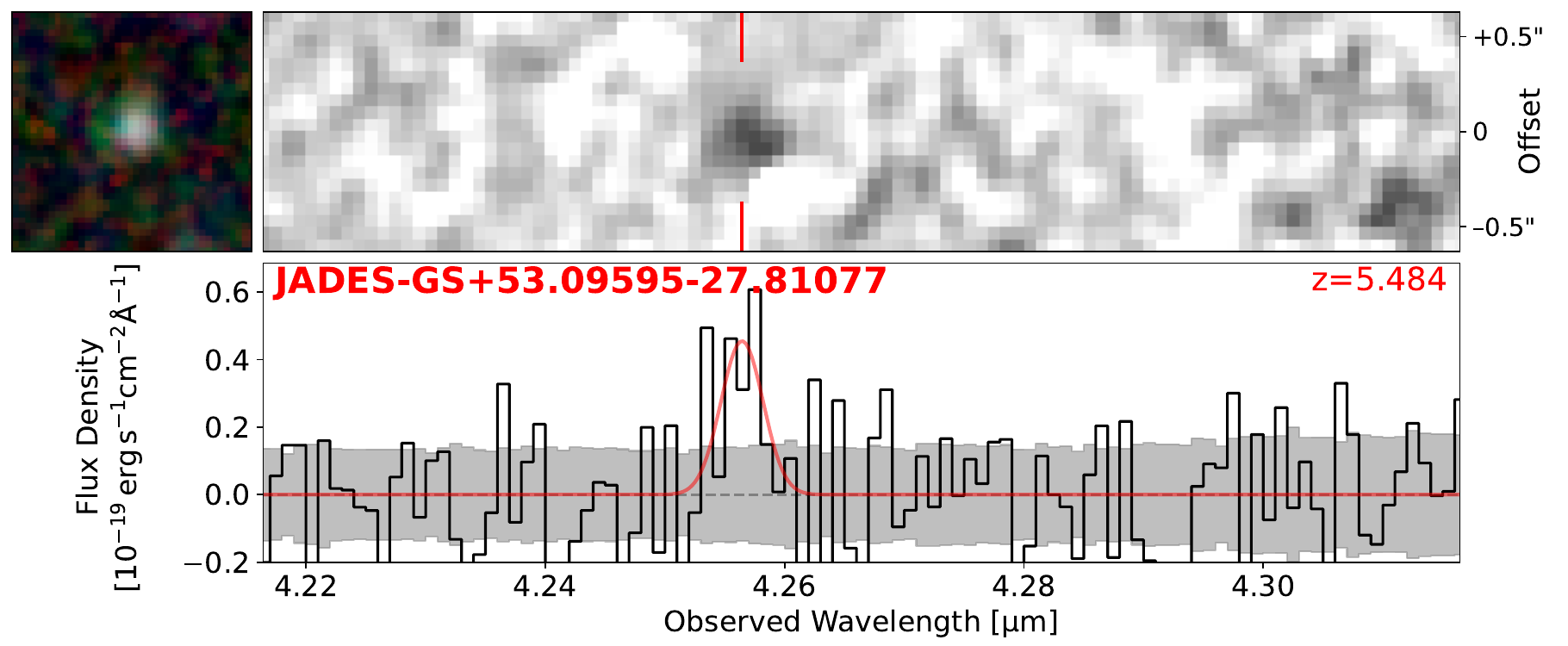}
\figsetgrpnote{NIRCam cutout images alongside the continuum-subtracted 2d and 1d grism spectra of JADES-GS+53.09595-27.81077 at $z = 5.484$, with $\mathrm{H} \alpha$ detected at $4.6\sigma$.}
\figsetgrpend

\figsetgrpstart
\figsetgrpnum{A2.16}
\figsetgrptitle{Ha}
\figsetplot{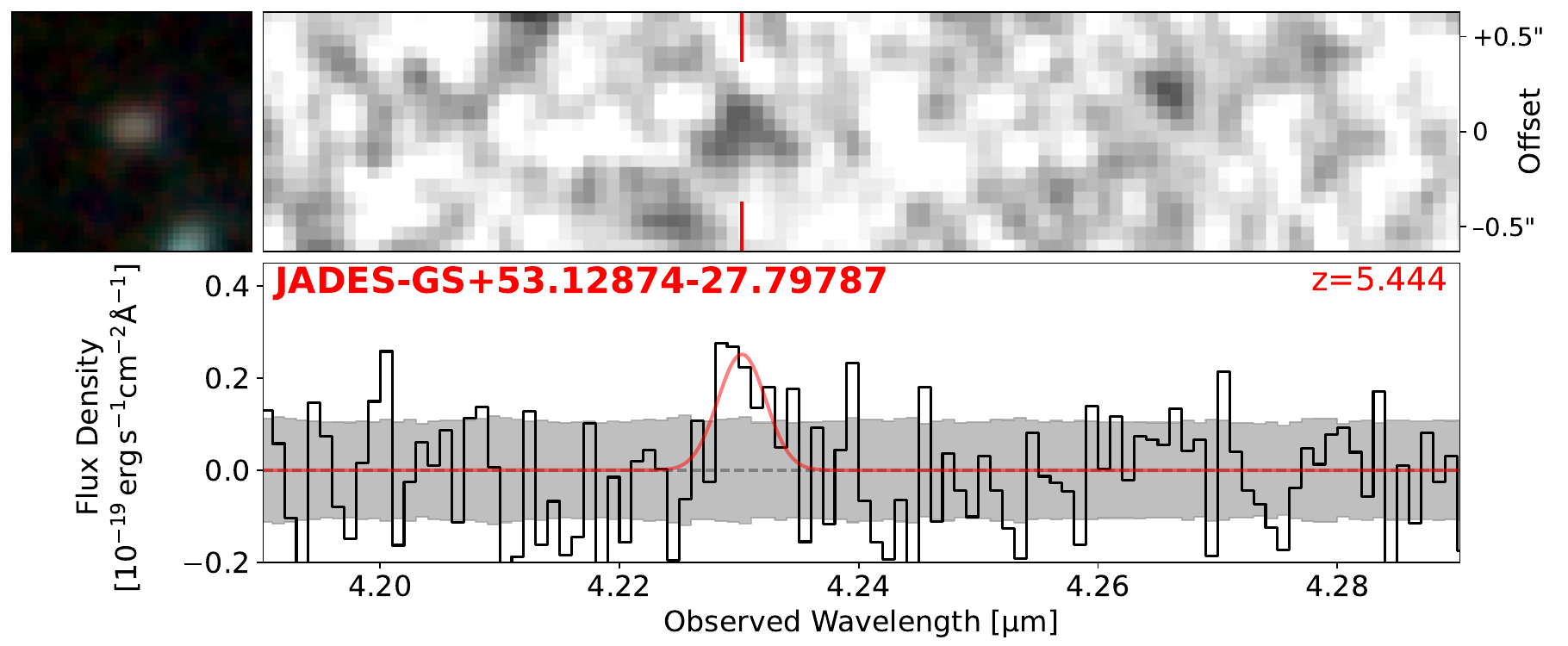}
\figsetgrpnote{NIRCam cutout images alongside the continuum-subtracted 2d and 1d grism spectra of JADES-GS+53.12874-27.79787 at $z = 5.444$, with $\mathrm{H} \alpha$ detected at $3.6\sigma$.}
\figsetgrpend

\figsetgrpstart
\figsetgrpnum{A2.17}
\figsetgrptitle{Ha}
\figsetplot{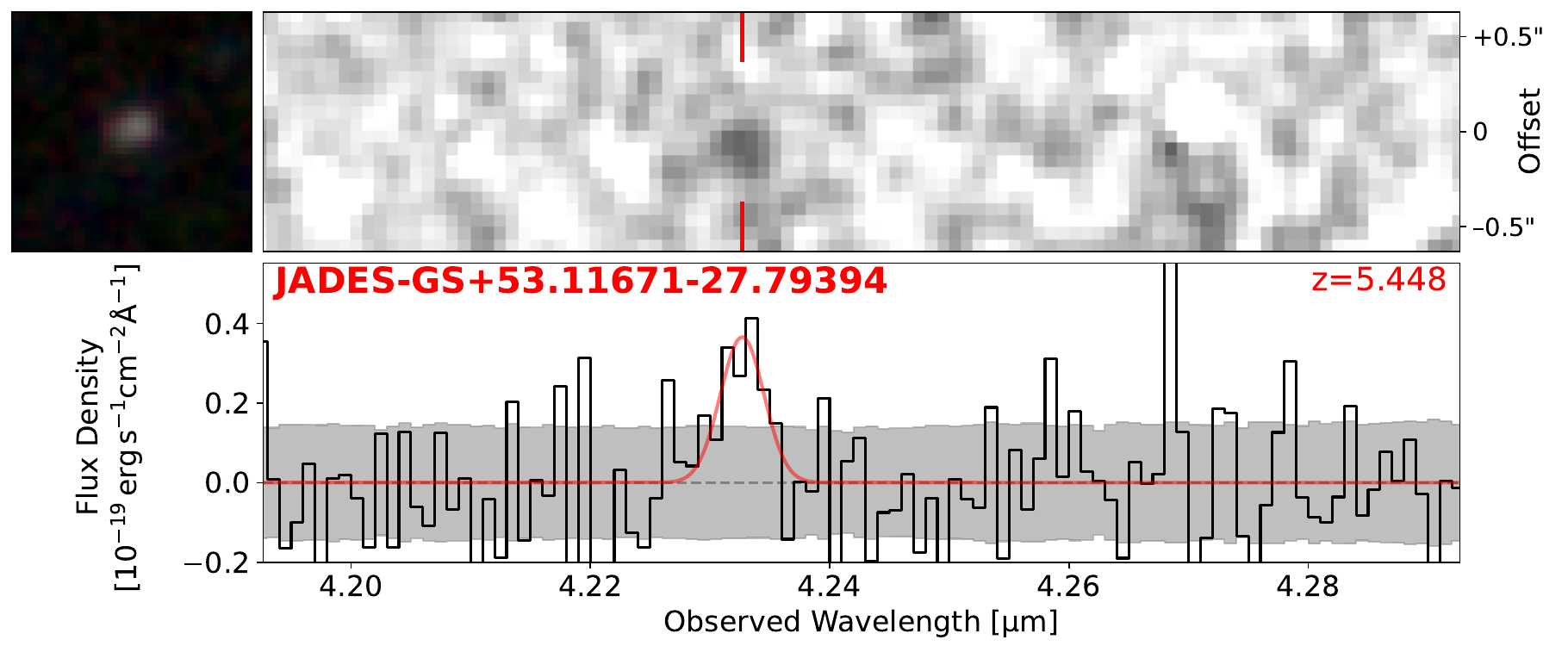}
\figsetgrpnote{NIRCam cutout images alongside the continuum-subtracted 2d and 1d grism spectra of JADES-GS+53.11671-27.79394 at $z = 5.448$, with $\mathrm{H} \alpha$ detected at $3.8\sigma$.}
\figsetgrpend

\figsetgrpstart
\figsetgrpnum{A2.18}
\figsetgrptitle{Ha}
\figsetplot{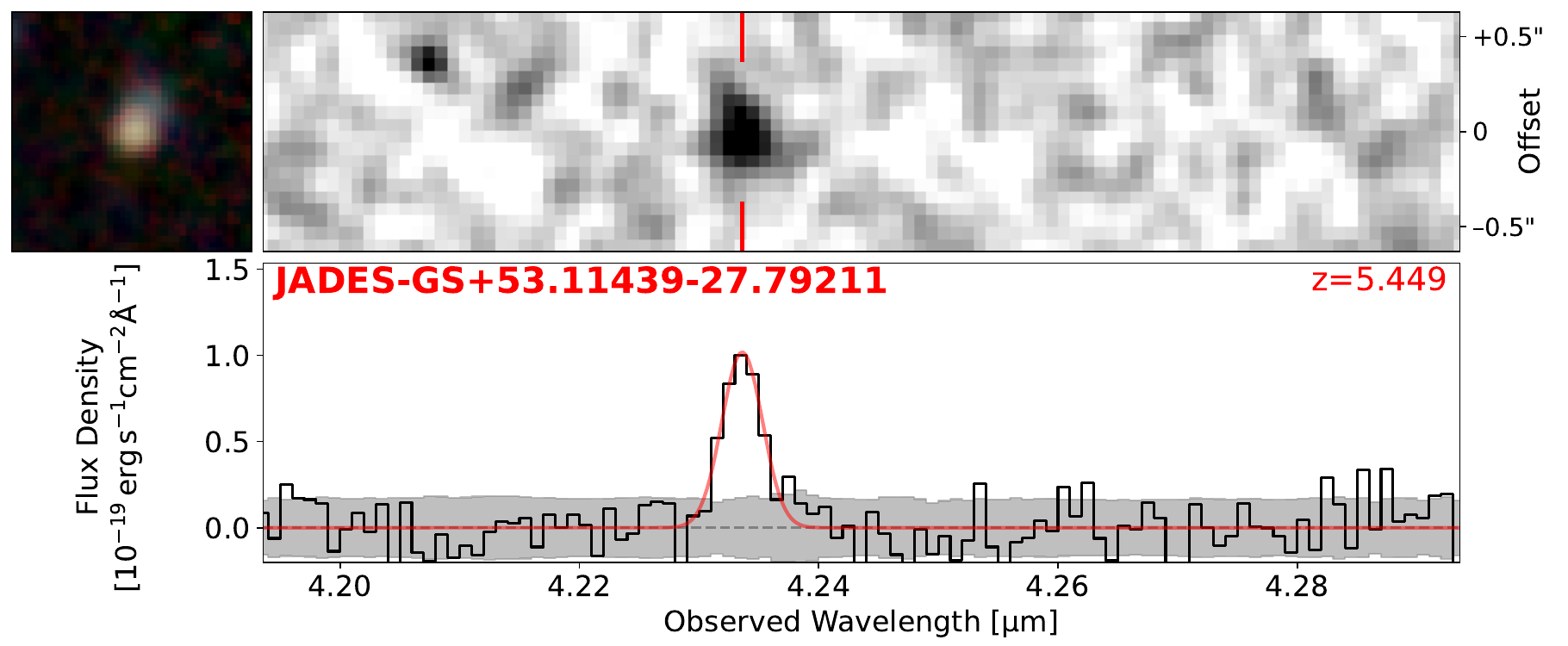}
\figsetgrpnote{NIRCam cutout images alongside the continuum-subtracted 2d and 1d grism spectra of JADES-GS+53.11439-27.79211 at $z = 5.449$, with $\mathrm{H} \alpha$ detected at $8.4\sigma$.}
\figsetgrpend

\figsetgrpstart
\figsetgrpnum{A2.19}
\figsetgrptitle{Ha}
\figsetplot{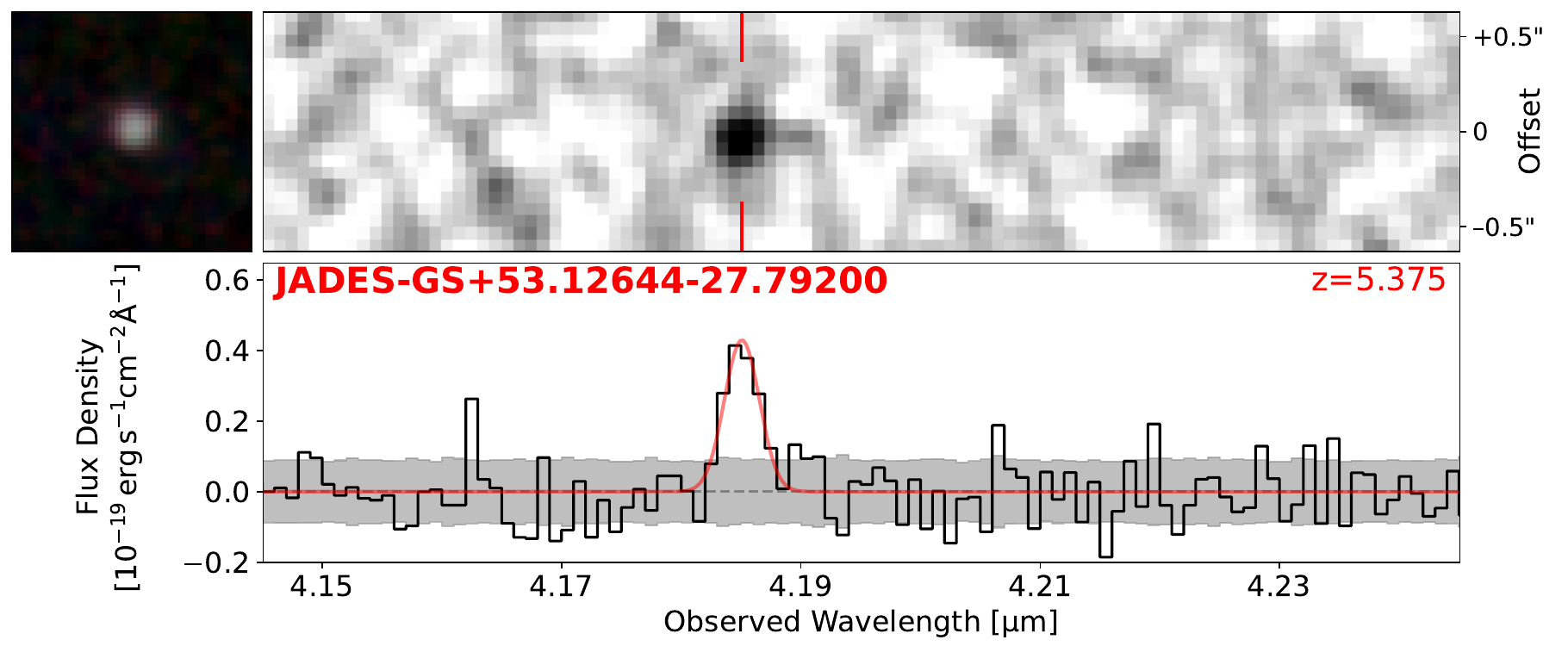}
\figsetgrpnote{NIRCam cutout images alongside the continuum-subtracted 2d and 1d grism spectra of JADES-GS+53.12644-27.79200 at $z = 5.375$, with $\mathrm{H} \alpha$ detected at $6.2\sigma$.}
\figsetgrpend

\figsetgrpstart
\figsetgrpnum{A2.20}
\figsetgrptitle{Ha}
\figsetplot{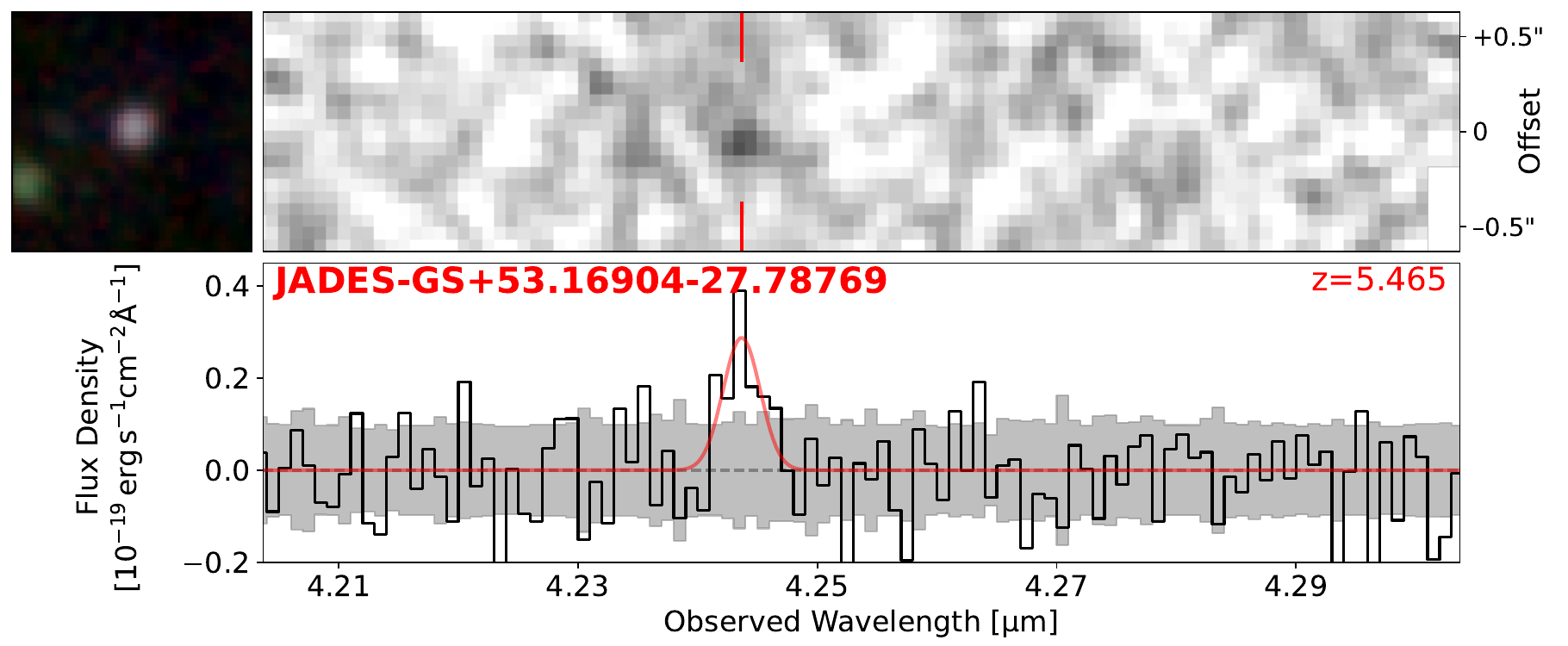}
\figsetgrpnote{NIRCam cutout images alongside the continuum-subtracted 2d and 1d grism spectra of JADES-GS+53.16904-27.78769 at $z = 5.465$, with $\mathrm{H} \alpha$ detected at $3.6\sigma$.}
\figsetgrpend

\figsetgrpstart
\figsetgrpnum{A2.21}
\figsetgrptitle{Ha}
\figsetplot{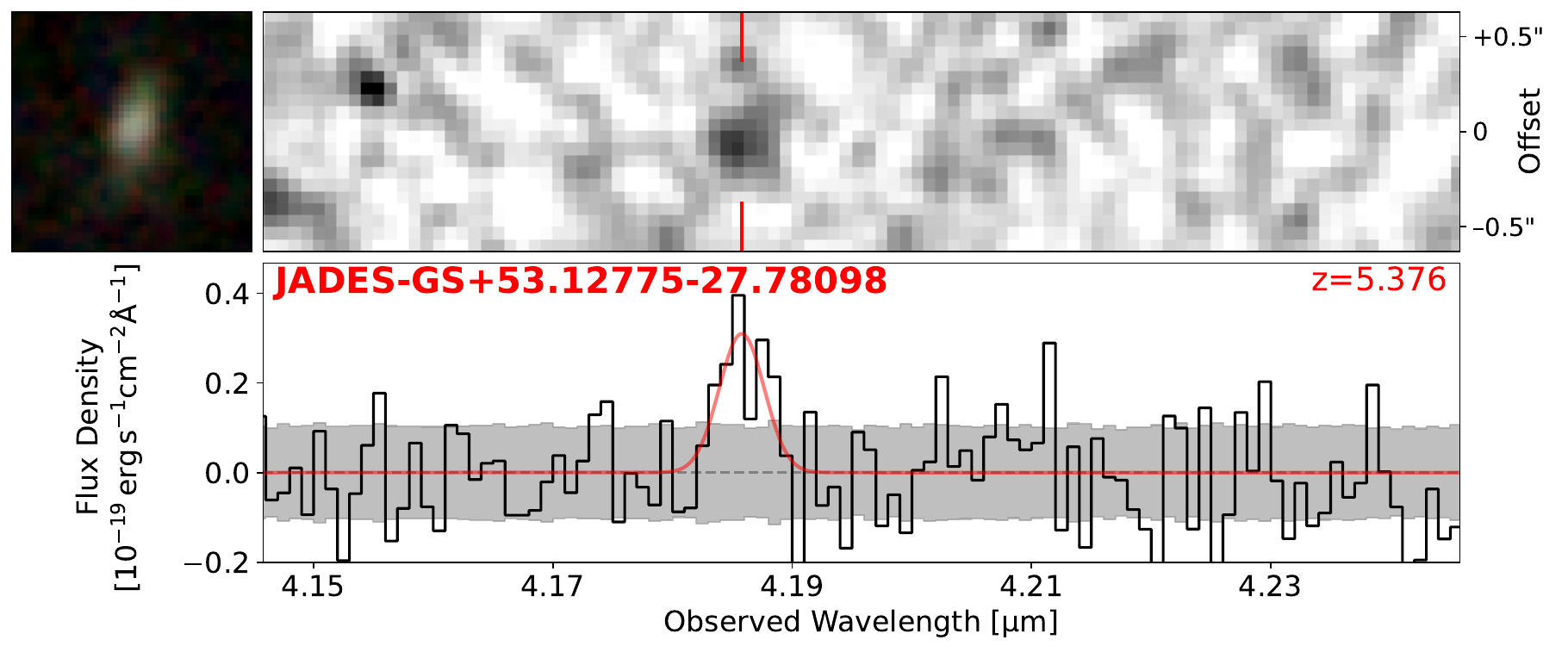}
\figsetgrpnote{NIRCam cutout images alongside the continuum-subtracted 2d and 1d grism spectra of JADES-GS+53.12775-27.78098 at $z = 5.376$, with $\mathrm{H} \alpha$ detected at $4.3\sigma$.}
\figsetgrpend

\figsetgrpstart
\figsetgrpnum{A2.22}
\figsetgrptitle{Ha}
\figsetplot{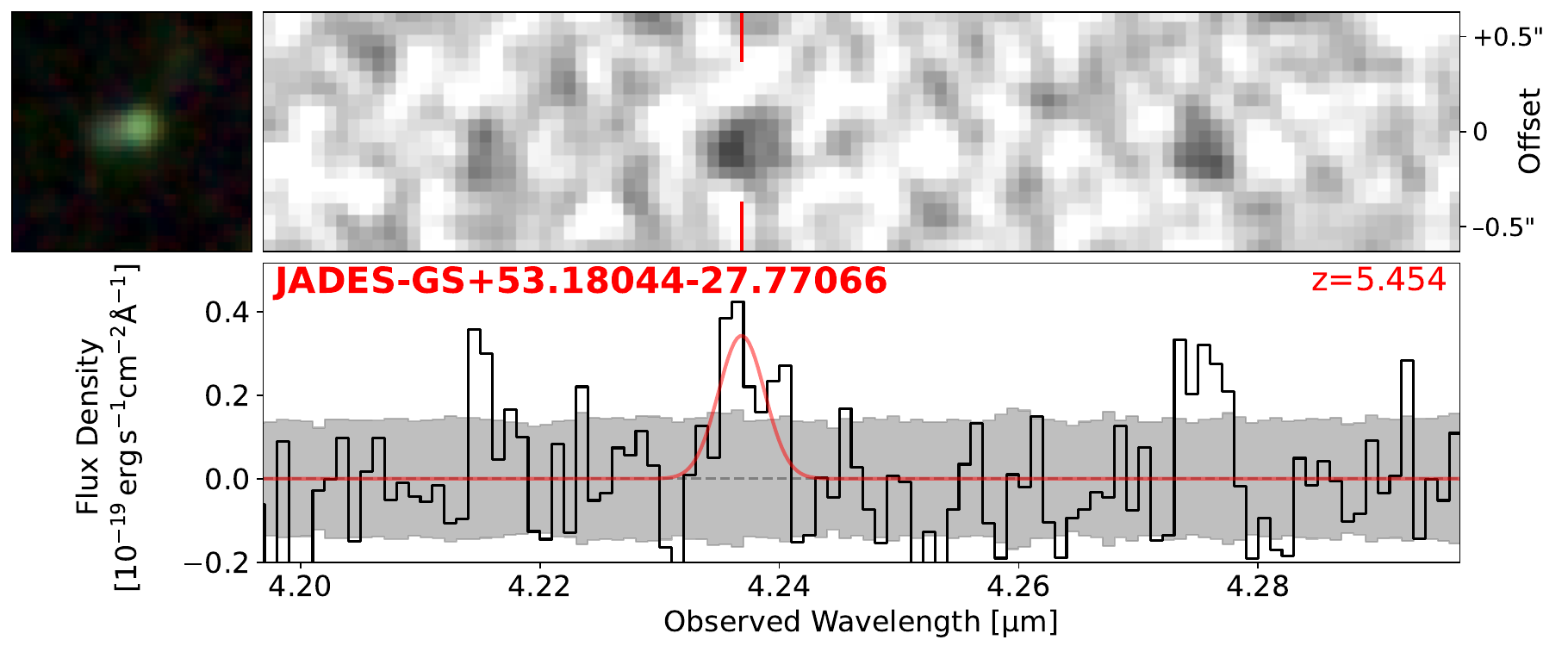}
\figsetgrpnote{NIRCam cutout images alongside the continuum-subtracted 2d and 1d grism spectra of JADES-GS+53.18044-27.77066 at $z = 5.454$, with $\mathrm{H} \alpha$ detected at $3.5\sigma$.}
\figsetgrpend

\figsetgrpstart
\figsetgrpnum{A2.23}
\figsetgrptitle{Ha}
\figsetplot{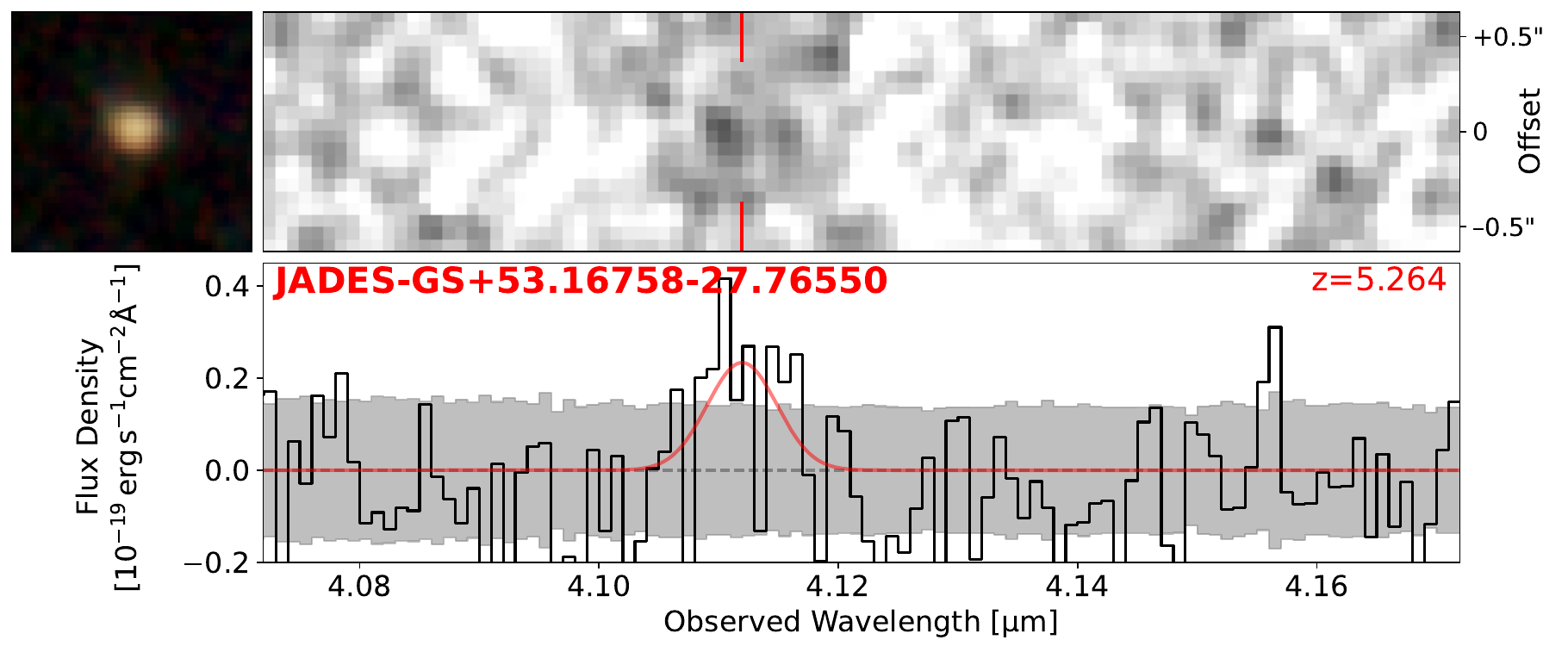}
\figsetgrpnote{NIRCam cutout images alongside the continuum-subtracted 2d and 1d grism spectra of JADES-GS+53.16758-27.76550 at $z = 5.264$, with $\mathrm{H} \alpha$ detected at $3.1\sigma$.}
\figsetgrpend

\figsetgrpstart
\figsetgrpnum{A2.24}
\figsetgrptitle{Ha}
\figsetplot{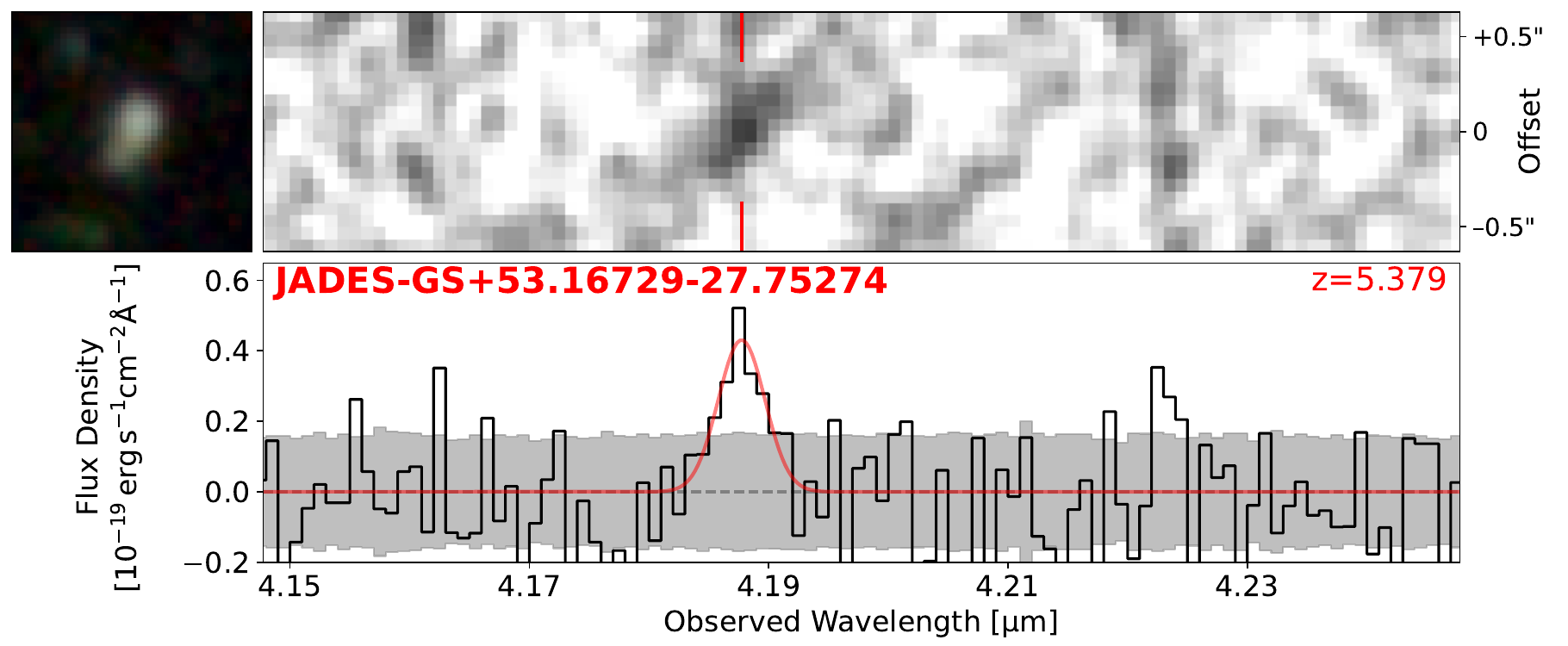}
\figsetgrpnote{NIRCam cutout images alongside the continuum-subtracted 2d and 1d grism spectra of JADES-GS+53.16729-27.75274 at $z = 5.379$, with $\mathrm{H} \alpha$ detected at $4.0\sigma$.}
\figsetgrpend

\figsetgrpstart
\figsetgrpnum{A2.25}
\figsetgrptitle{Ha}
\figsetplot{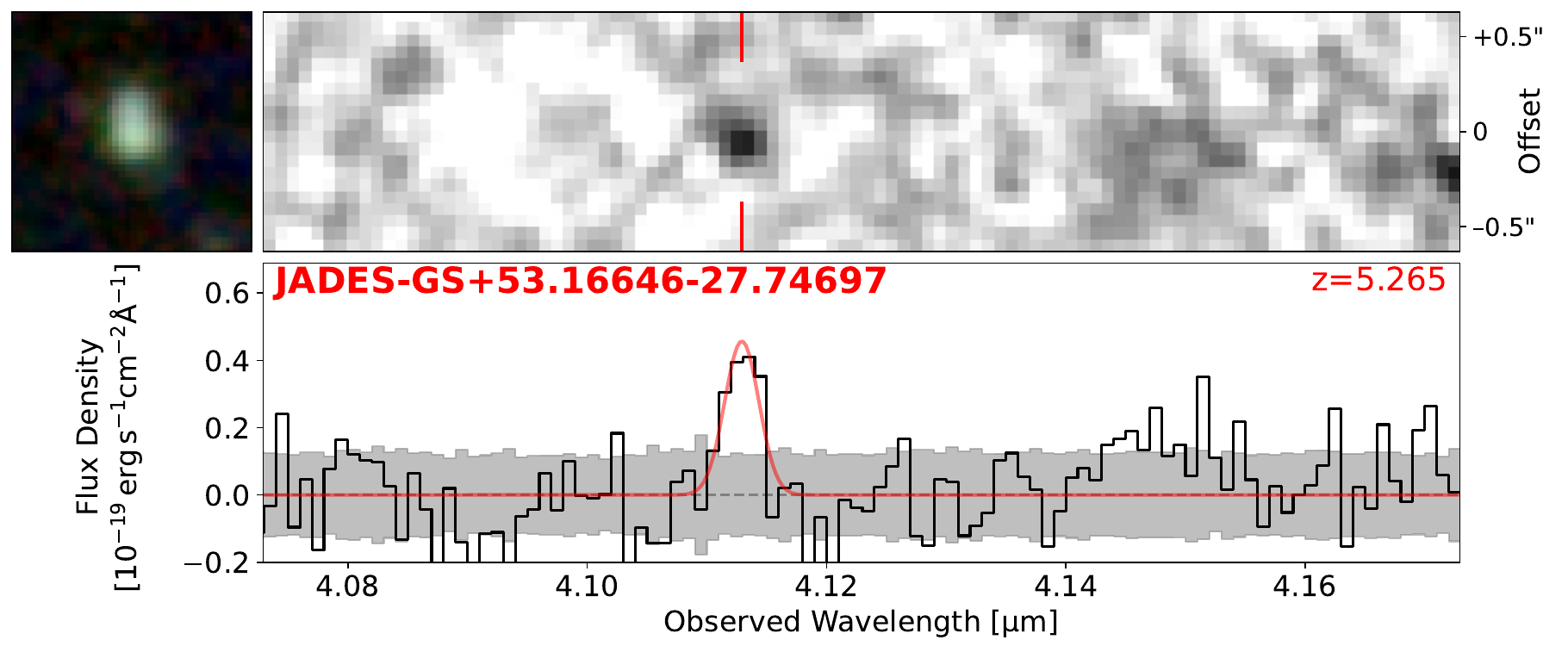}
\figsetgrpnote{NIRCam cutout images alongside the continuum-subtracted 2d and 1d grism spectra of JADES-GS+53.16646-27.74697 at $z = 5.265$, with $\mathrm{H} \alpha$ detected at $4.8\sigma$.}
\figsetgrpend

\figsetgrpstart
\figsetgrpnum{A2.26}
\figsetgrptitle{Ha}
\figsetplot{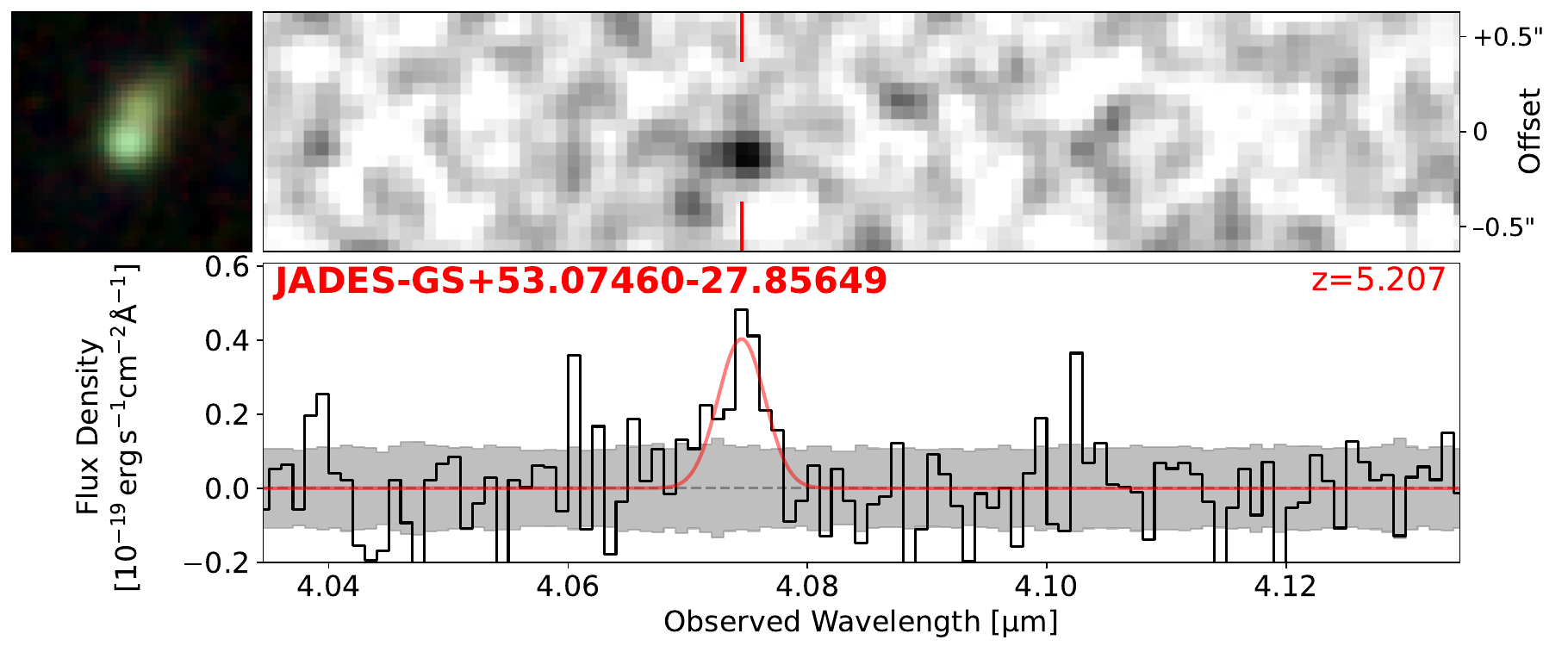}
\figsetgrpnote{NIRCam cutout images alongside the continuum-subtracted 2d and 1d grism spectra of JADES-GS+53.07460-27.85649 at $z = 5.207$, with $\mathrm{H} \alpha$ detected at $5.3\sigma$.}
\figsetgrpend

\figsetgrpstart
\figsetgrpnum{A2.27}
\figsetgrptitle{Ha}
\figsetplot{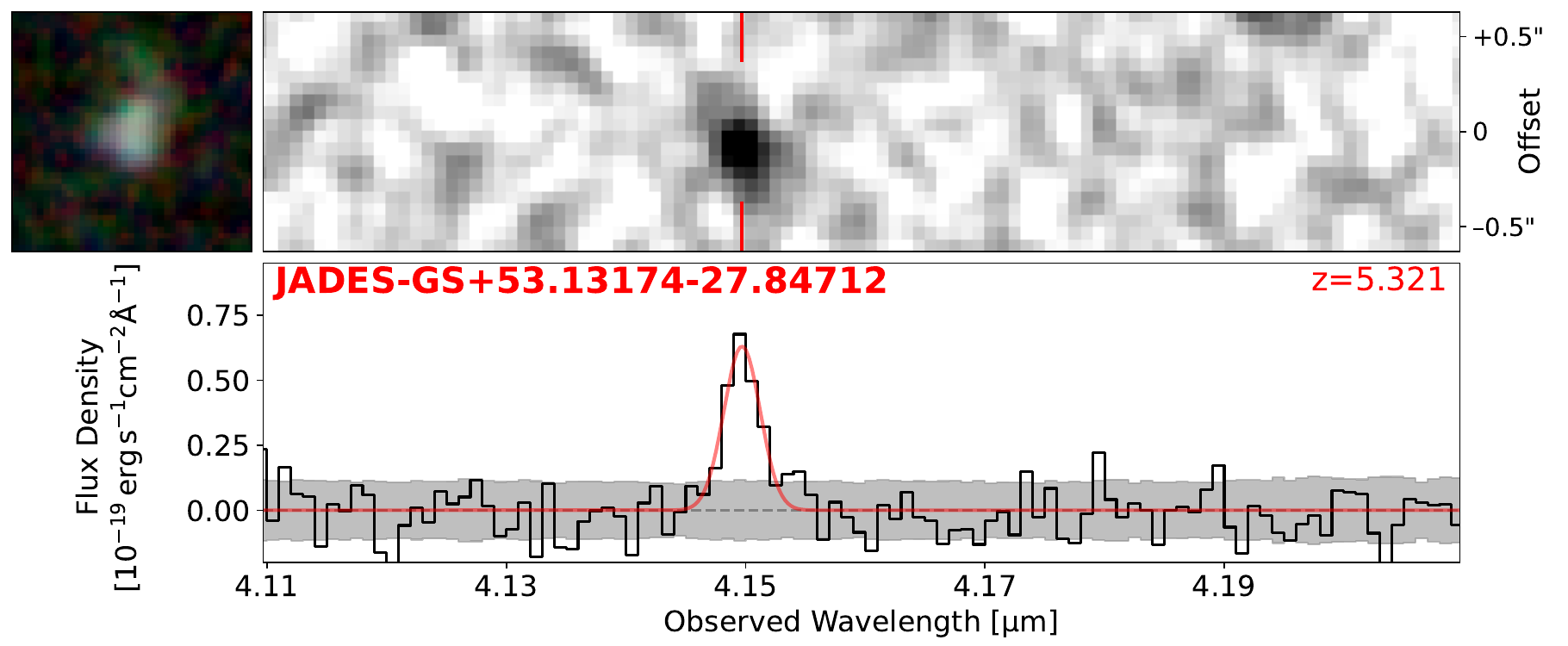}
\figsetgrpnote{NIRCam cutout images alongside the continuum-subtracted 2d and 1d grism spectra of JADES-GS+53.13174-27.84712 at $z = 5.321$, with $\mathrm{H} \alpha$ detected at $7.6\sigma$.}
\figsetgrpend

\figsetgrpstart
\figsetgrpnum{A2.28}
\figsetgrptitle{Ha}
\figsetplot{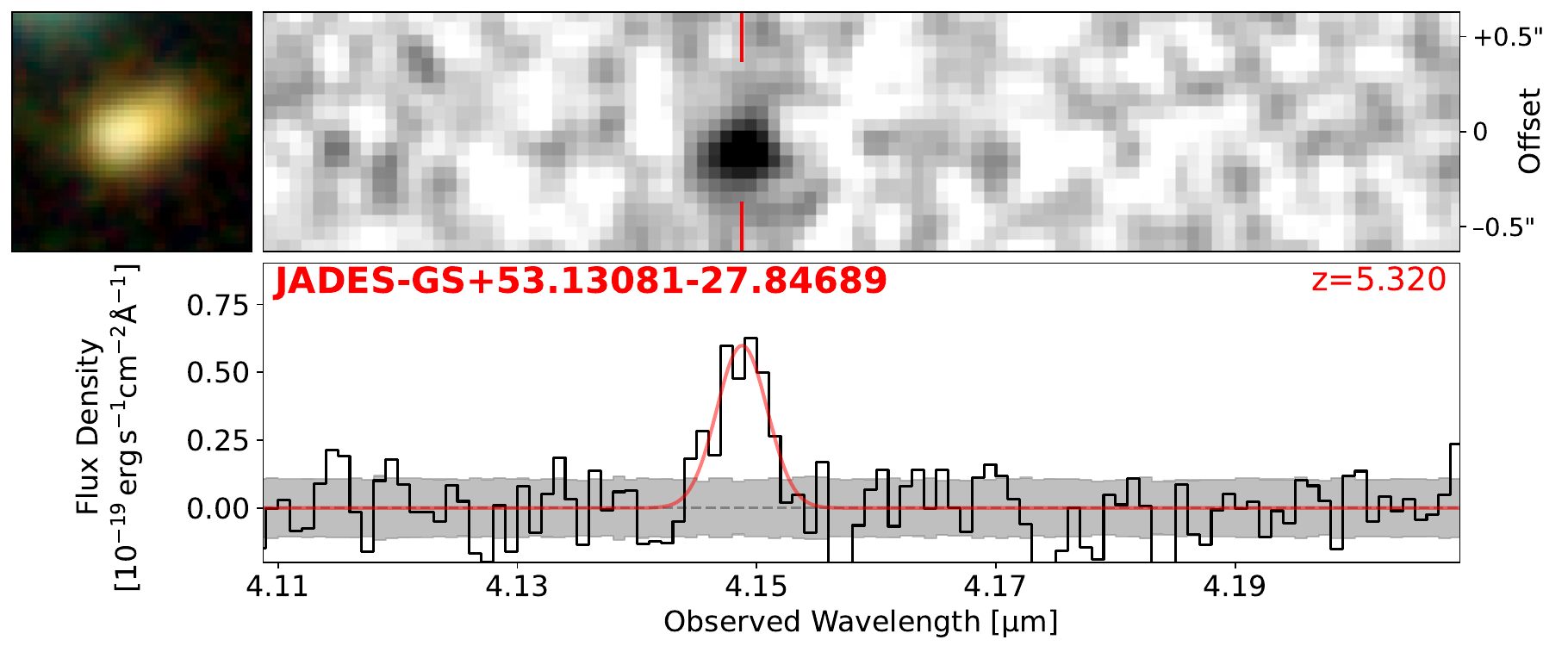}
\figsetgrpnote{NIRCam cutout images alongside the continuum-subtracted 2d and 1d grism spectra of JADES-GS+53.13081-27.84689 at $z = 5.320$, with $\mathrm{H} \alpha$ detected at $8.9\sigma$.}
\figsetgrpend

\figsetgrpstart
\figsetgrpnum{A2.42}
\figsetgrptitle{Ha}
\figsetplot{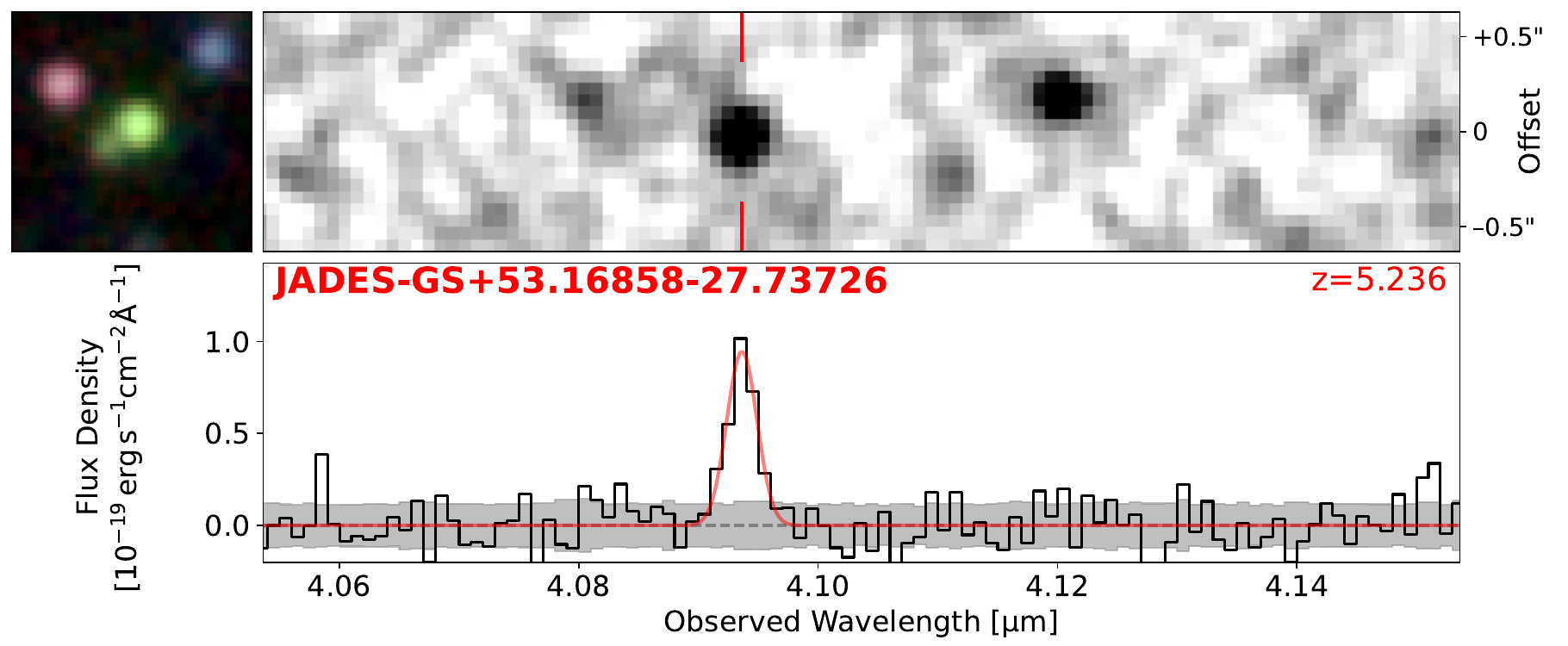}
\figsetgrpnote{NIRCam cutout images alongside the continuum-subtracted 2d and 1d grism spectra of JADES-GS+53.16858-27.73726 at $z = 5.236$, with $\mathrm{H} \alpha$ detected at $9.2\sigma$.}
\figsetgrpend

\figsetend

\section{Assumed Physical Model for Stellar Population Modeling}
\label{AppendixTwo}

\setcounter{table}{0}
\renewcommand{\thetable}{\thesection\arabic{table}}

Table~\ref{tab:prospector_model} gives a summary of the parameter and priors assumed in the \texttt{Prospector} model that was used for the stellar population modeling described in Section~\ref{SectionThreeTwo}.

\begin{table*}
	\caption{A summary of the parameters and priors used in our \texttt{Prospector} model (see Section~\ref{SectionThreeThree}).}
	\label{tab:prospector_model}
	\hspace*{-6mm}
        \makebox[\textwidth]{
	\begin{threeparttable}
	\begin{tabular}{lll} 
		\hline
		\hline
		Parameter & Description & Prior \\
		\hline
            $z$ & Redshift & Fixed\tnote{a}  \\
            $\mathrm{log}_{10}( Z_{\ast}/Z_{\odot} )$ & Stellar metallicity & Uniform\tnote{b} \\
            $\mathrm{log}_{10}( M_{\ast}/M_{\odot} )$ & Total stellar mass formed & Uniform\tnote{c} \\
            Non-Parametric SFH (Secondary) & Ratio of the SFRs in adjacent time bins with & Student’s t-distribution\tnote{d} \\
            & five ($N_{\mathrm{SFR}} - 1$) free parameters & \\
            Non-Parametric SFH (Primary) & Ratio of the SFRs in adjacent time bins with & Student’s t-distribution\tnote{e} \\
            & five ($N_{\mathrm{SFR}} - 1$) free parameters & \\
            Parametric SFH & Delayed-tau model with one free parameter & Log uniform\tnote{f} \\
            $n$ & Power-law modifier to the shape of the & Uniform\tnote{g} \\
            & \citet{Calzetti:2000} diffuse dust & \\
            & attenuation curve & \\
            $\tau_{\mathrm{dust},\,1}$ & Birth-cloud dust optical depth & Clipped normal\tnote{h} \\
            $\tau_{\mathrm{dust},\,2}$ & Diffuse dust optical depth & Clipped normal\tnote{i} \\
            $\mathrm{log}_{10}( Z_{\mathrm{gas}}/Z_{\odot} )$ & Gas-phase metallicity & Uniform\tnote{j}  \\
            $\mathrm{log}_{10}( U )$ & Ionization parameter for nebular emission & Uniform\tnote{k} \\
            $f_\mathrm{IGM}$ & Scaling of the IGM attenuation curve & Clipped normal\tnote{l} \\
		\hline
	\end{tabular}
	\begin{tablenotes}
	    \footnotesize
            \item \textbf{Notes}
            \item[a] Fixed value at $z = z_{\mathrm{spec}}$.
            \item[b] Uniform prior with min$= -2.00$, max$= +0.19$.
            \item[c] Uniform prior with min$= +6$, max$= +12$.
            \item[d] Student's t-distribution prior with $\sigma = +1.0$, $\nu = +2.0$.
            \item[e] Student's t-distribution prior with $\sigma = +0.3$, $\nu = +2.0$.
            \item[f] Log uniform prior with min$= +0.1$, max$= +30.0$.
            \item[g] Uniform prior with min$= -1.0$, max$= +0.4$.
            \item[h] Clipped normal prior in $\tau_{\mathrm{dust},\,1}/\tau_{\mathrm{dust},\,2}$ with min$= +0$, max$= +2$, $\mu = +1.0$, $\sigma = +0.3$.
            \item[i] Clipped normal prior with min$= +0$, max$= +4$, $\mu = +0.3$, $\sigma = +1.0$.
            \item[j] Uniform prior with min$= -2.0$, max$= +0.5$.
            \item[k] Uniform prior with min$= -4.0$, max$= -1.0$.
            \item[l] Clipped normal prior with min$= +0$, max$= +2$, $\mu = +1.0$, $\sigma = +0.3$.
        \end{tablenotes}
	\end{threeparttable}
	}
	\hspace*{+4mm}
\end{table*}

\section{Physical Properties of Final Spectroscopic Sample}
\label{AppendixThree}

\setcounter{table}{0}
\renewcommand{\thetable}{\thesection\arabic{table}}

Table~\ref{tab:physical_properties} gives a summary of the physical properties for the $81$ objects that are part of the final spectroscopic sample described in Section~\ref{SectionTwoThree}.

\begin{table*}
	\caption{A summary of the physical properties for the $81$ objects in our final spectroscopic sample.}
	\label{tab:physical_properties}
	\hspace*{-12mm}
        \makebox[\textwidth]{
	\begin{threeparttable}
	\begin{tabular}{cccccccccc} 
		\hline
		\hline
		Index & R.A. & Decl. & $z_{\mathrm{spec}}$ & $M_{\mathrm{UV}}$ & $f_{\mathrm{H}\alpha}$ & $\mathrm{log}_{10}(M_{\ast}/M_{\odot})$ & $\mathrm{SFR}_{0-100\,\mathrm{Myr}}$ & Type \\
		& (J2000) & (J2000) & & (mag) & ($10^{-19}\ \mathrm{erg/s/cm^{2}}$) & & ($M_{\odot}/\mathrm{yr}$) & \\
		\hline
            1 & 53.06177 & -27.84238 & $5.207$ & $-18.57 \pm 0.04$ & $14.2 \pm 2.2$ & $8.21 \pm 0.21$ & $0.8 \pm 0.6$ & Field \\
            2 & 53.07461 & -27.85649 & $5.207$ & $-19.43 \pm 0.07$ & $19.7 \pm 3.7$ & $8.64 \pm 0.24$ & $2.1 \pm 1.7$ & Field \\
            3 & 53.11271 & -27.83827 & $5.213$ & $-20.34 \pm 0.02$ & $24.9 \pm 3.9$ & $9.17 \pm 0.15$ & $11.1 \pm 6.0$ & Field \\
            4 & 53.16858 & -27.73726 & $5.236$ & $-18.76 \pm 0.03$ & $29.9 \pm 3.2$ & $7.85 \pm 0.05$ & $0.8 \pm 0.1$ & Field \\
            5 & 53.15489 & -27.81150 & $5.247$ & $-19.00 \pm 0.07$ & $24.8 \pm 3.5$ & $9.33 \pm 0.18$ & $5.6 \pm 3.4$ & Field \\
            6 & 53.18831 & -27.81283 & $5.248$ & $-19.44 \pm 0.03$ & $14.1 \pm 2.1$ & $8.01 \pm 0.12$ & $0.6 \pm 0.3$ & Field \\
            7 & 53.15105 & -27.78294 & $5.260$ & $-17.96 \pm 0.09$ & $27.1 \pm 2.8$ & $8.29 \pm 0.33$ & $1.5 \pm 0.9$ & Field \\
            8 & 53.17023 & -27.76296 & $5.263$ & $-19.57 \pm 0.04$ & $198.6 \pm 5.7$ & $8.99 \pm 0.33$ & $5.3 \pm 5.5$ & Field \\
            9 & 53.16758 & -27.76550 & $5.264$ & $-18.59 \pm 0.04$ & $17.1 \pm 5.6$ & $9.43 \pm 0.09$ & $29.6 \pm 5.4$ & Field \\
            10 & 53.17080 & -27.76230 & $5.264$ & $-19.34 \pm 0.05$ & $245.5 \pm 5.4$ & $9.08 \pm 0.13$ & $8.3 \pm 3.7$ & Field \\
            11 & 53.16646 & -27.74697 & $5.265$ & $-19.70 \pm 0.04$ & $16.4 \pm 3.4$ & $8.53 \pm 0.20$ & $1.9 \pm 1.1$ & Field \\
            12 & 53.17183 & -27.73771 & $5.268$ & $-17.98 \pm 0.10$ & $28.9 \pm 4.1$ & $8.76 \pm 0.19$ & $2.2 \pm 1.4$ & Field \\
            13 & 53.17531 & -27.84117 & $5.292$ & $-20.49 \pm 0.02$ & $53.0 \pm 5.0$ & $8.67 \pm 0.13$ & $2.9 \pm 1.3$ & Field \\
            14 & 53.13173 & -27.84999 & $5.312$ & $-18.54 \pm 0.09$ & $31.9 \pm 3.7$ & $8.97 \pm 0.10$ & $2.8 \pm 1.4$ & Field \\
            15 & 53.08803 & -27.81320 & $5.316$ & $-19.86 \pm 0.05$ & $44.0 \pm 4.1$ & $8.41 \pm 0.25$ & $1.4 \pm 1.0$ & Field \\
            16 & 53.13081 & -27.84689 & $5.320$ & $-18.75 \pm 0.10$ & $31.7 \pm 3.6$ & $9.85 \pm 0.13$ & $15.6 \pm 10.8$ & Field \\
            17 & 53.13174 & -27.84711 & $5.321$ & $-18.75 \pm 0.09$ & $24.0 \pm 3.2$ & $8.70 \pm 0.11$ & $3.2 \pm 1.0$ & Field \\
            18 & 53.15685 & -27.86069 & $5.327$ & $-19.18 \pm 0.02$ & $11.1 \pm 2.2$ & $8.20 \pm 0.15$ & $0.6 \pm 0.3$ & Field \\
            19 & 53.18328 & -27.77894 & $5.333$ & $-17.06 \pm 0.07$ & $14.8 \pm 4.4$ & $8.77 \pm 0.15$ & $0.8 \pm 0.7$ & Field \\
            20 & 53.15584 & -27.76672 & $5.348$ & $-20.66 \pm 0.02$ & $21.1 \pm 3.0$ & $7.96 \pm 0.12$ & $0.7 \pm 0.3$ & Field \\
            21 & 53.08698 & -27.84807 & $5.358$ & $-19.67 \pm 0.02$ & $39.3 \pm 5.1$ & $7.85 \pm 0.11$ & $0.7 \pm 0.3$ & Field \\
            22 & 53.07408 & -27.80401 & $5.374$ & $-19.57 \pm 0.04$ & $71.6 \pm 3.8$ & $8.70 \pm 0.19$ & $3.9 \pm 3.5$ & Group 1 \\
            23 & 53.12644 & -27.79200 & $5.375$ & $-18.07 \pm 0.08$ & $15.7 \pm 2.6$ & $7.63 \pm 0.11$ & $0.4 \pm 0.1$ & Field \\
            24 & 53.12775 & -27.78098 & $5.376$ & $-18.60 \pm 0.02$ & $14.6 \pm 3.4$ & $8.30 \pm 0.17$ & $1.1 \pm 0.6$ & Field \\
            25 & 53.07486 & -27.80461 & $5.378$ & $-18.01 \pm 0.05$ & $18.9 \pm 5.0$ & $9.00 \pm 0.20$ & $4.7 \pm 2.4$ & Group 1 \\
            26 & 53.07444 & -27.80484 & $5.378$ & $-21.28 \pm 0.02$ & $29.6 \pm 5.2$ & $9.31 \pm 0.21$ & $6.0 \pm 4.5$ & Group 1 \\
            27 & 53.06799 & -27.80816 & $5.378$ & $-19.71 \pm 0.02$ & $23.0 \pm 4.2$ & $8.99 \pm 0.14$ & $1.7 \pm 1.8$ & Group 1 \\
            28 & 53.07497 & -27.80445 & $5.378$ & $-17.06 \pm 0.07$ & $18.4 \pm 4.4$ & $8.52 \pm 0.17$ & $2.1 \pm 1.3$ & Group 1 \\
            29 & 53.07500 & -27.80421 & $5.378$ & $-20.66 \pm 0.02$ & $33.3 \pm 6.6$ & $9.58 \pm 0.12$ & $9.3 \pm 6.8$ & Group 1 \\
            30 & 53.06784 & -27.81850 & $5.379$ & $-19.44 \pm 0.03$ & $33.3 \pm 3.0$ & $8.93 \pm 0.17$ & $4.8 \pm 2.6$ & Group 1 \\
            31 & 53.07483 & -27.80478 & $5.379$ & $-18.88 \pm 0.05$ & $260.0 \pm 6.2$ & $9.61 \pm 0.17$ & $25.0 \pm 14.7$ & Group 1 \\
            32 & 53.16729 & -27.75273 & $5.379$ & $-18.82 \pm 0.04$ & $21.1 \pm 5.3$ & $8.48 \pm 0.26$ & $1.2 \pm 1.0$ & Field \\
            33 & 53.07625 & -27.80607 & $5.380$ & $-20.08 \pm 0.04$ & $39.5 \pm 7.0$ & $8.28 \pm 0.12$ & $1.9 \pm 0.9$ & Group 1 \\
            34 & 53.08113 & -27.82613 & $5.380$ & $-20.34 \pm 0.02$ & $9.8 \pm 1.9$ & $7.88 \pm 0.20$ & $0.4 \pm 0.3$ & Group 2 \\
            35 & 53.07353 & -27.81488 & $5.381$ & $-19.00 \pm 0.07$ & $18.2 \pm 3.9$ & $9.22 \pm 0.24$ & $16.0 \pm 6.4$ & Group 1 \\
            36 & 53.09642 & -27.85309 & $5.381$ & $-19.86 \pm 0.03$ & $21.5 \pm 2.9$ & $9.20 \pm 0.15$ & $17.6 \pm 2.3$ & Group 2 \\
            37 & 53.07495 & -27.80481 & $5.381$ & $-19.33 \pm 0.03$ & $104.7 \pm 3.0$ & $9.23 \pm 0.13$ & $13.8 \pm 5.8$ & Group 1 \\
            38 & 53.07497 & -27.80453 & $5.382$ & $-17.96 \pm 0.09$ & $20.5 \pm 6.1$ & $8.81 \pm 0.15$ & $3.0 \pm 2.1$ & Group 1 \\
            39 & 53.12557 & -27.86563 & $5.382$ & $-17.92 \pm 0.09$ & $17.7 \pm 4.3$ & $8.15 \pm 0.16$ & $1.3 \pm 0.7$ & Field \\
            40 & 53.07421 & -27.80500 & $5.383$ & $-19.79 \pm 0.02$ & $35.5 \pm 4.7$ & $9.57 \pm 0.27$ & $13.6 \pm 9.9$ & Group 1 \\
		\hline
	\end{tabular}
	\end{threeparttable}
	}
	\hspace*{+4mm}
\end{table*}
\addtocounter{table}{-1}
\begin{table*}
	\caption{Continued.}
	\hspace*{-12mm}
        \makebox[\textwidth]{
	\begin{threeparttable}
	\begin{tabular}{cccccccccc} 
		\hline
		\hline
		Index & R.A. & Decl. & $z_{\mathrm{spec}}$ & $M_{\mathrm{UV}}$ & $f_{\mathrm{H}\alpha}$ & $\mathrm{log}_{10}(M_{\ast}/M_{\odot})$ & $\mathrm{SFR}_{0-100\,\mathrm{Myr}}$ & Type \\
		& (J2000) & (J2000) & & (mag) & ($10^{-19}\ \mathrm{erg/s/cm^{2}}$) & & ($M_{\odot}/\mathrm{yr}$) & \\
		\hline
            41 & 53.07877 & -27.79750 & $5.384$ & $-19.34 \pm 0.05$ & $12.4 \pm 4.0$ & $8.11 \pm 0.19$ & $0.5 \pm 0.4$ & Group 1 \\
            42 & 53.10304 & -27.85386 & $5.385$ & $-17.92 \pm 0.09$ & $13.0 \pm 3.8$ & $8.87 \pm 0.28$ & $3.1 \pm 2.6$ & Group 2 \\
            43 & 53.09525 & -27.82278 & $5.385$ & $-19.67 \pm 0.03$ & $7.6 \pm 2.0$ & $9.15 \pm 0.17$ & $4.5 \pm 3.0$ & Group 2 \\
            44 & 53.10221 & -27.82234 & $5.387$ & $-19.86 \pm 0.05$ & $17.8 \pm 3.5$ & $8.63 \pm 0.21$ & $1.8 \pm 1.2$ & Group 2 \\
            45 & 53.08534 & -27.83268 & $5.387$ & $-19.43 \pm 0.07$ & $39.8 \pm 6.6$ & $7.89 \pm 0.14$ & $0.7 \pm 0.2$ & Group 2 \\
            46 & 53.10413 & -27.82042 & $5.388$ & $-18.75 \pm 0.10$ & $39.4 \pm 3.9$ & $8.52 \pm 0.13$ & $2.4 \pm 1.2$ & Group 2 \\
            47 & 53.10605 & -27.83743 & $5.388$ & $-18.60 \pm 0.02$ & $18.3 \pm 3.6$ & $8.79 \pm 0.15$ & $3.5 \pm 1.9$ & Group 2 \\
            48 & 53.10665 & -27.82834 & $5.389$ & $-18.39 \pm 0.04$ & $36.9 \pm 3.0$ & $8.30 \pm 0.10$ & $1.5 \pm 0.5$ & Group 2 \\
            49 & 53.10604 & -27.83732 & $5.390$ & $-19.70 \pm 0.04$ & $29.4 \pm 6.0$ & $9.68 \pm 0.19$ & $13.5 \pm 8.3$ & Group 2 \\
            50 & 53.10905 & -27.83919 & $5.390$ & $-19.67 \pm 0.02$ & $10.0 \pm 3.2$ & $8.12 \pm 0.19$ & $0.8 \pm 0.5$ & Group 2 \\
            51 & 53.06866 & -27.83498 & $5.390$ & $-18.57 \pm 0.04$ & $82.6 \pm 5.0$ & $9.35 \pm 0.27$ & $9.0 \pm 6.6$ & Group 2 \\
            52 & 53.10537 & -27.83920 & $5.390$ & $-18.60 \pm 0.05$ & $102.8 \pm 1.3$ & $8.69 \pm 0.21$ & $2.0 \pm 1.5$ & Group 2 \\
            53 & 53.10435 & -27.84056 & $5.390$ & $-18.07 \pm 0.08$ & $18.7 \pm 3.4$ & $8.87 \pm 0.08$ & $7.7 \pm 1.8$ & Group 2 \\
            54 & 53.08870 & -27.83335 & $5.391$ & $-19.70 \pm 0.04$ & $54.3 \pm 4.8$ & $9.17 \pm 0.20$ & $7.5 \pm 6.3$ & Group 2 \\
            55 & 53.10661 & -27.82919 & $5.392$ & $-20.49 \pm 0.02$ & $22.2 \pm 5.1$ & $8.99 \pm 0.11$ & $3.0 \pm 1.6$ & Group 2 \\
            56 & 53.10431 & -27.84020 & $5.392$ & $-18.54 \pm 0.09$ & $168.7 \pm 7.1$ & $9.34 \pm 0.16$ & $15.4 \pm 7.3$ & Group 2 \\
            57 & 53.08068 & -27.83515 & $5.392$ & $-18.82 \pm 0.04$ & $15.8 \pm 3.5$ & $7.83 \pm 0.01$ & $0.7 \pm 0.0$ & Group 2 \\
            58 & 53.10476 & -27.83229 & $5.392$ & $-18.75 \pm 0.09$ & $69.8 \pm 3.4$ & $8.68 \pm 0.15$ & $2.8 \pm 1.5$ & Group 2 \\
            59 & 53.11063 & -27.83967 & $5.392$ & $-18.50 \pm 0.03$ & $26.0 \pm 7.2$ & $9.35 \pm 0.17$ & $5.9 \pm 4.6$ & Group 2 \\
            60 & 53.07701 & -27.83456 & $5.393$ & $-19.86 \pm 0.02$ & $19.4 \pm 6.1$ & $8.76 \pm 0.12$ & $2.0 \pm 1.2$ & Group 2 \\
            61 & 53.08831 & -27.84042 & $5.393$ & $-19.18 \pm 0.02$ & $69.7 \pm 4.5$ & $8.70 \pm 0.16$ & $3.4 \pm 1.6$ & Group 2 \\
            62 & 53.10879 & -27.81817 & $5.393$ & $-20.01 \pm 0.05$ & $26.0 \pm 8.0$ & $8.14 \pm 0.13$ & $1.1 \pm 0.7$ & Group 2 \\
            63 & 53.08071 & -27.83532 & $5.396$ & $-18.59 \pm 0.04$ & $53.3 \pm 5.2$ & $9.04 \pm 0.19$ & $4.8 \pm 2.7$ & Group 2 \\
            64 & 53.08903 & -27.84190 & $5.398$ & $-18.13 \pm 0.10$ & $19.0 \pm 4.6$ & $9.35 \pm 0.12$ & $9.3 \pm 5.9$ & Group 2 \\
            65 & 53.14764 & -27.84205 & $5.401$ & $-19.86 \pm 0.02$ & $45.0 \pm 3.0$ & $9.10 \pm 0.14$ & $14.0 \pm 1.3$ & Field \\
            66 & 53.11184 & -27.84066 & $5.411$ & $-18.39 \pm 0.04$ & $19.3 \pm 4.1$ & $8.38 \pm 0.26$ & $1.4 \pm 0.9$ & Field \\
            67 & 53.12247 & -27.79652 & $5.442$ & $-19.79 \pm 0.02$ & $20.5 \pm 5.2$ & $7.75 \pm 0.14$ & $0.5 \pm 0.3$ & Field \\
            68 & 53.16407 & -27.79972 & $5.444$ & $-20.01 \pm 0.05$ & $80.7 \pm 3.5$ & $9.28 \pm 0.06$ & $19.0 \pm 4.6$ & Field \\
            69 & 53.12874 & -27.79788 & $5.444$ & $-20.08 \pm 0.04$ & $12.6 \pm 3.5$ & $8.10 \pm 0.13$ & $0.6 \pm 0.3$ & Field \\
            70 & 53.11671 & -27.79395 & $5.448$ & $-19.86 \pm 0.03$ & $16.7 \pm 4.4$ & $7.62 \pm 0.16$ & $0.4 \pm 0.2$ & Field \\
            71 & 53.11439 & -27.79211 & $5.449$ & $-18.13 \pm 0.10$ & $43.3 \pm 5.1$ & $8.01 \pm 0.14$ & $1.0 \pm 0.4$ & Field \\
            72 & 53.18044 & -27.77066 & $5.454$ & $-19.70 \pm 0.04$ & $16.6 \pm 4.7$ & $7.69 \pm 0.08$ & $0.5 \pm 0.2$ & Field \\
            73 & 53.16577 & -27.78490 & $5.464$ & $-18.01 \pm 0.05$ & $18.4 \pm 4.4$ & $7.83 \pm 0.11$ & $0.6 \pm 0.2$ & Field \\
            74 & 53.16904 & -27.78769 & $5.465$ & $-18.50 \pm 0.03$ & $11.2 \pm 3.1$ & $7.85 \pm 0.15$ & $0.6 \pm 0.4$ & Field \\
            75 & 53.16611 & -27.78574 & $5.466$ & $-18.88 \pm 0.05$ & $50.6 \pm 3.1$ & $8.84 \pm 0.12$ & $3.7 \pm 1.7$ & Field \\
            76 & 53.13859 & -27.79025 & $5.480$ & $-21.28 \pm 0.02$ & $105.9 \pm 4.9$ & $10.59 \pm 0.08$ & $434.8 \pm 35.4$ & Field \\
            77 & 53.12819 & -27.78769 & $5.481$ & $-19.33 \pm 0.03$ & $11.5 \pm 2.2$ & $7.79 \pm 0.15$ & $0.5 \pm 0.2$ & Field \\
            78 & 53.09595 & -27.81077 & $5.484$ & $-19.71 \pm 0.02$ & $20.0 \pm 4.4$ & $8.47 \pm 0.19$ & $1.4 \pm 1.0$ & Field \\
            79 & 53.11543 & -27.83347 & $5.484$ & $-19.67 \pm 0.03$ & $21.6 \pm 3.3$ & $8.06 \pm 0.09$ & $1.2 \pm 0.2$ & Field \\
            80 & 53.06055 & -27.84840 & $5.496$ & $-18.60 \pm 0.05$ & $19.4 \pm 2.4$ & $9.41 \pm 0.13$ & $28.8 \pm 1.5$ & Field \\
            81 & 53.13767 & -27.75528 & $5.499$ & $-20.54 \pm 0.03$ & $29.6 \pm 4.5$ & $8.95 \pm 0.02$ & $11.2 \pm 0.2$ & Field \\
		\hline
	\end{tabular}
	\end{threeparttable}
	}
	\hspace*{+4mm}
\end{table*}

\addtocounter{table}{-1}

%% End of the document.
\end{document}